\documentclass[final,3p,times,twocolumn]{elsarticle}

\usepackage{graphicx}
\usepackage{amssymb}
\usepackage{amsthm}
\usepackage{amsmath}
\usepackage{mathrsfs}
\usepackage{lineno}
\usepackage{xcolor}
\usepackage[flushleft]{threeparttable}

\def\etal{{\it et al.\/}}
\def \postrevisionbf{\bf}

\journal{Icarus}

\begin{document}

\begin{frontmatter}


\title{Constraints on the Initial Mass, Age and Lifetime of Saturn's Rings from 
Viscous Evolutions that Include Pollution and Transport due to 
Micrometeoroid Bombardment}



\author[Ames,SETI]{Paul R. Estrada}
\author[IndU]{Richard H. Durisen}

\address[Ames]{NASA Ames Research Center, Moffett Field, CA 94035}
\address[SETI]{Carl Sagan Center, SETI Institute, Mountain View CA 94043}
\address[IndU]{Department of Astronomy, Indiana University, Bloomington, IN 47405}

\begin{abstract}

The Cassini spacecraft provided key measurements during its more than twelve 
year mission that constrain the absolute age of Saturn's rings. These include the extrinsic
micrometeoroid flux at Saturn, the volume fraction of non-icy pollutants in the rings, and a measurement
of the ring mass. These observations taken together limit the ring {\{{exposure}} age to be $\lesssim$ a few 100 Myr 
if the flux was persistent over that time \citep{Kem23}. 
In addition, Cassini observations during the Grand Finale further indicate the rings 
are losing mass \citep{Hsu18,Wai18} suggesting the rings are ephemeral
as well. In a companion paper \citep{DE23}, we show that the effects of micrometeoroid bombardment and
ballistic transport of their impact ejecta can account for these loss rates for reasonable parameter
choices. In this paper, we conduct numerical simulations of an evolving ring in a systematic way in 
order to determine initial conditions that are consistent with these observations.

We begin by revisiting the ancient massive ring scenario of \citet[][Icarus 209, 771-785]{Sal10}.
Here,  we model not just the viscous evolution, but we subject the ring to pollution by micrometeoroid 
bombardment over the age of the Solar System. 
We find that regardless of initial mass, the ring always ends up with 
more pollutant than is currently observed, because the ring spends the majority of its lifetime
at relatively low mass where it is most susceptible to darkening. We then show that models with initial disk masses 
of $\sim 1-3$ Mimas masses reach volume fractions of pollutant consistent with the observed volume fractions 
of non-icy material in the A and B rings within a time scale of $\sim$ a few 100 Myr. 

Finally, we use the analysis of \citet{DE23} to add 
the dynamical effects of meteoroid bombardment into the evolution equations, 
namely, mass loading and ballistic transport. The treatment 
of mass loading is exact, while ballistic transport is handled in an approximate way. Simulations show
that: 
(1) mass loading and ballistic 
transport applied to an initially high optical depth annulus inevitably produce a lower density 
C ring analog interior to the annulus; and (2) high density rings subject to persistent micrometeoroid bombardment 
do not have an asymptotic mass but instead have an asymptotic lifetime much shorter than the
age of the Solar System. This is because micrometeoroid bombardment and ballistic transport drive the dynamical evolution of the ring once viscosity weakens, indicating that the exposure age of the rings and their dynamical age are connected. 


%

\end{abstract}


\end{frontmatter}


\section{Introduction}
\label{sec:intro}

All of the giant planets of the Solar System have ring systems, the majority of which are 
relatively low mass and composed of ringlets, arcs or gossamer rings and whose photometric 
characteristics indicate that they are spectrally dark and that water ice is essentially absent 
or not a dominant compositional constituent \citep{Nic84,Por87,Por95,deP18}. 
This is in stark contrast to the Saturnian rings which are relatively 
massive and composed of $> 95$\% water ice by mass \citep{Gro90,Doy89,CE98,Zha17a,Zha17b}. 
That Saturn's rings are so massive and icy is almost certainly a clue to their origin and age.
Let us briefly review some current ideas.

\subsection{Ring Origin}
\label{subsec:ringorigin}

Several models have been proposed for the origin of Saturn's rings. Primordial scenarios
consider the rings to be unaccreted remnants from the Saturnian subnebula
\citep{Pol76}, which requires the rings to survive gas
drag removal long enough for the subnebula to dissipate. A more recent idea envisions
that initially massive rings could have formed from the tidally stripped icy mantle of a Titan-sized 
differentiated moon at the tail end of satellite accretion \citep{Can10}. 
%
In this model, a Titan-sized, differentiated moon migrates via planetary tides and gas 
tidal-torques inside Saturn's Roche limit where its mantle is stripped, while its core is lost 
to the planet. The rubble disk derived from the mantle forms a new, primarily icy Rhea-sized 
moon {\it outside} the Roche limit (but inside the synchronous radius) that 
would eventually migrate back inward after the gas disk had mostly dissipated, but still billions of years ago, to be tidally disrupted into a massive ring \citep[see supp. material,][]{Can10}. 

Other ring origin scenarios require sizable interlopers of heliocentric origin. While these
events are not all ``primordial", they are usually envisioned to happen billions of years ago
and still lead to ``ancient" rings. 
The idea traces back to \citet{Har84}, who suggested that the rings 
were derived from a previously formed icy moon. The first variant considers the rings to be the 
remnant of a collisionally disrupted Mimas-mass moon that migrated inward due to  
gas tidal torque and/or gas drag and was left stranded in a nearly circular orbit near Saturn's Roche 
limit when the gas disk dissipated \citep[e.g.,][]{ME03a,ME03b}. Disruption of this moon
probably requires a heliocentric impactor of tens of kilometers in size \citep{Cha09},
because Saturn's tides alone may not be able to disrupt and grind down a cold and {\it solid} 
Mimas-mass moon unless it is very close to the planet \citep[within the C ring,][]{GT82,Dav99}. 
It has been shown \citep{Cha09} that a Mimas-sized moon located 
$10^5$ km from Saturn can be destroyed with significant probability during the Late Heavy Bombardment 
(LHB) some 700 Myr after Saturn formed \citep{Tsi05}. The key is keeping the moon there 
long enough for the disruption event to occur\footnote{Saturn's current synchronous orbit is well below the 
Roche limit, thus such a moon would avoid the fate of falling onto the planet. However, with Saturn's updated 
low tidal $Q$ values \citep{Lai12,Lai17}, it will migrate outwards rather quickly using traditional tidal theory which is problematic{{. \citet{Lai20} suggest these fast migration rates may be consistent with the resonance-lock mechanism \citep{Ful16}}}.}.
\citet{Dub19} recently
modeled this scenario in which he assumes that a Mimas-mass body remained close to Saturn because
it was trapped in mean motion resonance with Enceladus and Dione.

Another scenario suggests that the rings may be the remnants from a tidally disrupted comet \citep{Don91,Don07}.
In this picture, a large ($\gtrsim 200-300$ km) comet or Centaur that passes very 
close to Saturn may be tidally disrupted with the rings forming from some of the debris. \citet{Cha09} 
also considered this scenario at the LHB and showed that tens of Mimas masses could have been 
passed through Saturn's Hill sphere, with a fraction of this material ending up within Saturn's Roche radius. 
Although there are caveats \citep[see][]{Cha09,Cha18}, ideas like these are considered
the most plausible ``extrinsic" models for ring origin. A significant problem arises, however, 
for rings that form late in the arc of Solar System history, because the number of suitable interlopers drops off steeply after the LHB \citep{Zah03}, making ring formation more and more
implausible. 

\subsection{Ring Age}
\label{subsec:ringage}

Although the origin of Saturn's rings remains unclear, 
it now seems much more certain when they most likely formed  
thanks to several key observations provided by the Cassini mission 
(summarized in Table \ref{tab:observe}). 
First, from gravity field measurements in the final orbits of the Cassini Grand Finale, the mass of the 
rings has been determined to be $\sim 0.4$ Mimas masses \citep{Ies19}, lower than the 
post-Voyager estimate of a Mimas mass \citep{Esp84}, but similar to estimates from wavelet-based 
analyses of several density waves in the B ring \citep{HN16}. This mass is much lower 
than what had been envisioned earlier in the mission when it was discovered that the rings had a 
clump-and-gap structure. Numerical simulations of dense rings had suggested that density waves could hide 
large amounts of material leading to ring masses an order of magnitude or more than their current 
mass \citep{Ste07,Rob10}, but apparently this is not the case.

\begin{table*}
\renewcommand{\thetable}{\arabic{table}}
\centering
\caption{Summary of Relevant Cassini Observations} \label{tab:observe}
\vspace{0.1in}
\begin{threeparttable}
\begin{tabular}{lll}
\hline
\hline

Observation & Instrument & Derived Value \\
\hline
\tnote{a} Flux ($\dot{\sigma}_\infty$) & CDA & $2.2\cdot 10^{-16}$ kg m$^{-2}$ s$^{-1}$ \\
\tnote{b} Ring Mass   & RSS & $(1.54 \pm 0.49) \cdot 10^{19}$ kg\\
\tnote{c} Non-icy Fraction ($\upsilon$)   & radiometer & $\sim 1-2$\%, C ring/Cassini division \\
  &    &  $\sim 0.1 - 0.5$\%, B and A rings \\
Mass Inflow & \tnote{d} INMS & $4800 - 45000$ kg s$^{-1}$ \\
  &  \tnote{e} CDA  &  $\sim 320 - 1200$ kg s$^{-1}$ \\
  & \tnote{f} MIMI &  $\sim 5$ kg s$^{-1}$ \\

\hline

\end{tabular}
 \begin{tablenotes}
  \item[a] \citet{Kem23}
  \item[b] \citet{Ies19}; Cassini Radio Science Subsystem (RSS)
  \item[c] \citet{Zha17a,Zha17b}
  \item[d] \citet{Wai18}
  \item[e] \citet{Hsu18}
  \item[f] \citet{Mit18}; Magnetospheric Imaging Instrument (MIMI)
 \end{tablenotes}
\end{threeparttable}
 
\end{table*}

Second, \citet{Zha17a,Zha17b}, using Cassini radiometer microwave data, showed that 
the volume fraction of non-icy material in the rings, or ``pollutant'', was found to be
$\sim 1-2\%$ in the darker, low optical depth C ring and Cassini division and $\sim 0.1-0.5\%$ in the
brighter and optically thicker A and B rings. These fractions are consistent
with those derived from Voyager observations \citep{Doy89,CE98},
HST \citep{Cuz18} and VIMS results \citep{Cia19}, if the pollutant is volumetrically
intramixed within the icy grains as tiny inclusions\footnote{This appears consistent with observations 
using Cassini's Magnetospheric Imaging Instrument (MIMI) and the Cosmic Dust Analyzer, which 
found that the mass influx of material from the rings falling into Saturn when passing 
through the region between the D ring and the upper atmosphere was composed of nano-grain particles 
\citep{Mit18,Hsu18}.}. While visible and IR wavelengths only sample 
the surface layers of the particles, the microwave measurements sample the bulk of the ring particles, so 
that the radial variation of pollutants across the rings indeed indicates that the rings overall are not 
very polluted, and, moreover, have a distinct contrast in the level of pollution between low and high optical depth 
regions.

Third, observations using the Cassini Cosmic Dust Analyzer (CDA) alleviated the uncertainty 
about a critical parameter
for determining the ring age by measuring the extrinsic micrometeoroid flux at Saturn \citep{Kem23}.
CDA found that the flux entering the Hill sphere (which we refer to as ``at infinity'') smaller by a factor of two \citep{Alt15,Alt18,Kem17,Kem23} than the previously accepted value 
\citep{Gru85, CD90, CE98}. 
Moreover, orbit reconstruction of the collected impactors shows that the population 
is consistent with an origin in the Edgeworth-Kuiper belt (EKB) and not the Oort cloud, as  
previously assumed. Particles from this dynamical population have much lower mean velocities (at infinity) as they 
enter Saturn's Hill sphere 
than cometary particles and thus are about ten times more focused gravitationally by the planet \citep{Est18}.
This implies that 
the impact flux on the rings is 
an order of magnitude higher than previously estimated 
for cometary particles (See Sec. \ref{subsec:basiceq}).

Moreover, measurements during the Cassini Grand Finale orbits, in which the spacecraft flew 
through the 2000 km region between Saturn's D ring and the upper atmosphere, revealed that 
the rings are losing mass at a surprising rate \citep{Hsu18,Wai18,Mit18}. Some of the mass 
flux falls as ``ring rain'' at higher latitudes consistent with the H$_3^+$ infrared 
emission pattern thought to be produced by an influx of charged water products from the 
rings \citep{Odo19}. However, the contribution needed to account for the ring rain phenomenon 
is considerably less than the total measured mass influx using the Ion 
Neutral Mass Spectrometer (INMS) 
of $4800-45000$ kg s$^{-1}$ \citep{Wai18}, requiring mechanisms other than the loss of small 
charged particles to explain. 
These observations imply that the rings are ephemeral.

Because of their huge surface-area-to-mass ratio, about $\sim 10^4-10^5$ times more than a moon
of equal mass \citep{Est15}, the rings are continuously subjected to bombardment 
by extrinsic micrometeoroids. This has two main effects. First, micrometeoroid bombardment
(MB) can lead to structural evolution of the rings both due to direct deposition of meteoroids and due to the ejecta
produced from their impacts with ring particles. The ejecta cause mass and angular momentum transfer 
because they are often re-absorbed at different locations than where they originate \citep{Dur84,Dur89}, a
process referred to as ballistic transport \citep{Ip84}. 
\citet{DE23} have recently shown using the CDA measured flux \citep{Kem23} 
that these effects of MB 
are large enough to explain mass inflow rates of the magnitude measured by \citet{Wai18}.
Second, MB causes icy rings to become more and more darkened with time 
due to pollution from the non-icy component of the incoming meteoroids \citep{CE98}. 

Given these effects of MB, the Cassini measurements taken together constrain the ring age to be $\lesssim$ a few 100 Myr, 
provided the micrometeoroid flux has been sustained more or less at its currently observed value over that time \citep{Kem23,DE23}. The current flux of heliocentric bodies that are large enough 
to either destroy a resident, differentiated Mimas-mass moon, or to themselves become a ring parent body by
tidal disruption, gives a probability of only $\sim 10^{-4}$ for ring formation in the last $\sim 10^8$
years \citep[e.g.,][]{Lis88,Ip88,Don91,Cha09}, apparently ruling out such an origin for the rings. 
So none of the 
scenarios reviewed in 
Section \ref{subsec:ringorigin} seem viable, and we currently lack an accepted recent origin scenario \citep{Cri19}. The origin of Saturn's rings remains an intriguing problem.

\subsection{Ring Evolution}
\label{subsec:ringevolution}

The age estimates cited in Section \ref{subsec:ringage} are calculated for a static ring
and do not take into account ring dynamical evolution. It is important to determine whether age estimates
stand up to a proper evolutionary treatment. So this paper reports 
numerical simulations of a ring evolving under the influence of viscosity and subject to the various effects of MB.
Previously, \citet{Sal10} investigated the global and long-term 1D viscous evolution of an initially 
massive ring over the age of the Solar System. They used 
a realistic viscosity model that accounted for enhancement of the viscosity 
by gravitational instabilities \citep{Dai01}. We employ a
similar model here, 
%
but we also follow the evolution of the volume fraction of pollutants over time as
the ring evolves. We investigate both primordial/ancient ring scenarios 
and more recent, lower initial mass rings in an effort to determine which are more 
consistent with the observations. In these first models, we only consider the 
direct mass deposition (indicated as DD) of the impactors in addition to the pollutant they deliver, 
and thus these models evolve dynamically solely due to viscosity.

We then add the dynamical effects of mass loading (ML) and ballistic transport 
(BT) to the numerical evolutions. In this context, by mass loading we
mean the mass deposition {\it and} the subsequent angular momentum changes that result as a consequence of the impacting meteoroid. Both ML and BT cause an inexorable
inward radial drift and thus an influx of ring material toward the central planet. 
We find that this eventually drives the rings' dynamical evolution over viscosity for rings whose masses 
become below a few to several Mimas masses. While ML can be included exactly in the simulations,
modeling BT in global simulations is complicated \citep{Dur92,Est15,Est18} 
and beyond the scope of this initial study. Instead, we introduce BT approximately in a few
of our models
by adopting an approach similar to that of \citet{DE23}. Even though mass is conserved in 
this approach, it cannot capture the structural changes imposed by BT \citep[e.g., see][]{Est15,Est18} due to 
the mass exchanges between different ring regions, which are required for angular momentum conservation. 
Thus when BT effects are included herein, they should only be considered qualitatively as an indication of the magnitude and direction of its effects, and should be taken with 
caution. We plan to include BT properly in future work. 

This paper is organized as follows. In Section 2, we describe our numerical model by first introducing the
basic equations of ballistic transport and deriving the radial drift velocity components. We describe
the form of the viscosity law for conditions where the disk is and is not self-gravitating. We
also describe the numerical methods including how we follow the evolution of pollutants. In Section 3, we 
present our simulations. We first model a massive ring under the influence of {\postrevisionbf{only}} viscosity over the age of the Solar System, but include the consequences of pollution by micrometeoroids. 
We follow this with similar lower mass cases that are evolved over only a few 100 Myr. Finally,
we add the dynamical effects of ML (and BT) to the viscous simulations, and show how ML (and BT)
profoundly affect ring evolution. In Section 4, we discuss the implications of our findings.
Section \ref{sec:mainconclusions} provides a brief summary of our main conclusions.

\section{Ring Evolution Model}
\label{sec:model}

In the following sections, we describe the set of equations used to follow the physical
and compositional evolution of the rings due to viscosity and micrometeoroid bombardment. 
%
Our plan in this paper 
is to simulate the rings first using viscosity and direct deposition of polluting 
material only, and later to add the dynamical effects of mass loading and ballistic transport to demonstrate their influence on how fast the rings evolve. A summary of the simulations 
performed for this work is presented in Table \ref{tab:models}.
  
\begin{table*}
\renewcommand{\thetable}{\arabic{table}}
\centering
\caption{Summary of Simulations} \label{tab:models}
\vspace{0.1in}
\begin{tabular}{lcccccc}
\hline
\hline

Model type & Eq. solved & $M$(M$_{\rm{Mimas}}$) & $a$ (m) &  $F_{\rm{g}}$
& $\eta$ (\%) & Figure \\
\hline
V & (7), $\dot{\sigma}_{\rm{im}}=0$ & $1-100$ & 1 & ... &  ... & 1 \\
DD & (7) & 100 & 1 & 3 & 10 & 3\\
DD & (7) & 100 & 1 & 3 & $10,100$ & 4,6a,7a \\
DD & (7) & 100 & 1 & 30 & $10,100$ & 5,6b,7b \\
DD & (7) & $1-100$ & 1 & $3,30$ & $10,100$ & 8 \\
ML & (13), $x{\cal{R}}{\cal{P}}=0$ & 1 & 1 & 30 & 10 & 9\\
ML & (13), $x{\cal{R}}{\cal{P}}=0$ & 2 & 1 & 30 & 10 & 10\\
ML & (13), $x{\cal{R}}{\cal{P}}=0$ & 3 & 1 & 30 & 10 & 11\\
ML & (13), $x{\cal{R}}{\cal{P}}=0$ & 1 & 5 & 30 & 10 & 12\\
ML & (13), $x{\cal{R}}{\cal{P}}=0$ & $1-2$ & $1-15$ & 30 & 10 & 13\\
ML[+BT] & (13) & 1 & 1 & 30 & 10 & 14 \\
ML[+BT] & (13) & 2 & 1 & 30 & 20 & 16 \\ 
ML[+BT] & (13) & $1-1000$ & 1 & 30 & 10 & 17,18 \\
 
\hline
\end{tabular}
\end{table*}

\subsection{Basic Equations}
\label{subsec:basiceq}

Our approach in this paper is strictly 1D. We only address the radial evolution of
ring material and neglect azimuthal and vertical variations. The equation governing
the time-variation of the surface mass density $\Sigma$ under the influence of micrometeoroid
bombardment and viscosity is given by \citep{Est18}

\begin{equation}
\label{equ:mequ}
    \frac{\partial \Sigma}{\partial t} + \frac{1}{r}\frac{\partial}{\partial r}\left(r\Sigma 
    v_{\rm{r}}\right) = \Gamma_{\rm{m}}-\Lambda_{\rm{m}} + \dot{\sigma}_{\rm{im}},
\end{equation}

\noindent
where $\Gamma_{\rm{m}}$ is the net gain rate of mass per unit area at some radius $r$ due to
absorption of micrometeoroid impact ejecta emitted from other radial locations, $\Lambda_{\rm{m}}$ is the 
net loss rate of mass per unit area at $r$ due to ejecta emitted at $r$ and absorbed at other radial locations,
and $v_{\rm{r}}$ is the total radial velocity of ring material due to viscosity and the effects of MB. 

The two-sided micrometeoroid flux 
$\dot{\sigma}_{\rm{im}}$ that impacts the rings at $r$ is

\begin{equation}
\label{equ:sigim}
    \dot{\sigma}_{\rm{im}} \simeq 2 \dot{\sigma}_\infty {\mathscr{A}} F_{\rm{g}}(r/r_0)^{-0.8},
\end{equation}

\noindent
where $\dot{\sigma}_\infty$ is the single-sided, flat plate flux entering Saturn's Hill sphere,
$F_{\rm{g}}$ accounts for the gravitational focusing of the micrometeoroid flux 
by Saturn \citep{CD90,CE98} normalized to a reference radius $r_0 = 1.8$ R$_S$ 
($R_{\rm{S}} = 60,268$ km is Saturn's equatorial radius), and the function 
${\mathscr{A}}(\tau) = \left[1 - e^{-(\tau/\tau_s)^p}\right]^{1/p}$ is a parameterized fit to the numerically determined impact probability \citep{CD90,Dur96,Est15,Est18} as a function of the optical depth $\tau$.
The fit parameters are $\tau_s=0.515$ and $p=1.0335$ \citep{CE98}. 
The $(r/r_0)^{-0.8}$ {{factor}} comes from a fit to the 
radial dependence of the focusing effect in \citet{CD90}.
Most all previous
modeling of the effects of MB on the rings \citep[e.g.,][]{Dur92,CE98,Est15} have assumed the 
population of impactors are of Oort cloud origin, and adopted the derived value of 
$\dot{\sigma}_\infty = 4.5\cdot 10^{-16}$ kg m$^{-2}$ s$^{-1}$ \citep[{{see}}][]{Ip84,Gru85,CD90,CE98}. This
population which is isotropic in the heliocentric frame is characterized by a large velocity on
entering the Hill sphere so that gravitational focusing is small with the value of $F_{\rm{g}}\approx 3$
\citep[e.g., see][]{CE98}.

The micrometeoroid flux as measured by Cassini is (see Table \ref{tab:observe}) $\dot{\sigma}_\infty = 2.2 \cdot 10^{-16}$ kg m$^{-2}$ s$^{-1}$ \citep{Kem23}.
%
\noindent
Although in the same ballpark as the previously derived flux, there are important 
differences. First,
the dynamical origin of the population is the EKB (Sec. \ref{subsec:ringage}), which have 
isotropic velocities in the frame of the planet. 
As a consequence their mean velocity when entering the Hill sphere is considerably 
smaller and $F_{\rm{g}} \approx 30$ \citep[][see also \citealt{Mor83}]{Kem23}. Thus, for a given $\dot{\sigma}_\infty$, ten times more mass is hitting the rings with time, and consequently the rings 
are more susceptible to pollution. 
Second, it has been previously assumed that the composition of the cometary grains
is half water ice. For the EKB population, the volatile component of most grains of the
radii detected ($\lesssim 20 \mu$m) are not expected to survive their journey to Saturn because
the sublimation temperature is reached at 10 AU \citep{MM02,Gri07}.
So we assume that all of the EKB micrometeoroids are composed of potential pollutant (Sec. \ref{subsubsec:polevol}). We noted above that the gravitational enhancement by the planet in Eq. (\ref{equ:sigim}) is a numerical fit to the radial dependence of the focusing derived from the calculations of \citet{CD90}, and is for an Oort cloud population. As pointed out in \citet{DE23}, a more careful treatment for the EKB population should be done along these lines, but the order of magnitude difference in focusing between the two populations remains valid, and has been understood for some time \citep{Mor83}. 

The continuity equation for the areal angular momentum density can be used to determine an
expression for the radial velocity $v_{\rm{r}}$ \citep{Est15}:

\begin{equation}
\label{equ:amequ}
    \frac{\partial}{\partial t}\left(h_c\Sigma\right) + \frac{1}{r}\frac{\partial}{\partial r}
    \left(r h_c\Sigma v_{\rm{r}}\right) = -\frac{1}{2\pi r}\frac{\partial g}{\partial r} + \Gamma_{\rm{h}} -
    \Lambda_{\rm{h}} + \dot{J}_{\rm{im}}.
\end{equation}
 
\noindent
Here, $h_c = r^2\Omega =(GMr)^{1/2}$ is the specific orbital angular momentum for circular motion, $M$ is the
central planet mass, and $G$ the gravitational constant. $\Gamma_{\rm{h}}$ and $\Lambda_{\rm{h}}$
are the gain and loss rates of angular momentum per unit area at $r$ due to ejecta gained and lost,
$\dot{J}_{\rm{im}}$ is the direct deposition rate of angular momentum  
per unit area by meteoroids \citep[{{which is}} generally small and taken to be zero here; see][and below]{Dur96}, and 
$g=-2\pi r^3\nu\Sigma(d\Omega/dr)$ is the viscous angular momentum flux in terms of the kinematic
viscosity $\nu$. From Equations (\ref{equ:mequ}) and (\ref{equ:amequ}), one can easily find the
radial velocity contributions due to viscosity, ballistic transport, and mass loading (plus associated
torques) $v_{\rm{r}}=v_{\rm{r}}^{\rm{visc}} + v_{\rm{r}}^{\rm{ball}} + v_{\rm{r}}^{\rm{load}}$, where

\begin{equation}
\label{equ:vvisc}
    v_{\rm{r}}^{\rm{visc}} = -\frac{1}{2\pi r\Sigma}\left[\frac{dh_c}{dr}\right]^{-1}\frac{\partial g}
    {\partial r} = -\frac{3}{r^{1/2}\Sigma}\frac{\partial}{\partial r}\left(r^{1/2}\Sigma \nu \right),
\end{equation}
 
\begin{equation}
\label{equ:vball}
    v_{\rm{r}}^{\rm{ball}} = \frac{1}{\Sigma}\left[\frac{dh_c}{dr}\right]^{-1}\left[\Gamma_{\rm{h}}-
    \Lambda_{\rm{h}} - h_c\left(\Gamma_{\rm{m}}-\Lambda_{\rm{m}}\right)\right],
\end{equation}

\begin{equation}
\label{equ:vload}
    v_{\rm{r}}^{\rm{load}} = \frac{1}{\Sigma}\left[\frac{dh_c}{dr}\right]^{-1}\left(\dot{J}_{\rm{im}}-
    h_c\dot{\sigma}_{\rm{im}}\right).
\end{equation}

As a first effort to model the simultaneous ring evolution due to viscosity and pollution by
extrinsic micrometeoroids, we will ignore those terms in Eq. (\ref{equ:mequ}) due to mass loading 
and ballistic transport and retain only the impact flux (Eq. [\ref{equ:sigim}]). As indicated previously, this means that for viscous only evolutions (DD), we will add the mass due to impacting meteoroids, but will ignore the radial drift due to mass loading (Eq. [\ref{equ:vload}]). 
Treatment of mass inflow due to ML (which includes DD and $v^{\rm{load}}_{\rm{r}}$) will be discussed in a subsequent section. Eq. (\ref{equ:mequ}) reduces to the more simplified form

\begin{equation}
\label{equ:viscmequ}
    \frac{\partial \Sigma}{\partial t} = \frac{3}{r}\frac{\partial}{\partial r}\left[r^{1/2}
    \frac{\partial}{\partial r}\left(r^{1/2}\Sigma \nu\right)\right] + \dot{\sigma}_{\rm{im}},
\end{equation}
 
\noindent 
for the time evolution of the total surface density, and with the exception of the $\dot{\sigma}_{\rm{im}}$ term is the same as that utilized by \citet{Sal10} in their simulations. Equation (\ref{equ:viscmequ}) can be solved using a variety of techniques, given a model for the viscosity which we describe next.

\subsection{Viscosity Model}
\label{subsec:viscosity}

The rings are treated as a Keplerian disk with negligible pressure 
where the viscosity leading to angular momentum transport arises from particle interactions. 
The various contributions to the viscosity can be divided into a local shear stress component due to particle 
random motions $\nu_{\rm{L}}$ \citep{GT78a}, a non-local component $\nu_{\rm{NL}}$ 
(also referred to as a collisional viscosity) that comes as a result of the momentum transferred 
through physical collisions {{across}} ring particles \citep{AT86}, and, when
the rings are sufficiently dense, a component $\nu_{\rm{grav}}$ due to gravitational scattering 
in self-gravity wakes \citep[e.g.,][]{Sal92,DI99}. The former two components are
comparatively weak and only dominate viscous evolution when the {{ring small-scale transport is not dominated by self-gravity wakes}}, 
whereas in
a dense, self-gravitating disk, the last component dominates and evolution can be extremely rapid. 
The gravitational state of the disk is described by the Toomre parameter $Q_{\rm{T}} = c\Omega/3.36G\Sigma$, 
where $c$ is the radial velocity dispersion of ring particles. The transition from gravitating to non-self
gravitating occurs as $Q_{\rm{T}}$ decreases below $\sim 2$ \citep[e.g.,][]{Sal95}.

A simple prescription employed in this work \citep[][for a detailed discussion, see 
\citealt{Sch09}]{Sal10} is given by 

\begin{equation}
\label{equ:visc}
\begin{split}
    \nu = \nu_{\rm{L}} + \nu_{\rm{NL}} + \nu_{\rm{grav}} = \;\;\;\;\;\;\;\;\;\\
    \begin{cases}
    \frac{c^2}{2\Omega}\frac{0.46\tau}{1+\tau^2} + a^2\Omega \tau & \, Q_{\rm{T}} > 2;\\
    26\left(r_h^*\right)^5\frac{G^2\Sigma^2}{\Omega^3} + a^2\Omega \tau & \, Q_{\rm{T}} \leq 2\\
    \end{cases},
    \end{split}
\end{equation}

\noindent
where $r_h^* = r_h/2a = r(\pi \rho/9M)^{1/3}$ is the ratio of the mutual Hill sphere $r_h$ of two 
identical particles of mass $m$ and density $\rho$ to their mutual radii $a$ \citep{Dai01}. 
The velocity dispersion $c$ depends on the value of $r_h^*$: if $r_h^* \leq 1/2$, the velocity dispersion
is regulated by collisions and is $c \sim 2a\Omega$, whereas when $r_h^* > 1/2$ the dispersion
velocity is determined by the escape velocity from the particle surface, $c \sim (Gm/a)^{1/2}$
\citep{Sal95,DI99,Oht99}. A slight discontinuity can occur in the transition between the self-gravitating and non-self-gravitating viscosity laws about
the value of $Q_{\rm{T}} = 2$ in Eq. (\ref{equ:visc}), 
so we make the choice of using a simple linear smoothing scheme to transition between them 
over a range of $\Delta Q_{\rm{T}} \pm 0.5$. However, the absence of smoothing does not affect the results.

%

The viscosity will depend on a variety of ring parameters, {{the}} most obvious of which is the mass surface
density and its radial variation. Equation (\ref{equ:visc}) is generally applied
to an ensemble of same-sized particles, so that the dynamical optical depth $\tau = 3\Sigma/4\rho a$.
Given the relative simplicity of the the simulations we present herein, we will only explore cases that
employ a single particle size and leave more complicated treatments for the future.
The viscosity
also depends on the porosity $\phi$ of the ring particles through their density, which itself evolves 
as the rings become more polluted. In this paper, we will 
model cases in which the ring particles are solid ($\phi = 0$), but more generally we will 
choose $\phi = 0.75$ in agreement with fits to ring thermal emission \citep{Zha17a,Zha17b}.  

We recognize that there may be uncertainties associated with 
adoption of this particular viscosity formulation, but our intention here is to use essentially the
same prescription as in \citet{Sal10} so that our results are 
directly comparable, and we can better discern effects of the additional physics of pollution, ML,
and BT.

\subsection{Mass Loading and Ballistic Transport}
\label{subsec:MLBT}

As discussed above in Sec. \ref{subsec:basiceq}, mass loading (ML) due to micrometeoroid 
bombardment not only deposits mass and with it pollutant, it also causes radial {{drifts}}  
due to the change it produces in the specific angular momentum of a ring region. These drifts are inward 
\citep{DE23}. 

Furthermore, MB leads to ballistic transport (BT) due to the copious ejecta from impacts 
that cause exchange of mass and angular momentum between ring regions. 
Because meteoroid impacts preferentially occur on the leading hemispheres
of ring particles, the ejecta from the impacts tend to be prograde and thus decrease
the specific orbital angular momentum of a ring region, causing it to drift inwards
\citep{Dur89,Dur92,Dur95,CD90,Est15,Est18}. This inward drift is derived in detail for a uniform ring 
evolved under BT alone in \citet{DE23} using Eqs. (\ref{equ:mequ}) and (\ref{equ:vball}).
$\Sigma$ stays constant to high order, but $v_{\rm{r}}^{\rm{ball}}$ is negative. The small negative 
direct mass exchange in Eq. (\ref{equ:mequ}) represented by $\Gamma_{\rm{m}}-\Lambda_{\rm{m}}$ 
is balanced to high-order accuracy by the divergence of the flux term on the LHS of  Eq. (\ref{equ:mequ}) 
contributed by the non-zero, negative $v_{\rm{r}}^{\rm{ball}}$.

Thus a direct consequence of micrometeoroid bombardment is that both ML and BT lead to mass inflow in the rings. As shown in \citet{DE23}, this inward transport of material from the B ring to the C ring is probably large enough 
to account for the mass influx of material observed by Cassini \citep{Wai18,Hsu18}.

Inclusion of ML is straightforward. In addition to the nonzero $\dot{\sigma}_{\rm{im}}$ 
in Eq. (\ref{equ:mequ}), we include the radial drift $v_{\rm{r}}^{\rm{load}}$
given by Eq. (\ref{equ:vload}). For this paper, as in \citet{DE23}, we make the simplification
in Eq. (\ref{equ:vload}) that {{the}} torque due to meteoroid deposition $\dot{J}_{\rm{im}}$ is zero. 
The actual value of $\dot{J}_{\rm{im}}$ is likely to be negative due to aberration and slant
path effects, but the total contribution of angular momentum deposition to radial drift is relatively 
modest \citep[see, for example, Fig. 1 of][]{Dur96}. For the viscosity plus ML models discussed
in Section \ref{subsec:mlandbt}, all the $\Gamma$'s and $\Lambda$'s plus $v_{\rm{r}}^{\rm{ball}}$ are set
to zero in the evolution equations.

BT is more complex. The $\Gamma$'s and $\Lambda$'s due to ejecta exchanges are 
multiple integrals that depend on ejection direction and speed and contain several functions that
depend on both $r$ and $\tau$ both at the point of ejection and at the point of absorption \citep[see, for example,][]{Est15}. 
Typical distances over which ejecta are thrown range from
10's to 1,000's of kilometers.  The multiple integrals increase the expense of calculations
and the smaller scales involved are not well represented by our adopted grid-spacing of
78 km (see \ref{subsubsec:gensol}). Full implementation of BT is left to future work. Instead,
we adopt {{a}} simplified approach that captures the scale and direction of BT's effect
on global ring evolution.

\citet{DE23} considered a quasi-steady uniform ring where $\Sigma$ and $\tau$ are constant {{\citep[see also][]{Dur89,Est15,Est18}}} 
and simplified the loss and gain integrals using approximations previously 
made by \citet{Lis84} and \citet{Dur95}. They found that the
induced inward radial drift due to ballistic transport (Eq. [\ref{equ:vball}]) can be expressed as 

\begin{equation}
\label{equ:vballDE}
 \Sigma v_{\rm{r}}^{\rm{ball}} = -2 x r {\cal{R}}{\cal{P}},
\end{equation}

\noindent
where $x = v_{\rm{ej}}/v_{\rm{K}}$ is the ratio of ejecta velocity $v_{\rm{ej}}$ to the local 
orbital speed $v_{\rm{K}}$.  ${\cal{R}}$ and ${\cal{P}}$ are the $\tau$-dependent 
local ejecta mass emission rate, in units of mass per unit area per unit time, and the ejecta 
absorption probability\footnote{As discussed in \citet{Dur95} and \citet{Lat12}, Eq. (\ref{equ:Pemit}) for the relative probabilities of absorption at the radius of ejection, or at some other radius of interception is a good approximation for a uniform ring of optical depth $\tau$. Making this approximation simplifies the linear analysis which we utilize in this paper \citep{DE23}. In the limit of $\tau \rightarrow 0$, Eq. (\ref{equ:Pemit}) says that absorption at either location is equally probable.}. These are given by

\begin{equation}
\label{equ:Remit}
    {\cal{R}} = {\cal{R}}_\infty \left(1 - e^{-\tau/\tau_0} + \frac{\tau}{\tau_0}
    e^{-\tau/\tau_0}\right).
\end{equation}

\begin{equation}
\label{equ:Pemit}
    {\cal{P}} = \frac{1 - e^{-\tau/\tau_p}}{1 - e^{-2\tau/\tau_p}};
\end{equation}

\noindent
In the above equations, $\tau_p \approx 0.5$ represents a typical slant path through the rings, $\tau_0 = 0.28$ and 

\begin{equation}
\label{equ:scriptr0}
  {\cal{R}}_\infty = 2\dot{\sigma}_\infty Y F_{\rm{g}}(r/r_0)^{-0.8}.
\end{equation} 

\noindent
The functional form of Eq. (\ref{equ:scriptr0}) was determined originally by
\citet{CD90}. 

It should be noted that $v_{\rm{r}}^{\rm{ball}}$ (similarly for 
$v_{\rm{r}}^{\rm{load}}$) is asymptotically non-zero as $\tau \rightarrow 0$ with both approaching a finite, but
{\it maximally negative} value for the drift velocity. 
That is to say, the fastest inward drifts occur at the lowest $\tau$ \citep[see also,][]{Dur96,
DE23}. This can easily be seen, for instance, with $v_{\rm{r}}^{\rm{ball}}$ in Eq. (\ref{equ:vballDE}).
As $\tau \rightarrow 0$, the $\cal{R}$ on the RHS is proportional to $\tau$. For constant opacity, 
$\tau$ is proportional to $\Sigma$. So, as $\tau \rightarrow 0$, the $\Sigma$ on the RHS is canceled
by a $\Sigma$ on the LHS.

The magnitude of BT effects is governed by the parameters in 
Eq. (\ref{equ:scriptr0}). The CDA results reported by \citet{Kem23}
constrain both $F_{\rm{g}}$ and $\dot{\sigma}_\infty$ (Sec. \ref{subsec:basiceq}) with
the latter being a factor of two lower than
the value $4.5 \cdot 10^{-16}$ kg m$^{-2}$ s$^{-1}$
adopted in \citet{Dur92} and \citet{Est15,Est18}. 
The most uncertain parameter is the ejecta yield $Y$ 
from a hypervelocity cratering event on a ring particle, defined as the ratio of the 
ejecta mass to the impactor mass. For the impact speeds expected in Saturn's rings,
$Y$ can typically range from $10^4$ for impacts into solid targets,
to $10^6$ for granular or powdery surfaces. \citep[see][and references therein]{Est15}. 
For this work, we use a value of $Y=10^4$, though structural simulations using full BT indicate values as high as $Y \simeq 10^5$ are needed to maintain the inner B ring edge
\citep{Est15,Est18}. Thus our choice may be conservative. We refer the reader to \citet{DE23} for discussion of the uncertainties in $Y$, especially as it may pertain to particle 
porosity.

With this approximation for $v_{\rm{r}}^{\rm{ball}}$, 
and including mass loading (Eq. [\ref{equ:vload}]) as well, we can rewrite Eq. (\ref{equ:mequ}) as

\begin{equation}
\label{equ:btmequ}
\begin{split}
    \frac{\partial \Sigma}{\partial t} = \frac{3}{r}\frac{\partial}{\partial r}\left[r^{1/2}
    \frac{\partial}{\partial r}\left(r^{1/2}\Sigma \nu\right)\right] + \;\;\;\;\;\;\;\;\; \\ 
    \frac{2}{r}\frac{\partial}{\partial r}\left[r^2\left(x{\cal{R}}{\cal{P}} 
    + \dot{\sigma}_{\rm{im}}\right)\right] 
    + \dot{\sigma}_{\rm{im}}.
    \end{split}
\end{equation}

\noindent
Equations (\ref{equ:vballDE}) and (\ref{equ:btmequ}) are valid for a single $x$, though ejecta will have a distribution of velocities. We thus choose an average value based on assuming 
an $x^{-3}$ power-law speed distribution from some
lower bound $x_b$ to an upper bound $x_t \gg x_b$, as used in previous models of ballistic transport 
\citep{Dur92,Est15}. The {{weighted average}} $x = 2x_b$, with typically $x_b = 10^{-4}$ 
\citep[see][]{Est15}. {{This $x_b$ corresponds to an ejecta velocity of}} $2$ m s$^{-1}$ at the current radial location
of the inner B ring edge. The upper bound on ejecta velocities {{for}} $x_t$ are typically up to a few 100 m s$^{-1}$
\citep[see][for more discussion]{DE23}.

There are aspects of this BT approximation that need to be explained. We have omitted the net 
loss and gains due to mass exchanges between different ring regions $\Gamma_{\rm{m}}-\Lambda_{\rm{m}}$ 
(first term on the RHS of Eq. [\ref{equ:mequ}])\footnote{Contained in these terms are all the aspects of BT that lead to the formation, or maintenance of quasi-steady-state structures in the rings \citep[e.g., see][]{Est18}, which nonetheless occur over the global inward transport of material which we demonstrate in this paper.}. We cannot meaningfully compute those terms
here without treating boundary regions properly and without a much higher grid resolution. 
On the other hand, the prograde nature of the ejecta must inexorably cause {{an}} inward drift of 
ring material, and this is important to 
include, even in an approximate way. Using Eq. (\ref{equ:vballDE}) is similar to using
the steady-state accretion disk inflow rate $\dot{M} = 3\pi\nu\Sigma$ to estimate {{the}} viscous
flow in an accretion disk. 
This is only strictly applicable to a steady-state disk \citep[see, for example,][]{Har09} 
but is useful as a rough estimate for the mass  inflow in the non-steady accretion case. 
In the same spirit, Eq. (\ref{equ:vballDE}) provides a ballpark of the disk inflow 
even when a BT-active disk is not in a steady state, but we cannot properly account for the 
detailed mass restructuring. Setting $\Gamma_{\rm{m}}-\Lambda_{\rm{m}}$ to zero
at least ensures mass conservation. But angular momentum is not conserved using 
Eq. (\ref{equ:vballDE}) except for the idealized case of a strictly uniform ring.
So our treatment of BT, while capturing the magnitude and
$\tau$-dependence of the mass influx, is incomplete, and thus results 
including BT in this paper should be viewed with some caution.

In summary, it should be emphasized that our treatment of ML is 
physically exact and {\it does} conserve mass and angular momentum properly as long as 
the the external deposition torque $\dot{J}_{\rm{im}}$ is zero. Typically, $\dot{J}_{\rm{im}} < 0$
can arise from the asymmetric absorption of micrometeoroids by ring particles or from
asymmetric mass loss due to impacts, for example by production of hot vapor \citep{Dur96}. 
The latter could even influence $v_{\rm{r}}^{\rm{ball}}$. 
%
Apart from this external torque issue, ML is not only modeled in a
physically precise way, but the parameters that determine its magnitude are much
better constrained than for BT. 
For BT, the treatment we employ here is informative and effectively 
captures the influence of BT in a approximate way. While the magnitude of BT-induced 
radial drifts is more uncertain, primarily due to the uncertainty in impact yield $Y$, 
the drifts are likely to be greater than those induced by ML \citep{DE23}.

\subsection{Numerical Methods}
\label{subsec:nummethod}

\subsubsection{General Solution}
\label{subsubsec:gensol}

We solve Eq. (\ref{equ:viscmequ}) using a variable time-step, semi-implicit finite-differencing scheme with
source terms which are second-order in space and time. The method is only semi-implicit because viscosity
is a complicated function of the surface density and optical depth so that expressing these time-variable
expressions in tri-diagonal form is not possible. For this reason, we employ a variable time step that
is determined from satisfying a Courant-like stability condition that depends on the viscous diffusion time 
across a bin, namely

\begin{equation}
\label{equ:stabcrit}
    \Delta t = 0.2 \frac{(\Delta r)^2}{3 \nu_{\rm{max}}},
\end{equation}

\noindent
where $\Delta r$ is the uniform grid spacing and
$\nu_{\rm{max}}$ is the maximum value of the viscosity at any location in the grid. This step-size criterion
provides satisfactory stability with time steps ranging from a few days (from the initial state) to
up to thousands of years at the later stages when {{the}} disk viscosity is weak. Our simulations are
conducted over the age of the Solar System for massive ring cases, and over several 100 Myr for rings
of recent origin.

We follow closely the work of \citet{Sal10} and set our radial grid size to $N=1001$ bins over a radial domain ranging from $r_{\rm{inner}} = 61,000$ km 
to $r_{\rm{outer}} = 139,000$ km so that our grid bin size is $\Delta r = 78$ km. The initial annulus of 
ring material can vary in width. For models in which we make direct comparisons to the results of \citet{Sal10}, we choose their initial annulus width of $\sim 3000$ km centered about $110,000$ km.
For our massive disk cases, we choose an 
initial radial extent that also lies well within the grid, typically from $100,000$ to $120,000$ km. 
For our less massive, recent ring scenarios, the annuli may be thinner and varied in location. 
These will be pointed out when discussing the particular models. 


Once the ring material reaches the inner and outer boundary, it is (1) assumed to be lost to the
planet at the inner boundary and (2) assumed to contribute to the formation of an icy moon
when lost at the outer boundary \citep{Cha10}. In either case, the boundary condition is determined by continuation
of the mass flux (Neumann boundary condition) across the boundary using ghost points that are
extrapolated from the grid values. Since mass is removed, we specifically choose $\Sigma = 0$ there. 

\begin{figure}
 \resizebox{\linewidth}{!}{%
 \includegraphics{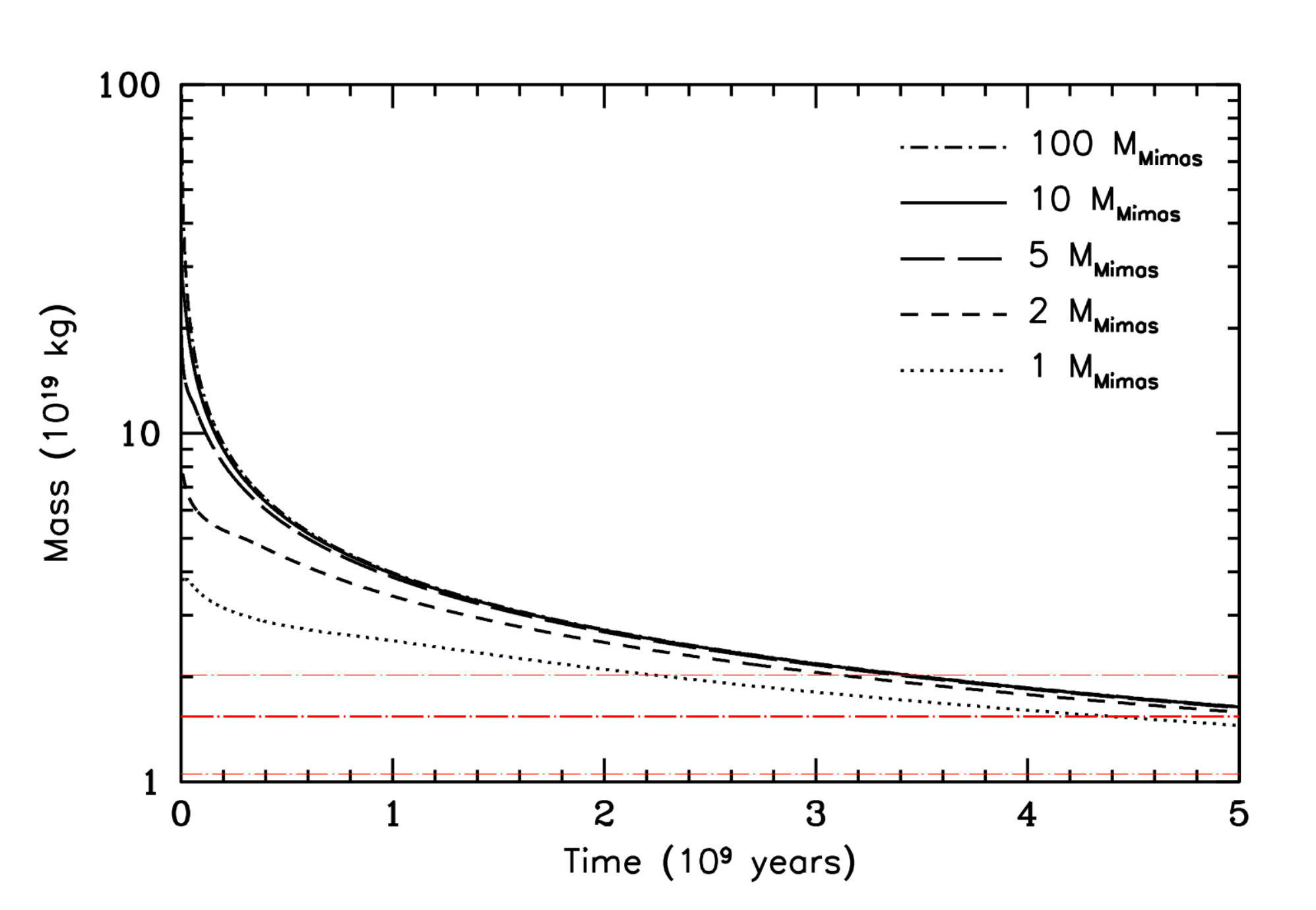}}
 \vspace{-0.3in}
\caption{Ring mass as a function of time for different initial masses, which can be 
compared with Fig. 10 of \citet{Sal10}. The mass loss for the most massive cases 
($M \geq 5$ M$_{\rm{\rm Mimas}}$) occurs very rapidly, indicating that the rings spend most of 
their lifetime at low mass. As demonstrated by \citet{Sal10}, initially massive rings evolve 
viscously to an asymptotic mass near the currently observed mass of $1.54 \pm 0.49$ 
kg ($0.41 \pm 0.13$ M$_{\rm{Mimas}}$) of 
the rings \citep[bold dot-dashed red line;][]{Ies19}. We plot the corresponding error bars given 
by the bracketing red dot-dashed lines for completeness. There is good agreement between our 
computational approach 
and that \citet{Sal10}. }
\label{fig:salmon}
\vspace{-0.1in}
\end{figure}

Our solution method differs from \citet{Sal10}, who used a 1D finite
elements code on a staggered mesh combined with
an explicit, second-order Runge-Kutta scheme. As a test of our code, we reproduce the results
from Fig. 10 of \citet{Sal10} in {\bf Figure \ref{fig:salmon}} where we plot the evolution of disk mass
over 5 Gyr for initial disk masses of 1, 2, 5, 10 and also for 100 $M_{\rm{\rm Mimas}}$ under viscosity alone (labeled as model V, see Table \ref{tab:models}). 
For these simulations, the mass deposition term  $\dot{\sigma}_{\rm{im}}$ is set to
zero in Eq. (\ref{equ:mequ}) and $v_r =  v_{\rm{r}}^{\rm{visc}}$ only. We have used a mean particle size 
$a = 1$ m, and the same initial annulus width and starting location ($\sim 3000$ km, centered at $110,000$ km) 
as in their simulations.

The agreement between the final masses in our evolutions and those in \citet{Sal10} 
are within a few percent, 
%
%
while the overall qualitative behavior is the same in both sets of simulations. 
The disk initially spreads very quickly, 
especially for a massive ring, filling the domain {{after which time the ring begins to rapidly lose}} 
mass through the 
boundaries. The most massive case (100 M$_{\rm{Mimas}}$, dash-dotted curve) has already 
lost 90\% of its mass after the first $\sim 10^7$ years and is $\sim 3.5$ M$_{\rm{Mimas}}$ by $10^8$ 
years. 
Similar rapid loss rates are observed for the $M \geq 5$ M$_{\rm{Mimas}}$ cases, indicating 
that even in a massive ancient ring scenario, the ring spends the {{majority}} 
of its lifetime at 
low mass, which has strong implications for rings that are subject to micrometeoroid pollution. 
We also see that for all of the chosen initial disk masses, the rings viscously evolve to an asymptotic mass 
as found in \citet{Sal10} over {{the age of the Solar System.}} 
Like these authors, we see little evolution after 4 Gyr and all ring masses are close to $\sim 1.54 \cdot 10^{19}$ kg after 5 Gyr which is their current mass determined by Cassini \citep[bold red dot-dashed line;][]{Ies19}. The corresponding error bars for their determined mass are given by the bracketing red dot-dashed lines.

\begin{figure}
 \resizebox{\linewidth}{!}{%
 \includegraphics{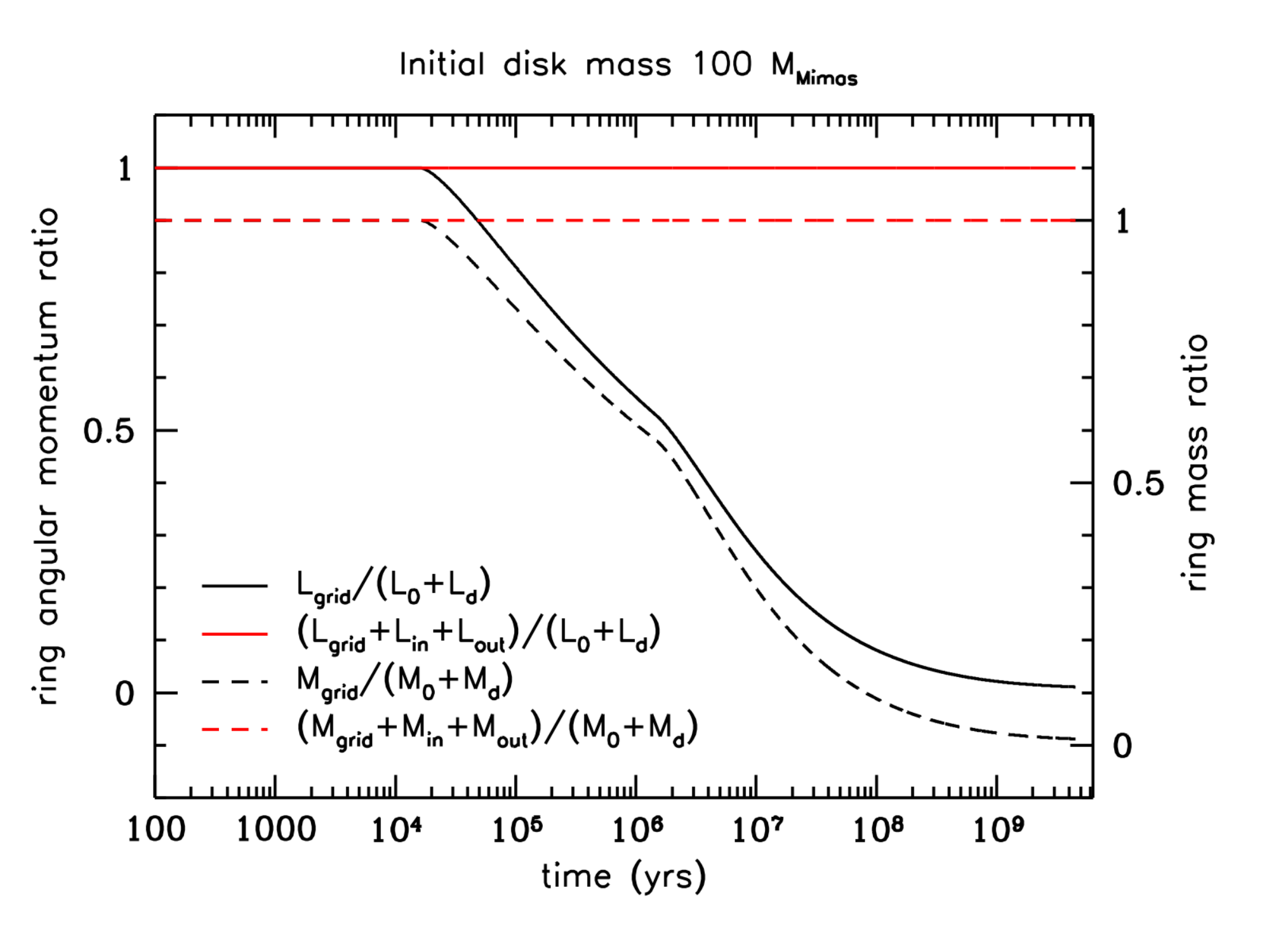}}
 \vspace{-0.3in}
\caption{Conservation of mass and angular momentum in a simulation
that includes the EKB flux (Sec. \ref{subsubsec:polevol}) for an initial disk mass of
100 M$_{\rm Mimas}$ {{using the same initial conditions as in Fig. \ref{fig:salmon}}}. Plotted are the evolving ring mass and angular momentum
$M_{\rm{grid}}$ and $L_{\rm{grid}}$ (black curves) and their sum with mass and angular momentum
lost through the inner and outer boundaries (red curves) in ratio to the total initial and deposited mass 
and angular momentum associated with micrometeoroid bombardment. }
\label{fig:conserve}
\vspace{-0.1in}
\end{figure}

Our numerical code conserves mass and angular momentum well when the ring material 
evolves only under the influence of viscosity. The total mass and angular momentum of the 
ring at all times is calculated from summing over all bins

\begin{equation}
    M_{\rm{grid}} = 2\pi \int_{r_{\rm{inner}}}^{r_{\rm{outer}}} \Sigma r\,dr = 2\pi \sum_{i=1}^{N}
    \Sigma_i r_i \Delta r,
\end{equation}

\begin{equation}
\begin{split}
   L_{\rm{grid}} = 2\pi \int_{r_{\rm{inner}}}^{r_{\rm{outer}}} \Sigma h_c r\,dr = \;\;\;\;\;\;\\
   2\pi \sum_{i=1}^{N}
    \Sigma_i \sqrt{GM} r_i^{3/2} \Delta r.
    \end{split}
\end{equation}

%

\noindent
In {\bf Figure \ref{fig:conserve}}, we show the conservation of mass and angular momentum for 
a calculation using Eq. (\ref{equ:viscmequ}) presented in Sec. \ref{subsec:basiceq} in which the initial disk 
mass is $100$ M$_{\rm Mimas}$, and we include the mass deposition due to the MB but not radial drifts (DD). 
We plot the ratio of the mass and angular momentum $M_{\rm{grid}},L_{\rm{grid}}$ in the ring (black 
curves), as well as sums that include the mass (e.g., $M_{\rm{in}} = \sum_j \dot{M}_{{\rm{grid}},j}(r_{\rm{inner}})\cdot \Delta t_j$) and angular momentum lost through the inner 
and outer boundaries (red curves), compared to the total (initial + deposited from bombardment),
$M_0+M_{\rm{d}},L_0+L_{\rm{d}}$\footnote{The angular momentum $L_{\rm{d}}$ 
of the deposited meteoroid mass is accounted for in the viscous only simulations, 
because, for that case, deposited meteoroid mass is assumed to join the 
rings  at $r$ with a specific angular momentum equal to the circular orbit value at $r$. 
In other words, $v_{\rm{r}}^{\rm{load}}$ is assumed
to be zero in viscosity only simulations. We tally accumulated meteoroid mass in order to
measure pollution rates, but we do not allow ML to produce a
radial drift.}. The outer edge of the ring encounters the 
outer boundary of the grid just after $10^4$ years, and begins to lose mass rapidly. The inner 
boundary is encountered after $10^6$ years. Despite loss of most of the mass and angular momentum, 
these losses are well followed over the age of the Solar System, as indicated by the red curves, 
where we tally the flow of these quantities across the boundaries. 

\subsubsection{Pollution evolution}
\label{subsubsec:polevol}

The rate per unit area at which a ring annulus of constant surface density $\Sigma$ at some location $r$ is 
polluted by direct deposition of incoming micrometeoroids is given by Eq. (\ref{equ:sigim}). 
In our code, we incorporate the amount of pollutant that accumulates with time and co-evolve it with the background total surface density $\Sigma = \Sigma_{\rm{pol}} + \Sigma_{\rm{ice}}$, where $\Sigma_{\rm{ice}}$ is the icy material surface density \citep[e.g., see Eqns. 19 and 20 of][]{Est18}.
{{Analogous}} to Eq. (\ref{equ:viscmequ}), the equations for $\Sigma_{\rm{pol}}$ and $\Sigma_{\rm{ice}}$ are

\begin{equation}
\label{equ:viscmequpol}
    \frac{\partial \Sigma_{\rm{pol}}}{\partial t} = \frac{3}{r}\frac{\partial}{\partial r}\left[f r^{1/2}
    \frac{\partial}{\partial r}\left(r^{1/2}\Sigma \nu\right)\right] + f_{\rm{ext}}\eta \dot{\sigma}_{\rm{im}},
\end{equation}

\begin{equation}
\label{equ:viscmequice}
\begin{split}
    \frac{\partial \Sigma_{\rm{ice}}}{\partial t} = \frac{3}{r}\frac{\partial}{\partial r}\left[(1-f) r^{1/2}
    \frac{\partial}{\partial r}\left(r^{1/2}\Sigma \nu\right)\right] + \\
    (1-f_{\rm{ext}}\eta) \dot{\sigma}_{\rm{im}},
    \end{split}
\end{equation}

\noindent
where $f = \Sigma_{\rm{pol}}/(\Sigma_{\rm{ice}} + \Sigma_{\rm{pol}})$ is the mass fraction of pollutant, $f_{\rm{ext}}$ is the non-icy fraction of the meteoroid and $\eta$ represents the fraction of non-icy 
pollutant post impact that remains as absorbing material, as assumed in previous ballistic transport modeling \citep{CE98,Est15}. We adopt two values for $f_{\rm{ext}}$: 50\%, which corresponds to
Oort cloud projectiles, and 100\% for EKB projectiles (Sec. \ref{subsec:basiceq}). Equation (\ref{equ:btmequ})
can be similarly expressed.
As the measured values from Cassini for the non-icy material in the rings are given in terms of
volume fraction $\upsilon$ \citep{Zha17a,Zha17b}, we choose to do the same in this paper, so in
terms of the mass fraction, $\upsilon = (\rho/\rho_{\rm{pol}})\,f$. We assume the
compacted material density of the micrometeoroids $\rho_{\rm{pol}} = 2800$ kg m$^{-3}$ as is assumed 
in \citet{Kem23}, which is a mixture of fraction of 70 vol\% silicate material, and 30 vol\%
carbonaceous material \citep{Woo17}. The solid density of a ring particle is then
$\rho = \rho_{\rm{pol}}\upsilon + \rho_{\rm{ice}}(1-\upsilon)$ where we take $\rho_{\rm{ice}} = 920$
kg m$^{-3}$.

At every time step, we calculate the mass fraction $f(r,t)$ from $\Sigma_{\rm{pol}}(r,t)$ and the
total mass surface density $\Sigma(r,t)$. The mass fraction of pollutant 
is then used to calculate the new particle density, and thus the volume fraction $\upsilon(r,t)$, 
which in turn can affect the
optical depth and viscosity at $r$ which must be updated accordingly. It should be noted that because the
pollutant is also evolving, this means that an initially 
icy ring becomes less icy, but as it loses mass at the boundaries, it also loses pollutant. We thus track
the total amount of pollutant that impacted the rings and compare it with what remains in the rings
as a function of time.

Equation (\ref{equ:mequ}) assumes that all of the impactor mass (as given in Eq. [\ref{equ:sigim}]) 
is retained as solid material, but we do not necessarily assume that all of the material remains
as a pollutant. 
We explore two values in our simulations regarding how much of the micrometeoroid 
retains its radiative 
absorptive properties after impact: $\eta = 10$\% and $\eta = 100$\%. This means that, depending on the choice of $\eta$, as much as 90\% of the icy and non-icy components of {\it impactors} are assumed to be vaporized and recondensed on the rings, as water ice or as non-absorbing material \citep{Doy89,CE98}.
We consider the lower 
value of $\eta$ to be a conservative lower limit because \citet{CE98} found 
that it was not possible 
to use lower values and simultaneously explain the relatively sharp contrast in 
darkening (due to direct deposition) between the low-$\tau$ C ring and high-$\tau$ inner B ring 
regions, and the much more gradual color transition that arises from
the advection of icier (and spectrally more red) material from the B ring to the C ring 
due to BT across the B-C ring boundary. 
When allowed to act long enough, BT homogenizes composition regionally, and also globally \citep[independent of $\eta$;][]{CE98}, 
but it can also eventually erase the contrast in darkening between the two ring regions, because the more polluted C ring becomes the dominant source of darkening of inner B ring material \citep{Est15,Est18}. 
On the other hand, too low an $\eta$ may never create a strong darkening contrast because the low-$\tau$ region would be overwhelmed by the advected icy material from the B ring and both regions would darken slowly at more similar rates. So the fact that a relatively sharp darkening contrast 
exists suggests both a lower limit on $\eta$, and that BT has not acted
long enough since the contrast was created to erase it. 

Finally, we also consider models in which the micrometeoroid flux was higher 
in the past. Although this may be a reasonable assumption, pinning an actual value on the flux and 
how it changes with time is speculative. A factor of ten higher at the LHB may be applicable 
(K. Zahnle, priv. comm.). Thus for specificity,
we consider a model in which the flux is a factor of ten higher up to the LHB 
($t_{\rm{LHB}} \sim 700$ Myr), and then let it exponentially decay to its current value as

\begin{equation}
    \label{equ:sdotx10}
    \dot{\sigma}_\infty(t) = \dot{\sigma}_\infty(t_{\rm{SS}}) 
    \cdot 10^{{\rm{MIN}}\left[1,(t_{\rm{SS}}/t - 1)t_{\rm{LHB}}/(t_{\rm{SS}}-t_{\rm{LHB}})\right]},
\end{equation}

\noindent
where $t_{\rm{SS}} = 4.5$ Gyr.

%

\section{Results}
\label{sec:results}

\subsection{Long-Term Viscous Evolution of Initially Massive Rings}
\label{subsec:longterm}
 
Let us first consider initially massive rings that are evolving viscously
over the age of the Solar System $t_{\rm{SS}}$ while also subject to
pollution by micrometeoroid bombardment. We conduct simulations using both
the previously assumed Oort cloud (cometary) population and the EKB population because we do not know
which may  have been dominant over the Solar System lifetime. This means that for the {{former}} we choose $\dot{\sigma}_{\infty} = 4.5\cdot 10^{-16}$ kg m$^{-2}$ s$^{-1}$ \citep{Gru85,CD90}, with a
gravitational focusing factor $F_{\rm{g}} \approx 3$, and $\dot{\sigma}_{\infty} = 2.2\cdot 10^{-16}$ kg m$^{-2}$ s$^{-1}$ \citep[][Sec. \ref{subsubsec:polevol}]{Kem23} with $F_{\rm{g}} \approx 30$ for the {{latter}}.
For a few cases, we consider the possibility of a much higher flux during the LHB 
(Eq. [\ref{equ:sdotx10}]). 

\begin{figure}
 \resizebox{\linewidth}{!}{%
 \includegraphics{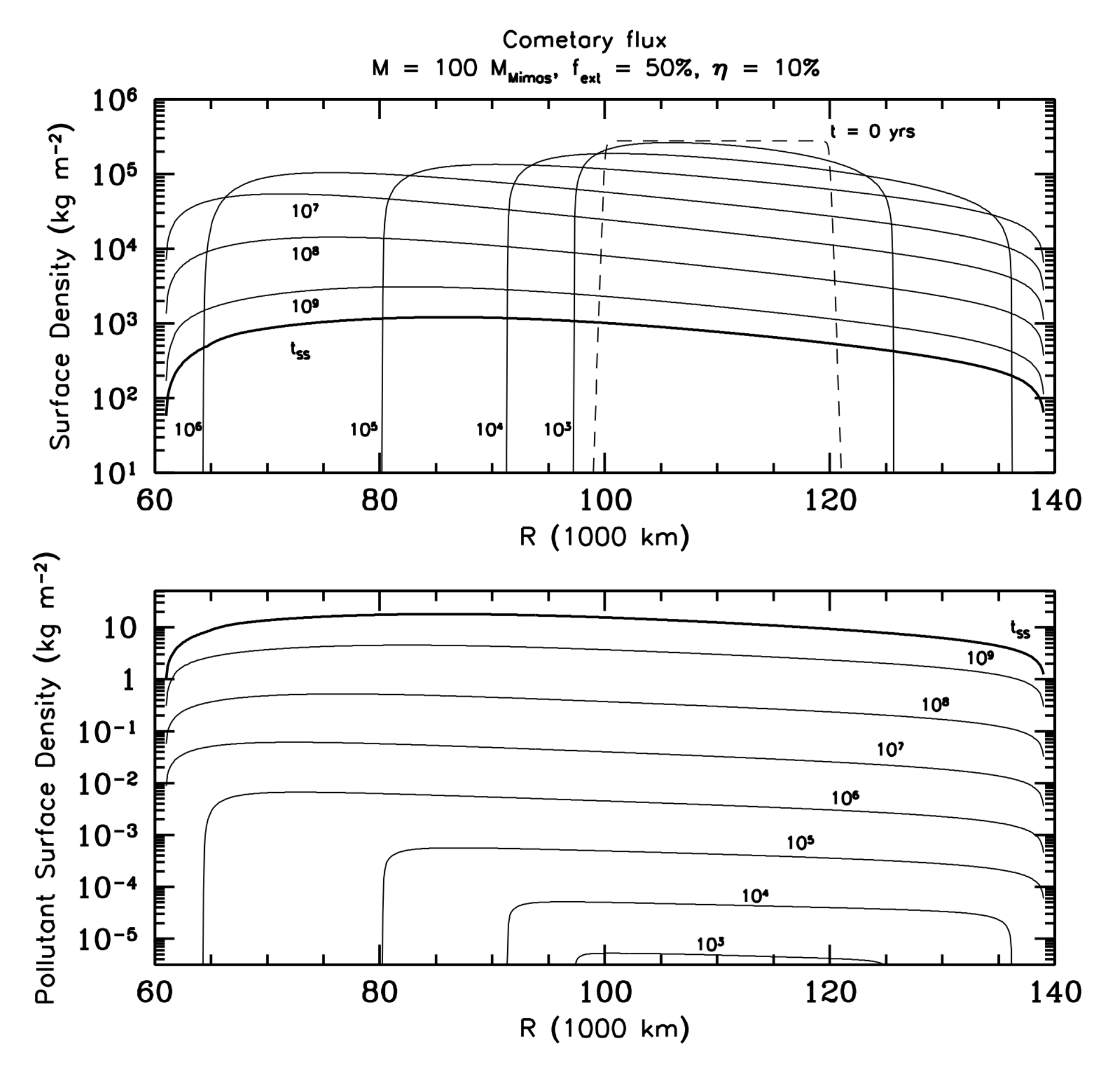}}
 \vspace{-0.3in}
\caption{Simulation for an initial 100 M$_{\rm Mimas}$ ring subject to the 
cometary micrometeoroid flux for $f_{\rm{ext}} = 50$\% and $\eta = 10$\%. \underline{upper panel}: 
Surface mass density plotted at various times over the age of the Solar System $t_{\rm{SS}}$. Such 
a massive ring has already spread to the outer boundary {{in}} $\lesssim 10^5$ years, while the inner 
edge is reached after $10^6$ years. The ring mass at $t=t_{\rm{SS}}$ is 0.94 M$_{\rm Mimas}$. 
\underline{lower panel}: Corresponding surface density of accumulated pollutant.}
\label{fig:evoltsteps}
\vspace{-0.1in}
\end{figure}

Our fiducial model will be an initial disk
mass of 100 M$_{\rm Mimas}$ ($\sim 1.6\times$ the mass of Rhea), but we will consider the long-term evolution of smaller initial disk masses as well. 
Naturally, the largest initial disk mass should be the most resistant to pollution. 
We assume that the ring begins as pure ice. While this seems unlikely, we
consider it a conservative assumption that will set an upper limit on the time it 
takes for the ring to be polluted to what we now observe. Although the current rings are characterized by
a particle size distribution with different powerlaw dependencies for different ring regions \citep[e.g, see][]{Cuz18}, the viscosity model of \citet[][Sec. \ref{subsec:viscosity}]{Dai01} is derived for a single particle size. We thus adopt a ring particle radius $a$ of one meter, which has been a common choice in $N$-body simulations \citep[e.g.,][]{Rob10,Sal18}, and is also a value similar to the effective radius for ring particles used in 
BT simulations \citep[e.g.,][]{Dur92,CE98,Est15}.  
%
%

An initial ring formed from larger chunks will likely grind itself down as it 
spreads, and thus the size distribution in the rings would change with time affecting the
viscosity, but we do not attempt to model this process here. {{The}} particle size would directly affect 
the non-wake viscosity, which increases with particle radius, and the wake viscosity indirectly 
through the Toomre parameter. \citet{Sal10} showed that the effect of larger particles is {{to make the ring}} 
evolve more quickly {{compared}} to the non-self gravitating regime. We will consider different 
effective particle sizes for 
our younger, lower mass ring models in the next subsection.

 \begin{figure}[t]
 \resizebox{\linewidth}{!}{%
 \includegraphics{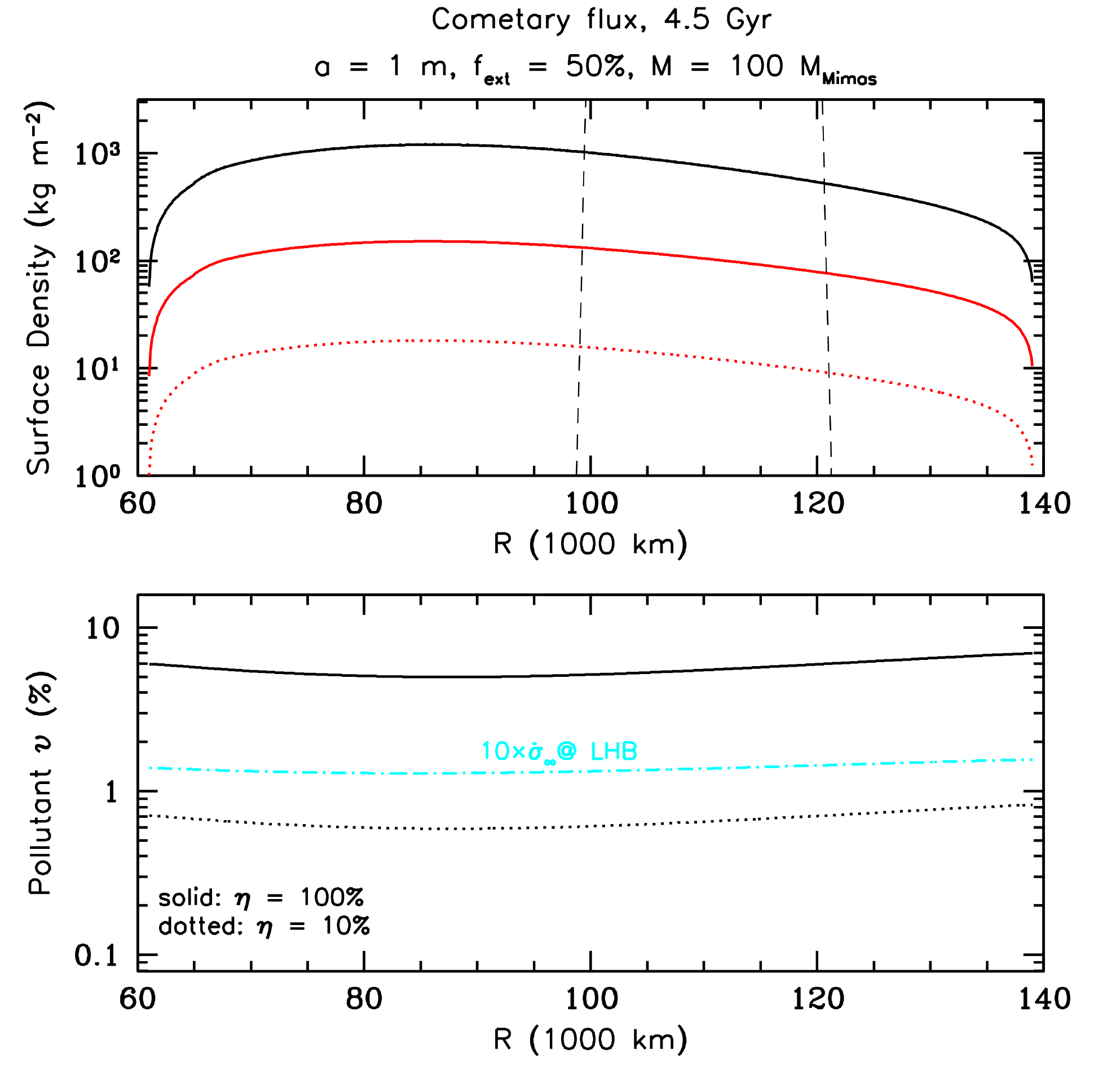}}
 \vspace{-0.3in}
\caption{Comparison of simulations for an initial 100 M$_{\rm Mimas}$ ring at
$t=t_{\rm{SS}}$ for the cometary flux with $f_{\rm{ext}} = 50$\%, and $\eta = 10$\% and 
$\eta = 100$\%. The former corresponds to the case in Fig. \ref{fig:evoltsteps}. 
\underline{upper panel}: Final surface mass densities for the $\eta = 10$\% (dotted curves) 
and $\eta = 100$\% (solid curves). The black curves correspond to the total surface density,
and the red curves to pollutant surface density. The dashed curve marks the radial extent of the initial 
annulus. Note that the black curves for the overall surface mass density lie on top of each other. 
\underline{lower panel}: Accumulated volume fraction of pollutant for both models. An
additional model is included in which the meteoroid influx was higher in the past (Sec. 
\ref{subsec:nummethod}) for the $\eta = 10$\% model. The difference leads to about a
factor of 2 increase in the non-icy fraction.}
\label{fig:longoort}
\vspace{-0.1in}
\end{figure}

{\bf Figure \ref{fig:evoltsteps}} shows an evolution of the rings for 
cometary micrometeoroids with $f_{\rm{ext}}=50$\% and $\eta = 10$\%. We plot the total (ice
+ pollutant, top panel) and pollutant mass surface densities (bottom panel) for several different 
times during the simulation. The initial ring annulus is given by the dashed curve. For this
configuration, the time for the ring outer edge to reach the outer grid boundary and begin to shed
ring mass is $\lesssim 10^5$ years, 
while the inner edge
of the grid is reached after $\gtrsim 10^6$ years. The characteristic shape of the curves
with higher densities inwards are characteristic of a self-gravitating disk. The relative
sharpness of the ring edges is {{mainly due to the viscosity being such a strong
function of the surface density. The higher surface density away from edges spreads much more quickly
than the lower surface density edge can manage, which further steepens the edge \citep{Sal10}.}}
The bottom panel shows the corresponding surface
density of accumulated pollutant. After a time $t_{\rm{SS}}$ (bold curves), the rings have
retained a peak value near $10$ kg m$^{-2}$ of non-icy material. The final ring mass is 0.94 
M$_{\rm Mimas}$.

\begin{figure}[t]
 \resizebox{\linewidth}{!}{%
 \includegraphics{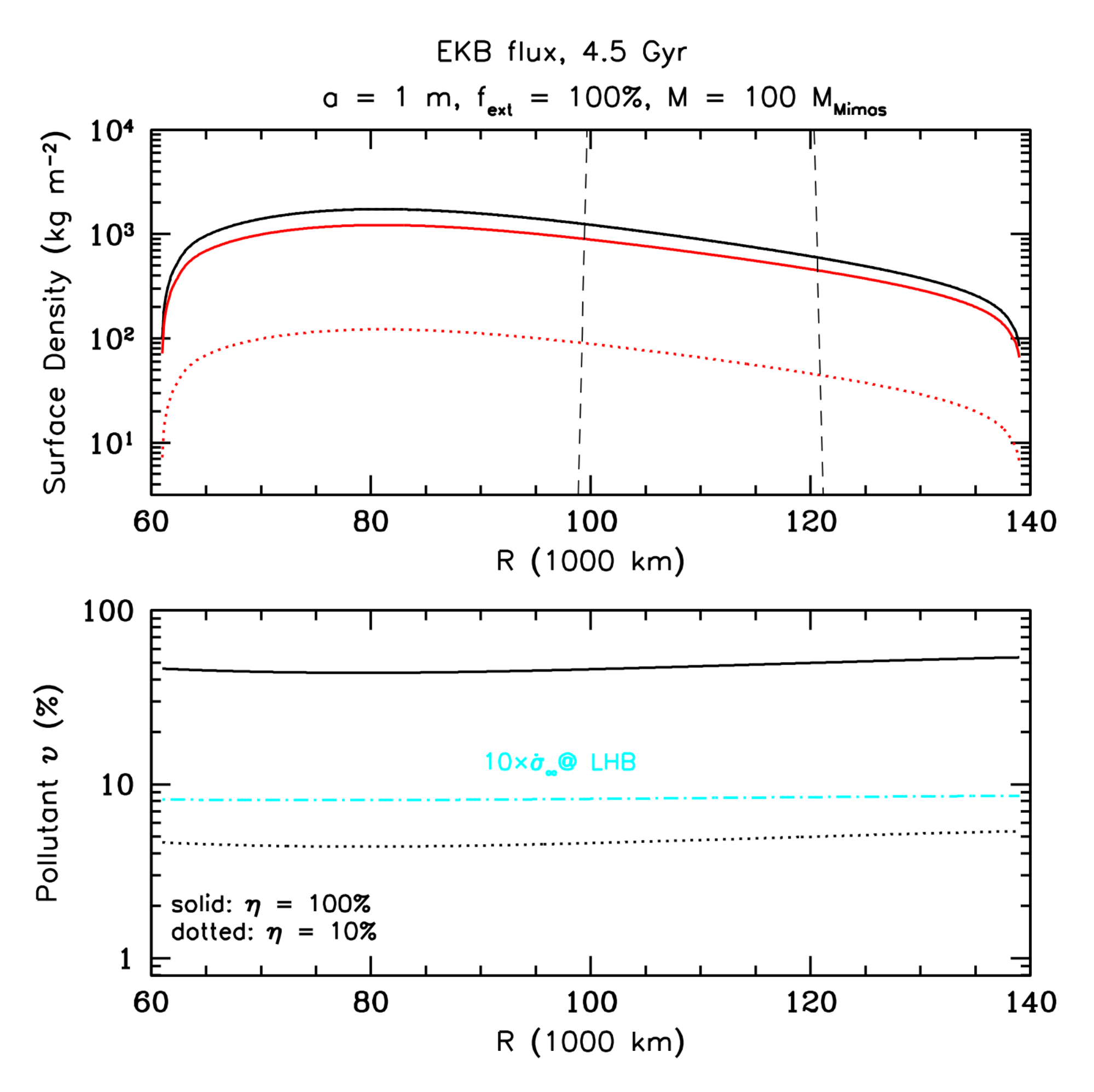}}
 \vspace{-0.3in}
\caption{ Descriptions are the same as Fig. \ref{fig:longoort}, except that
here the EKB meteoroid influx is  used, and $f_{\rm{ext}} = 100$\%. 
The flux being much higher leads to much darker rings
compared to the cometary case, and for the $\eta = 100$\% model the rings achieve a volume fraction of 
$\upsilon \sim 50$\%. The factor of increase in pollution using a higher meteoroid influx in the
past is not as large as it is for the cometary case (see text).}
\label{fig:longekb}
\vspace{-0.1in}
\end{figure} 
 
{\bf Figure \ref{fig:longoort}} compares the final state of the above simulation (dotted curves) with 
two additional cases: one where $\eta = 100$\% (solid curves) and a case where the flux was
ten times higher at LHB (dot-dashed, cyan curve) for $\eta = 10$\%.
The dashed curves in the upper panel show the ring edges at $t=0$.
The top panel shows the total final surface mass density (black curve) and the final mass
densities of pollutants (red curves) for the constant $\dot{\sigma}_{\infty}$ runs. The final total
densities curves cannot be distinguished and are generally consistent with surface densities 
one can infer from the current ring mass \citep{Ies19}. 
The bottom panel shows the volume fraction of accumulated 
pollutant. For the conservative case of $\eta = 10$\%, the pollutant volume fraction is 
$\sim 0.6-0.7$\%, which is consistent with, but slightly larger than current measured values for the 
A and B rings. The simulation in which the flux is higher in the past and exponentially decays
increases the fraction to roughly $1.4$\%, or by a factor of two. Thus, an initially massive ring 
evolving dynamically under viscosity alone for our conservative choice of $\eta$ and the cometary flux 
apparently allows for ancient rings. The final ring mass is predicted to be larger than the currently 
observed mass based on the accumulation of micrometeoroids over this time, but the difference in
mass is not significant enough to make it inconsistent (also see Fig. \ref{fig:finalmass}). On the other
hand, the volume fraction for the $\eta = 100$\% case is nearly ten times larger than measured values, and would be consistent with much younger rings.

\begin{figure}
 \resizebox{\linewidth}{!}{%
 \includegraphics{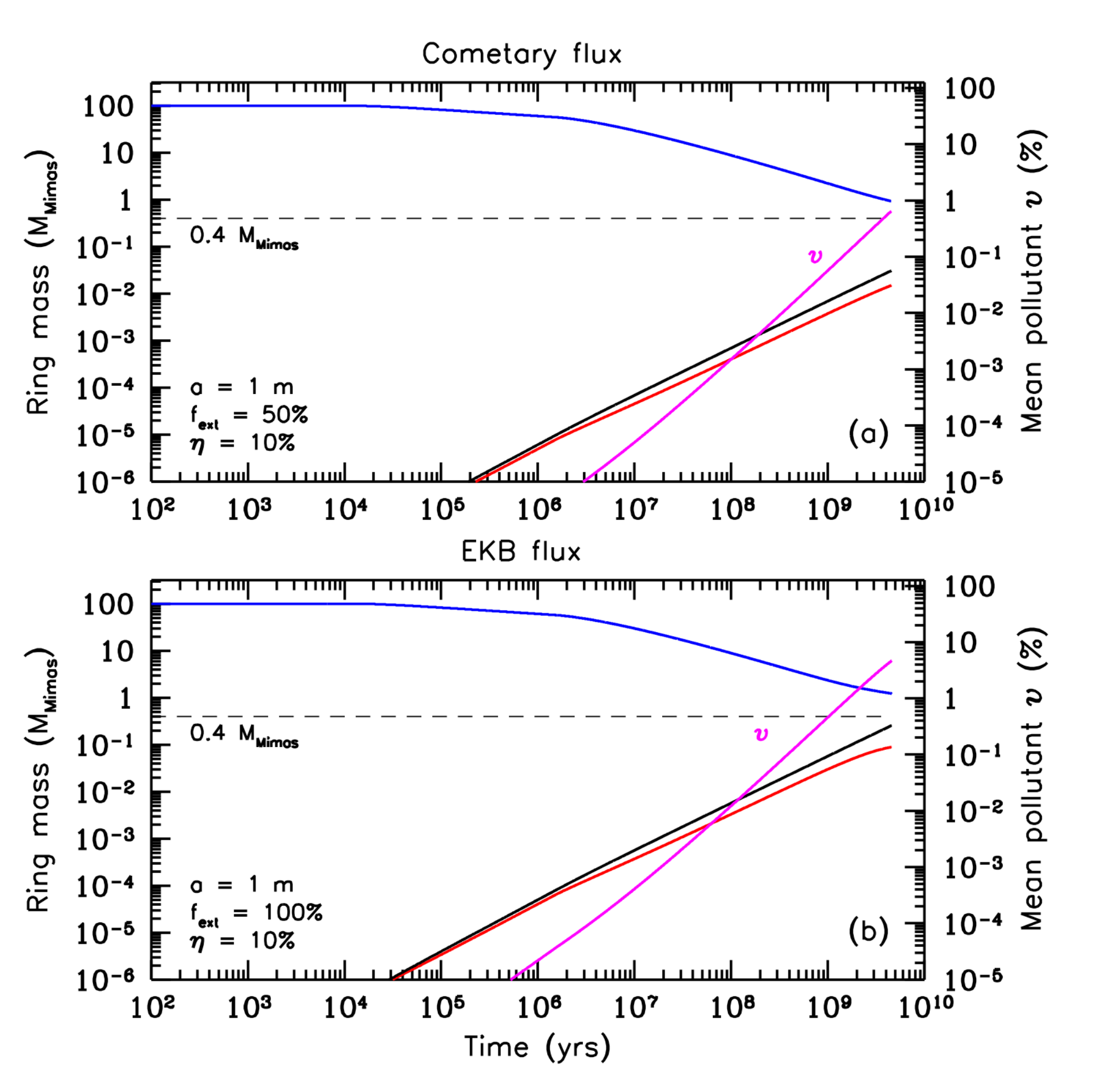}}
 \vspace{-0.3in}
\caption{Time evolution of the ring mass (blue curves), accumulated volume fraction of
pollutant (mass weighted mean, magenta curves), total mass of pollutant delivered (black
curves) and total mass of pollutant retained in the rings (red curves). Compared are
the cometary (upper panel) and EKB (lower panel) models show in Figs. \ref{fig:longoort}
and \ref{fig:longekb} for $\eta = 10$\%. The dashed line marks the current ring mass.
See text for details. }
\label{fig:mvst10}
\vspace{-0.1in}
\end{figure}

Redoing these simulations for the EKB flux provides a stark contrast, as shown
in {\bf Figure \ref{fig:longekb}}. Now, the $\eta = 10$\% simulation 
yields a volume fraction of pollutant of 
$\sim 4-5$\%, or a B ring that has more than ten times the pollution fraction currently observed. 
The higher flux in the past also increases the level of pollution by slightly less than a factor of two, thus lower {{than}} the cometary case, 
while the $\eta=100$\% case produces a 
volume fraction near 50\%, more than a factor of 100 more than currently observed in the B ring. 

\begin{figure}
 \resizebox{\linewidth}{!}{%
 \includegraphics{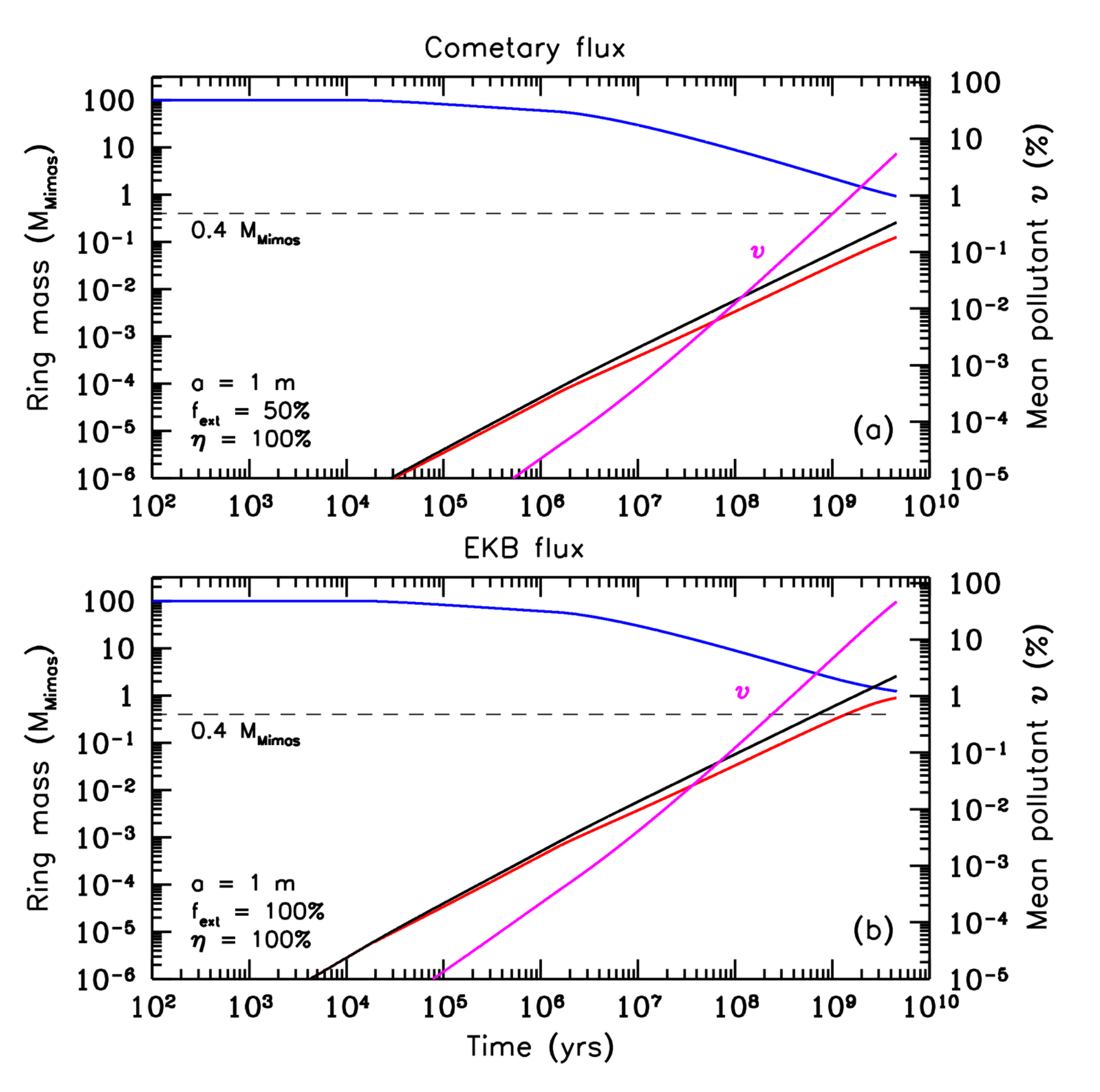}}
 \vspace{-0.3in}
\caption{Same descriptions as Fig. \ref{fig:mvst10}, except here we compare
the cometary (upper panel) and EKB (lower panel) models for $\eta = 100$\%. }
\label{fig:mvst100}
\vspace{-0.1in}
\end{figure}

\begin{figure}
 \resizebox{\linewidth}{!}{%
 \includegraphics{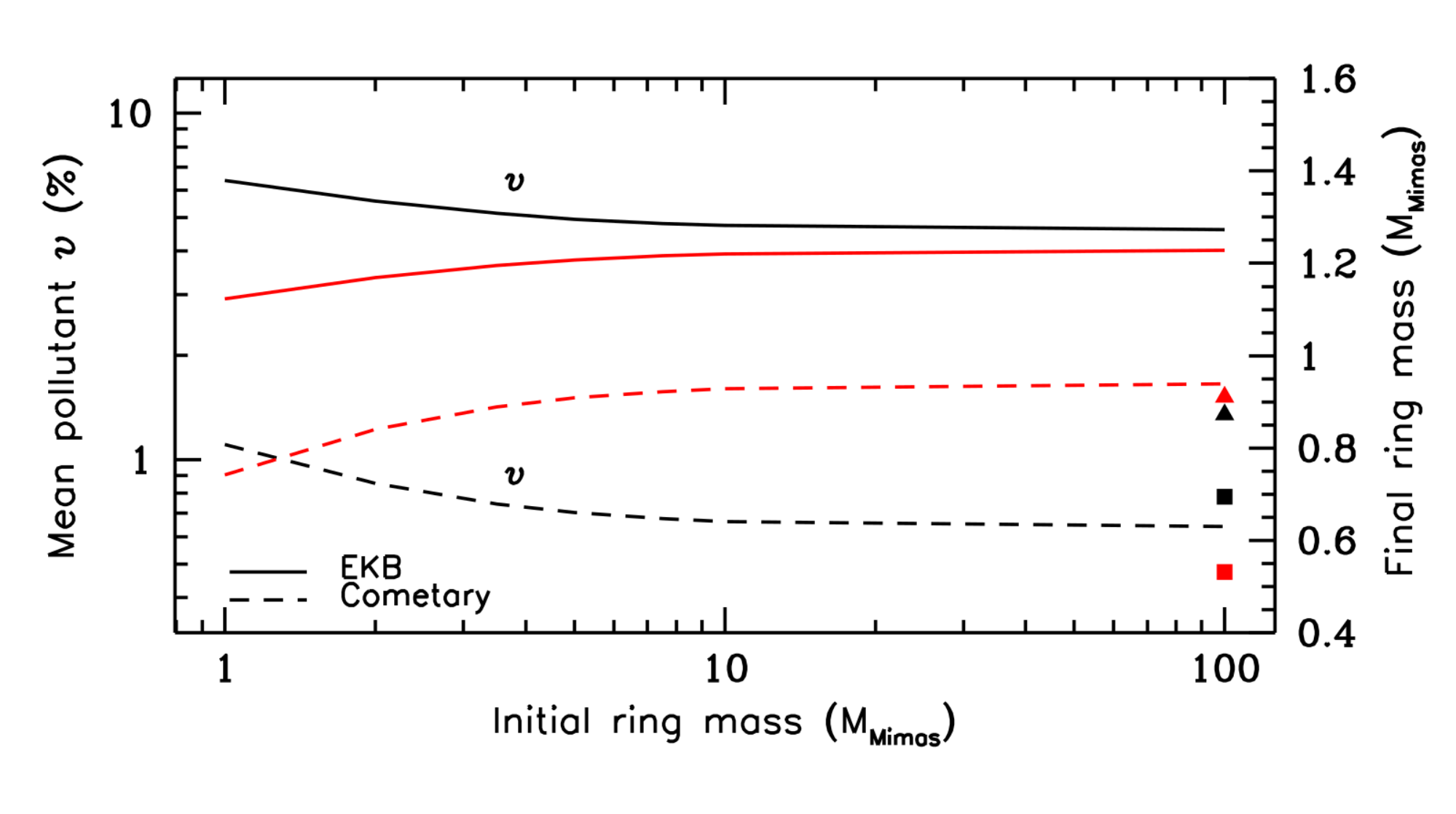}}
 \vspace{-0.3in}
\caption{ Plot of the overall level of pollutant ({\postrevisionbf{black}} curves), and final ring mass
({{red}} curves) as a function of the initial ring mass for a range of models spanning
$1-100$ M$_{\rm Mimas}$, for $\eta = 10$\%. The mean level of pollution over the age
of the Solar System shows little variation, except at the lowest masses, where rings
are darker. The final disk masses also show an asymptotic behavior. Also shown are
two additional models for zero porosity (squares), and for the higher flux in the past
(triangles).}
\label{fig:finalmass}
\vspace{-0.1in}
\end{figure}

{\bf Figure \ref{fig:mvst10}} and {\bf Figure \ref{fig:mvst100}}, for $\eta=10$\% and  100\%,
respectively, summarize the time evolution of the ring mass (blue curves), and accumulated volume 
fraction of pollutant (magenta curves) for the simulations presented above. The cometary cases 
are listed in panel (a), and EKB cases in panel (b). The dashed lines indicate
the current mass of the rings (Iess \etal, 2019). The black solid curves in these figures indicate
the total amount of pollutant in Mimas masses that were deposited in the rings, while the red
curves are the amount of pollutant that remains in the rings. That is to say, when ring mass is 
lost at the boundaries, it not only carries away ice, it carries away pollutant with it. The final
ring mass for the cometary cases is 0.94 M$_{\rm Mimas}$, and 1.23 M$_{\rm Mimas}$ for the EKB case.
The difference in final mass is due to a competition between the amount of mass that has been
deposited in the rings versus the more rapid viscous evolution that results from higher surface
mass densities. Over the course of the simulation, the rings in the EKB cases have been bombarded 
by $\gtrsim 2.5$ M$_{\rm Mimas}$, but only by $\sim 0.5$ M$_{\rm Mimas}$ in the cometary case. 

This competition between mass deposition and viscosity is also at the root of the 
differences seen in the LHB case. 
Because the meteoroid flux is higher, much more mass is deposited in the rings up to LHB 
so that the mass of the rings and their level of pollution are higher 
compared to the nominal constant $\dot{\sigma}_{\infty}$ assumption. 
The higher mass means that the rings are evolving faster due
to higher viscosity and the ring mass drops off more steeply, and with it, more pollutant is
removed. The combination of this and a decaying flux eventually closes the gap between the
level of pollution in both cases. In the cometary case, the final difference in pollution 
volume fraction is
about a factor of $\gtrsim 2$ (Fig. \ref{fig:longoort}), whereas it is $\sim 1.8$ in the 
EKB case (Fig. \ref{fig:longekb}). The final masses are 
smaller, 0.91 M$_{\rm Mimas}$ and 1.09 M$_{\rm Mimas}$ for the cometary and EKB cases, respectively.

In {\bf Figure \ref{fig:finalmass}}, we explore the effect of initial disk mass on the overall level of mean 
pollutant in the rings and on the final ring mass after the age of the Solar System for $\eta = 10$\%
with both cometary (dashed) and EKB (solid curves) fluxes for initial ring masses ranging 
from 1 to 100 M$_{\rm Mimas}$.
We find that there is very little difference in the final pollution state of the rings across
the full range of initial disk masses ({{black}} curves). 
%
This is because the rings spend the
majority of their lifetime at low mass. For instance, even in the 100 M$_{\rm Mimas}$ case, the
ring mass has dropped due to rapid viscous evolution to $\sim 8-9$ M$_{\rm Mimas}$ after $10^8$ years 
and to $\sim 2$ M$_{\rm Mimas}$ after $10^9$ years (cf. Fig. \ref{fig:mvst10}). The final disk
masses ({{red}} curves) also show asymptotic behavior with a final preferred ring mass consistent
with previous viscous evolution models \citep{Sal10}. The only differences occur for very
low initial ring masses where spreading is slower, and the extra time at low mass allows
for slightly higher levels of pollutant. We also show results for the higher past LHB flux (triangles)
and one with zero porosity (squares), both for 100 M$_{\rm Mimas}$ and the cometary flux. As noted
before, the LHB case leads to a slightly lower final ring mass and rings that are a factor
of two darker. The zero porosity case also leads to a much lower final ring mass ($\gtrsim 0.5$
M$_{\rm Mimas}$) and is only slightly darker than our nominal $\phi = 0.75$ {{porosity}} case.

\subsection{Short-Term Evolution of Initially Low-Mass Rings}
\label{subsec:shortterm}

Even in the case of an evolution dominated by cometary meteoroids with 
$\eta = 10\%$, the Solar System age simulations in the previous section give volume 
fractions of pollutant greater than $0.1 - 0.5$\%. These are larger than what  
Cassini measurements give in the A and B rings \citep{Zha17b}, and we know that cometary meteoroids  
are not dominant
at the present time \citep{Kem23}. The final masses of these simulations also 
do not match the current ring mass \citep{Ies19} even for the smallest initial masses considered due
to mass added to the rings from direct deposition over time (see also below), Furthermore, they do not {\it look} like the current rings, where the C ring is relatively low mass.

This motivates us to consider simulations 
starting with lower initial ring masses that are subject to a sustained pollution rate over 
shorter time periods. In this subsection, and as before, we present models that evolve due to viscosity only (DD),
\textit{i.e.}, radial drifts induced by mass loading or ballistic transport are still ignored
(Table \ref{tab:models}). 
Because the rings spend the majority of their lifetime at these lower masses, darkening 
is efficient and should not require long times to reach their current state,
even under the assumption that they began as pure water ice. 

Based on Cassini CDA measurements of the micrometeoroid flux at Saturn, the implied age of
the rings is no more than a few 100 Myr \citep{Kem23,DE23}. 
Moreover, as discussed in 
Sec. \ref{sec:intro}, the rings are losing mass at a surprising rate which is attributed 
in great part to effects caused by micrometeoroid bombardment \citep{Hsu18,Wai18,Odo19,DE23}. 
Implicit in these mass loss rates is that the rings were more massive in 
the past. From the results of INMS, which likely measures the total equatorial mass
inflow, a constant loss
rate would suggest an initial mass for the rings ranging from Mimas to somewhat larger than Enceladus, or
from $\sim$ 1 to 3 M$_{\rm Mimas}$, some 100's Myr ago. Adopting this range of initial masses, we will 
compare the rates of pollution using the 
recently measured 
micrometeoroid flux \citep{Kem23} and conservatively assuming a pollutant retention of 
$\eta = 10$\%. Our nominal evolutions again assume ring particles with radius $a = 1$ m, 
but we will also explore the effect of particle size on spreading rate (Table 
\ref{tab:models}). 
Finally, all models presented here begin with a ring annulus from
125,000 km to 130,000 km, but we will later discuss the effects of different initial conditions.

\begin{figure}[t]
 \resizebox{\linewidth}{!}{%
 \includegraphics{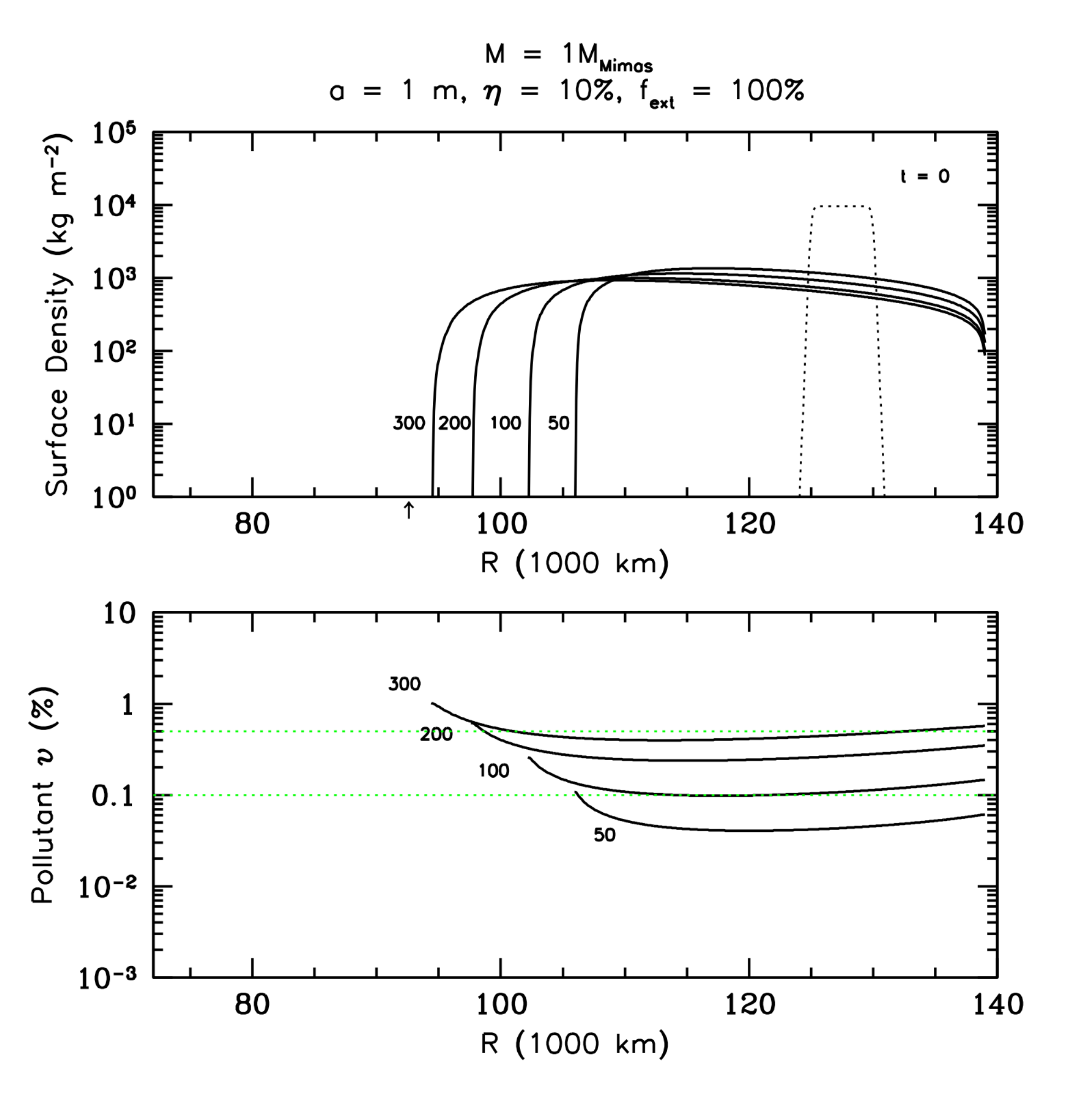}}
 \vspace{-0.3in}
\caption{Simulations of an initial 1 M$_{\rm Mimas}$ ring for $a = 1$ m, 
$f_{\rm{ext}}=100$\% and $\eta = 10$\% for the EKB flux. 
\underline{upper panel}: Surface mass density plotted at 50, 100, 200, and 300 Myr. 
The initial annulus at $t=0$ is shown by the dotted curve. The B-C ring boundary at 92,000 km is indicated by the small arrow. By 300 Myr, the inner edge of the annulus still lies outside the current B-C boundary. \underline{lower panel}: Evolution of the volume fraction of pollutant at the same times. 
The 100 and 200 Myr curves bracket the range of observed volume fractions from
Cassini for the A and B rings, as shown by the green dotted lines. }
\label{fig:1M_1m_LH}
\vspace{-0.1in}
\end{figure}

\begin{figure}[t]
 \resizebox{\linewidth}{!}{%
 \includegraphics{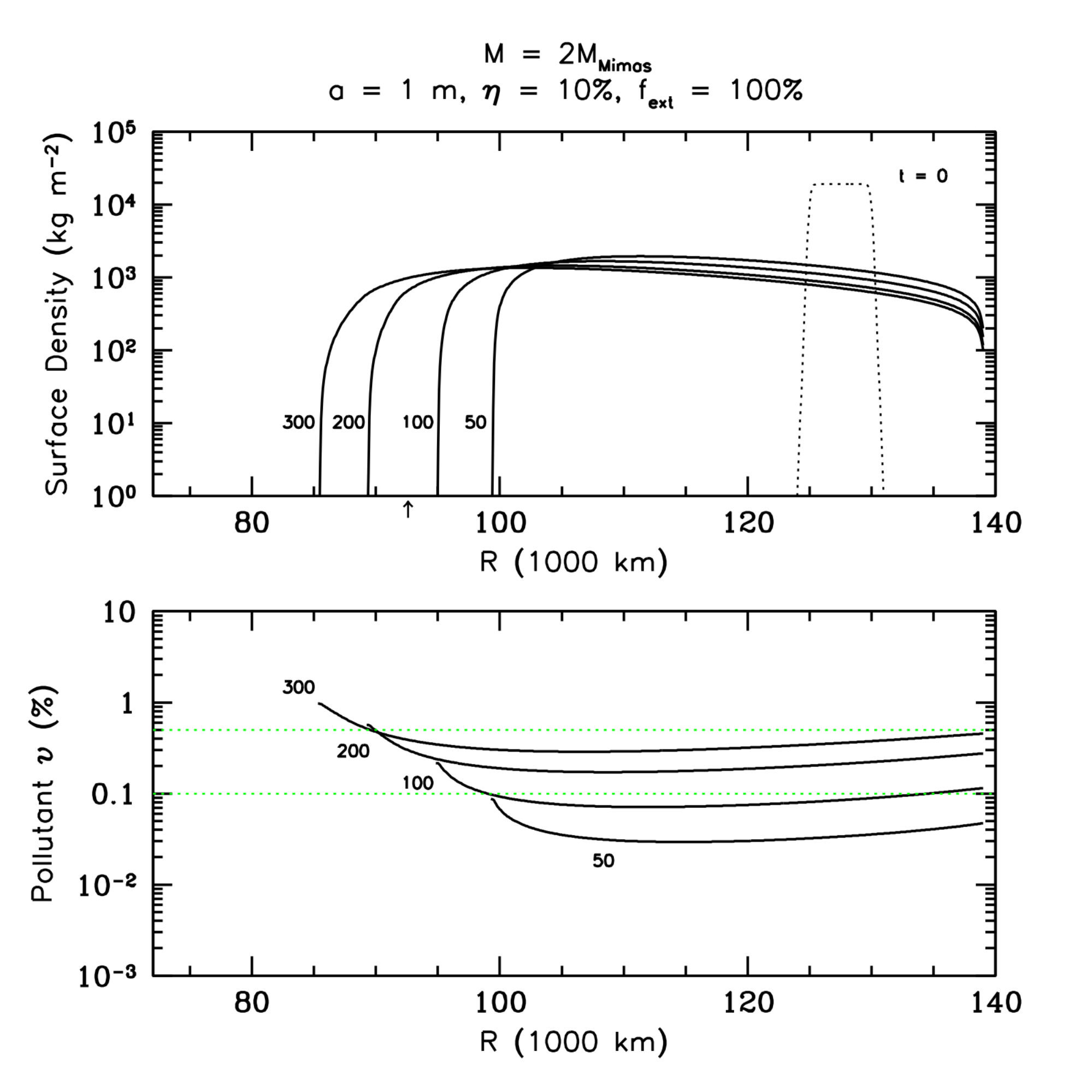}}
 \vspace{-0.3in}
\caption{Simulations of an initial 2 M$_{\rm Mimas}$ ring for $a = 1$ m, 
$f_{\rm{ext}}=100$\% and
$\eta = 10$\% for the EKB flux. 
\underline{upper panel}: Surface mass density plotted at 50, 100, 200 and 300 Myr. 
The higher mass ring spreads much further inward, with the 100 Myr case nearly reaching the
B-C ring boundary. These simulations shed about a Mimas mass beyond the A ring. \underline{lower panel}: 
Evolution of the volume fraction of pollutant at the same times. 
The 200 and 300 Myr cases 
bracket the observations, while the 100 Myr marginally does so. Definitions are the same 
as in Fig. \ref{fig:1M_1m_LH}. }
\label{fig:2M_1m_LH}
\vspace{-0.1in}
\end{figure}

\begin{figure}[t]
 \resizebox{\linewidth}{!}{%
 \includegraphics{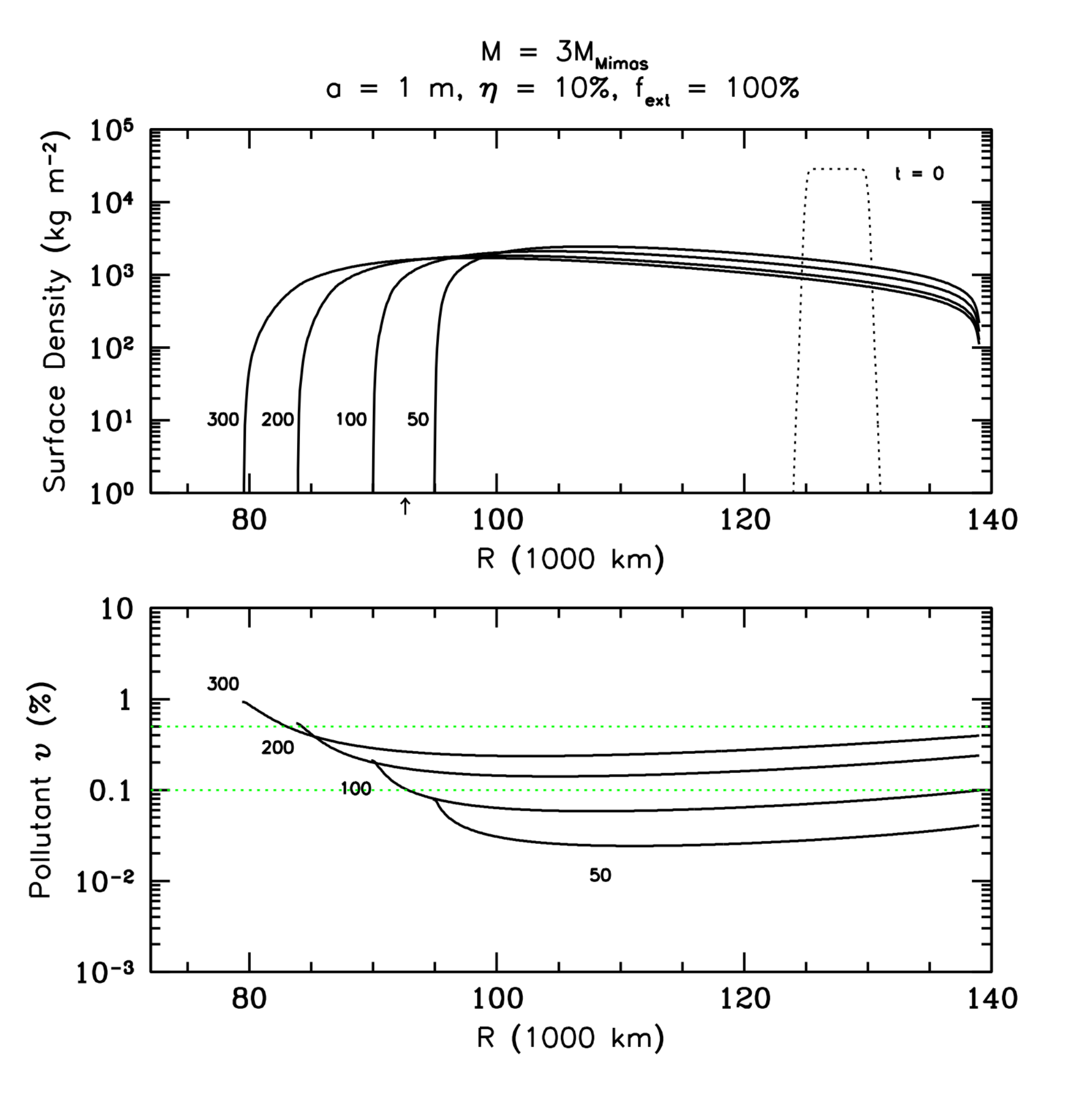}}
 \vspace{-0.3in}
\caption{Simulations of an initial 3 M$_{\rm Mimas}$ ring for $a = 1$ m, 
$f_{\rm{ext}}=100$\% and
$\eta = 10$\% for the EKB flux. 
\underline{upper panel}: Surface mass density plotted at 50, 100, 200, and 300 Myr. 
\underline{lower panel}: Evolution of the volume fraction of pollutant. The 200 and 300 Myr cases 
continue to bracket the observations, but their inner edges are far inside the B-C boundary. Definitions are the same 
as in Fig. \ref{fig:1M_1m_LH}.}
\label{fig:3M_1m_LH}
\vspace{-0.1in}
\end{figure}

{\bf Figures \ref{fig:1M_1m_LH}$-$\ref{fig:3M_1m_LH}} comprise a suite of simulations for initial disk
annuli of 1 to 3 M$_{\rm Mimas}$ \footnote{The mass of Enceladus is $\sim 2.8$ M$_{\rm Mimas}$.}. Shown
are a set of curves for 50, 100, 200, and 300 Myr. 
%
%
We here assume that,
as determined by the CDA analysis \citep{Kem23}, the influxing meteoroids are dominated by an EKB population
and that this population has dominated at a constant rate over the relatively short duration of these 
simulations. The top panels plot the ring mass surface density evolution, 
while the bottom panels give the corresponding level of pollution in the rings. In the
former, the dotted black curve shows the ring annulus at $t=0$. In the latter, the dotted green lines 
mark the range of currently observed non-icy volume fraction in the A and B rings, from $0.1-0.5$\% 
\citep{Zha17b}. 

In the top panel of Fig. \ref{fig:1M_1m_LH}, which is the lowest initial ring mass model, the spreading 
naturally occurs at the slowest rate due to the weaker viscosity. The ring encounters 
the outer edge of the grid and begins to shed mass there after 2 Myr. 
The disk {{mass}} (and masses lost) at 50, 100, 200 and 300 Myr are 0.67 (0.34), 0.62 (0.41), 0.57 (0.48), and 0.56 (0.52) M$_{\rm Mimas}$. 
The amount of mass lost once the outer edge is reached 
is initially steady, but then begins to decrease with time as viscosity weakens further.
The mass that is lost would be relatively icy, and could form its own moon,
or contribute to the formation of a moon outside the Roche limit. 
The retained disk mass at the same time begins to asymptote; however, we expect the disk mass to actually increase at later times 
because {{continued direct deposition of}} incoming material 
will eventually overwhelm the rate at which viscosity can remove mass from the 
disk \citep{DE23} through the outer edge, and, as the disk spreads inwards presenting a larger target for meteoroids, 
it will continue to accumulate even more mass. 
We note that in this lowest initial mass case, the inner edge of the annulus has not
quite reached
the current location of the B-C ring boundary at 92,000 km (marked by
the small arrow) even after 300 Myr. 

The trend of volume fraction of pollutant in the bottom panel of Fig. \ref{fig:1M_1m_LH} shows a
systematic increase 
in darkening over time, 
with the 100 and 200 Myr curves nicely bracketed by the range of observed volume fraction in the A and B rings. Thus, the CDA measured value of the micrometeoroid flux, 
combined with this conservative choice for $\eta$ would probably be an appropriate fit. 
The characteristic turn-up of the volume fraction curves shows that
lower optical depth regions (at the foot of the inner boundary) are darkening faster. This faster
darkening rate is what would be expected to explain the contrast between the A and B ring and the lower
optical depth C ring and Cassini division. 

For the 2 and 3 M$_{\rm Mimas}$ mass simulations (Figs. \ref{fig:2M_1m_LH} and \ref{fig:3M_1m_LH}), we see a 
similar trend in the surface mass density, 
but the spreading rate is systematically faster with increasing mass due to the higher viscosity,
with the B-C boundary location encountered correspondingly sooner. The outer edge of
the grid is reached at $\sim 5 \cdot 10^5$ and $\sim 2 \cdot 10^5$ yrs, respectively. In the 2 
M$_{\rm Mimas}$ mass case, the disk masses (and masses lost)
at 50, 100, 200 and 300 Myr are 1.09 (0.92), 0.99 (1.04), 0.91 (1.15) and 0.89 (1.22)
M$_{\rm Mimas}$. 
This
model would thus shed about a Mimas mass by itself, but the moon formed would still be mostly ice
by mass. The 3 M$_{\rm Mimas}$ model shows a similar behavior in retained disk mass and mass lost, 
though the amount of mass lost can be as high as 2 M$_{\rm Mimas}$ after 200 Myr. 
As one would expect, these higher mass models show the same behavior of the mass lost beyond the Roche limit decreasing with time, and the retained disk masses are leveling off as is the case in Fig. \ref{fig:1M_1m_LH}, though it will take more time for the direct deposition rate to eventually exceed the viscous mass loss rate.

In the bottom panels, we note that although there is an overall decrease in the volume fraction of
pollutant obtained over these time scales compared with the 1 M$_{\rm Mimas}$ mass model, the differences
are at most about a factor of 2 for the 3 M$_{\rm Mimas}$ model, so that the CDA measured 
flux
would still satisfy the observed non-icy volume fractions in the A and B rings after $\sim 200-300$ Myrs.
These models are conservative in the sense that we use a low value for $\eta$ 
and the rings are assumed to begin as pure ice. So, a plausible parameter space exists 
consistent with both very young rings and the current measured level of pollution in the
A and B rings where the rings have spread viscously and were polluted by 
micrometeoroid influx determined by the CDA. Note that a significant feature missing from these
evolutions is the development of a C ring structure. Recall, that in these type of models, 
the inner edge remains sharp {{because viscosity is such a strong function of the 
surface density \citep[Sec. \ref{subsec:longterm}][]{Sal10}.}} 
We will address the 
formation of C ring structure in Section \ref{subsec:mlandbt} below.

\begin{figure}
 \resizebox{\linewidth}{!}{%
 \includegraphics{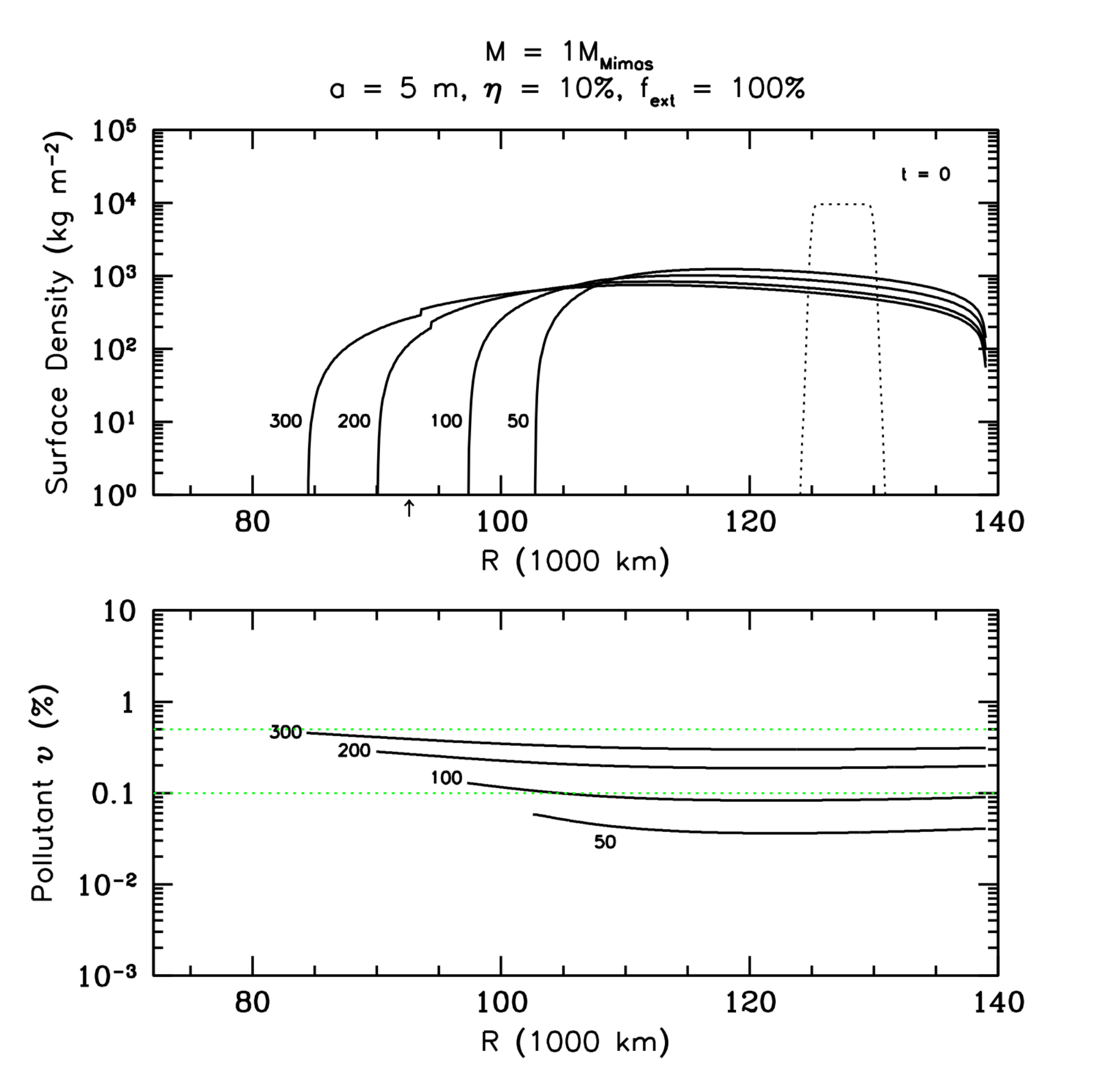}}
 \vspace{-0.3in}
\caption{Simulations of an initial 1 M$_{\rm Mimas}$ ring for $a = 5$ m, $f_{\rm{ext}}=100$\% and
$\eta = 10$\% for the EKB flux. 
\underline{upper panel}: Surface mass density plotted at 50, 100, 200, and 300 Myr. 
{{A}} larger particle size leads to considerably faster evolution, with this case
{{similar}} to the $a = 1$ m, 2 M$_{\rm Mimas}$ case. The kinks in the curves are due to the
ring transitioning from self-gravitating to non-self gravitating forms of viscosity.
\underline{lower panel}: Evolution of the volume fraction of pollutant. Again, similar to the 2 M$_{\rm{Mimas}}$ model (Fig. \ref{fig:2M_1m_LH}), the 200 and 300 Myr cases are 
bracketed by the observations, but the 100 Myr case only marginally so.  Definitions 
are the same as in Fig. \ref{fig:1M_1m_LH}.}
\label{fig:1M_5m_LH}
\vspace{-0.1in}
\end{figure} 

The initial boundary conditions will have little effect on these outcomes. A ring annulus that begins 
further inward would encounter the outer boundary at later times, and likely would lose less mass 
outside the Roche limit, whereas ring annuli that start far enough inward may begin to lose mass 
to the planet. The amount of mass (and thus pollutant) due to direct deposition that the ring will 
accumulate will be higher the longer it takes to evolve to a point where ring mass can be lost.
The higher ring mass will cause more rapid evolution due to higher viscosity. Regardless of 
{{the initial conditions}}, 
the darkening rate would be more or less the same as we found for the
case of ancient rings (cf. Fig. \ref{fig:finalmass}). {{An apparently}} 
robust outcome
of these models {{then}} is 
that the level of pollution will reach the observed values over short time scales.

In {\bf Figure \ref{fig:1M_5m_LH}}, we show an additional model for a 1 M$_{\rm Mimas}$ mass in which
we choose a ring particle radius of $a = 5$ m. As shown by \citet{Sal10}, 
the immediate effect of using a larger particle size
is that the rings spread more quickly. For this choice of $a$, the ring edge 
reaches the outer boundary only slightly more
quickly than the $a = 1$ m case at about $\sim 1.9$ Myr. Because of this, slightly more mass is 
shed from the outer boundary compared to the $a = 1$ m case (0.35, 0.43, 0.5, 0.55 M$_{\rm Mimas}$ for the
50, 100, 200, and 300 Myr curves, respectively). 
%

Although the surface mass densities are considerably lower, the overall spreading behavior for the $a = 5$ m 
case is similar in many respects to the 2 M$_{\rm Mimas}$ mass case (Fig. \ref{fig:2M_1m_LH}). In fact, 
a slight kink is visible near $95,000$ km in the 200 and 300 Myr curves 
because the ring is transitioning to the non-self gravitating regime. The volume fraction
of obtained pollutant (bottom panel) is also more similar and consistent with the 2 M$_{\rm{Mimas}}$ model. 
The lack of a turn up in the volume fraction curves seen in this model compared with the $a = 1$ m cases
is caused by the overall lower optical depth at the ring edge (1/5 lower for a given surface density), 
which lowers the probabilities of deposition at the ring inner edge.

%

\begin{figure}
 \resizebox{\linewidth}{!}{%
 \includegraphics{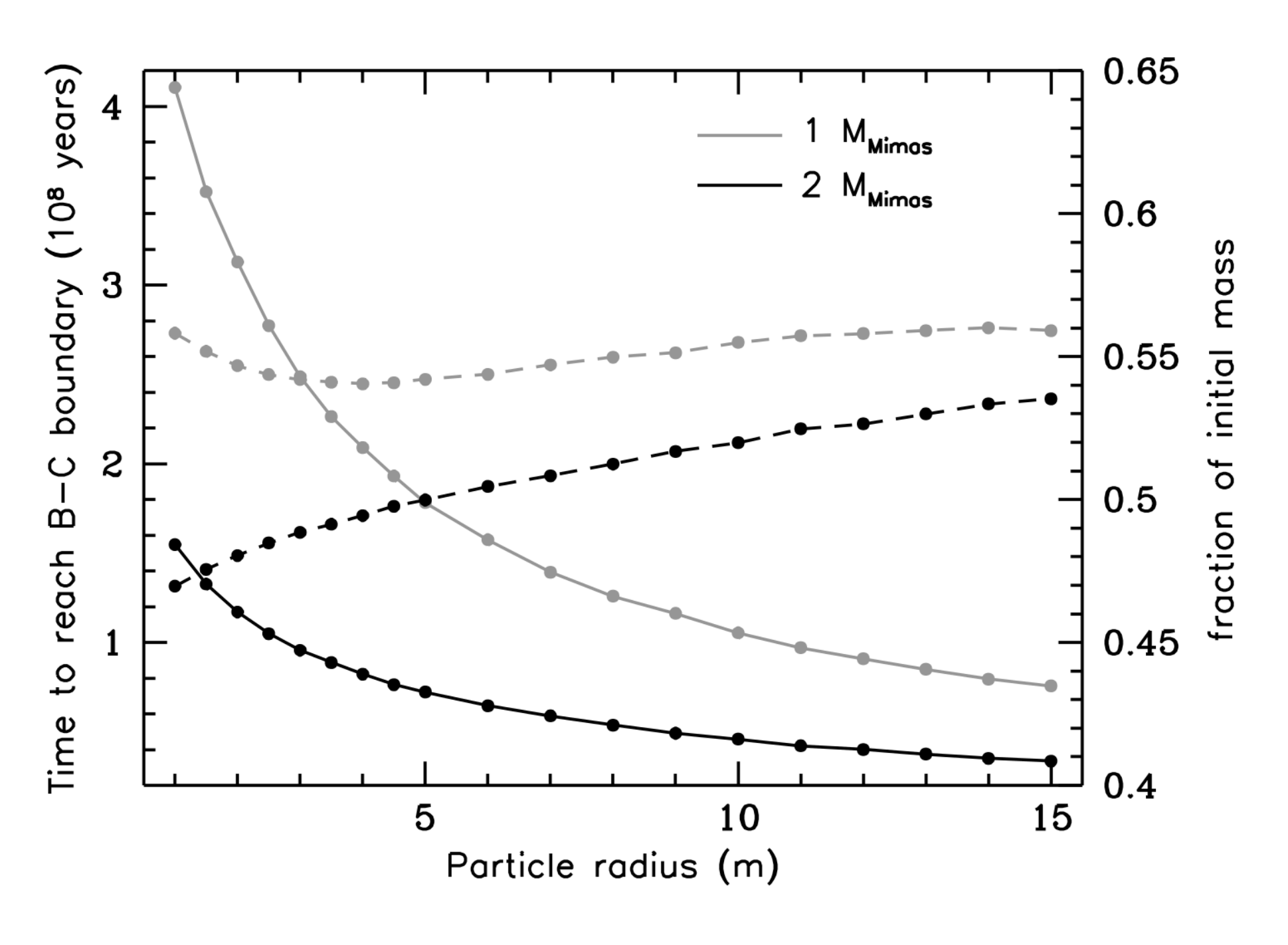}}
 \vspace{-0.3in}
\caption{The effects of particle size on the viscous spreading rate for a 1 M$_{\rm Mimas}$ (grey curves) 
and 2 M$_{\rm Mimas}$ (black curves) 
model. Plotted are the time to spread from the inner annulus boundary
{\postrevisionbf{at}} 125,000 km to the location of the B-C ring boundary at 92,000 km (solid curves), and the 
fraction of the initial mass retained (dashed curves) for particle sizes ranging {{(indicated by points)}} from $1-15$ m. 
The
general trend of faster spreading with larger particle size was shown by \citet{Sal10}. See text.}
\label{fig:tbc}
\vspace{-0.1in}
\end{figure}  

In {\bf Figure \ref{fig:tbc}}, we demonstrate the effect of different particle sizes on the viscous
evolution of our low mass rings by plotting the time it takes 
for the ring's inner edge to reach a reference location, 
which we take to be the current location of the B-C ring boundary at 92,000 km. Models are shown that correspond to 
1 (grey curves) and 2 (black curves) Mimas masses. We also plot the fraction of initial mass left in the ring
when the inner edge reaches the B-C boundary (dashed curves). 
The trends seen further {{confirm}} that spreading is much faster as a function of particle size \citep[e.g., see Fig. 13,][]{Sal10}. At 
$a \sim5$ m, though, there appears to be an inflection point where the spreading time for smaller sizes steeply increases, while dropping off gradually for larger sizes.
The spreading time to reach the B-C boundary can become quite long, over
400 Myr for the 1 Mimas mass, $a = 1$ m case. 
On the other hand, it takes $\lesssim 100$ Myr when $a = 15$ m for the same model.
The mass curves generally show a trend of increasing mass retention with increasing particle size. This is simply due to having more time for mass
to diffuse outside the Roche limit. However, an uptick is seen for the smallest sizes for the 1 Mimas mass
case because spreading is slow enough that the ring has time to acquire a significant amount of mass
by direct deposition.
The same trend is not seen in the 2 M$_{\rm Mimas}$ model as of yet because the spreading times to the B-C boundary remain short in comparison even for the smaller particle sizes. 

%
  
\subsection{Effects of Mass Loading and Ballistic Transport on Ring Evolution}
\label{subsec:mlandbt}

In the previous sections, we studied the effects on viscously-evolving rings of {{the addition of}} 
mass and pollutant. In this subsection, we demonstrate the dramatic effect of introducing mass loading (ML) {{due to direct deposition of bombarding micrometeoroids}} and {{a simplified treatment}} of {{the dynamical effects of}} 
ballistic transport (BT) {{of impact ejecta}} 
{{on the}} global evolution of {{the}} rings. 
As discussed in Section \ref{subsec:MLBT}, we are able to compute mass loading exactly but
here use an approximation for the inward drift due to ballistic transport that is strictly valid only for a uniform ring. We believe this gives us a correct sense for the direction and magnitude of BT over most of the rings but is inaccurate in the outer regions, over long times, and near edges. For all ML or ML+BT {{simulations}} shown in this subsection we assume the EKB population of impactors, and continue to 
adopt the CDA measured value for the micrometeoroid flux at infinity of $\dot{\sigma}_\infty = 2.2\cdot 10^{-16}$ kg m$^{-2}$ s$^{-1}$ 
(see Table \ref{tab:models}).
 
\begin{figure}
 \resizebox{\linewidth}{!}{%
 \includegraphics{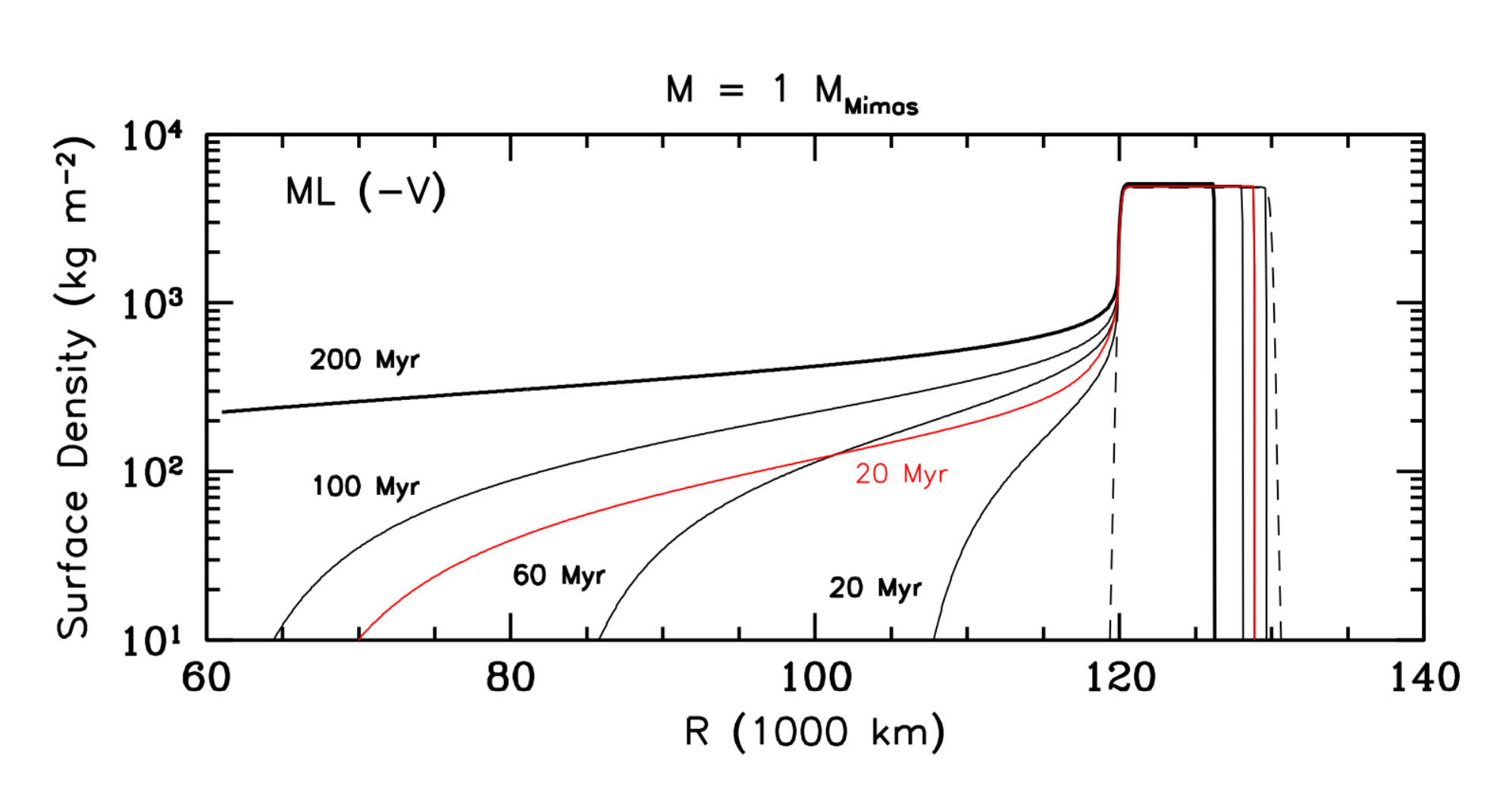}}
 \vspace{-0.3in}
\caption{The effect of mass loading (ML), in the absence of viscosity for a 1 
M$_{\rm Mimas}$ initial disk with $\eta = 10$\%, 
using the EKB flux
(black curves). Curves are plotted at times between $2\cdot 10^7-2\cdot 10^8$ years.
The evolution is characterized by a more
rapid flow of material at low optical depths, away from the inner edge of the annulus. A low
surface density ring interior to the high surface density annulus is created in a relatively
short amount of time. As the material builds up there, the mass influx and loss to the planet
becomes substantial. The red curve is a model that includes BT with $Y=10^4$ 
at $2\cdot 10^7$ years, which can be compared directly with its ML-only counterpart. 
The formation of the C ring feature is then much faster due to the much higher 
induced radial drift.}
\label{fig:MLBT}
\vspace{-0.1in}
\end{figure}  
 
{\bf Figure \ref{fig:MLBT}} shows the evolution of a 1 M$_{\rm Mimas}$ initial mass ring 
evolving under the effects of ML in the {\it absence} of viscosity. 
The initial conditions are the same as in Section \ref{subsec:shortterm}. The effect of
ML is characterized by the flow of material at low surface densities (optical depths) away
from the inner edge of the high surface density (optical depth) ring.
This inflow is relatively rapid with material reaching the inner edge in $\sim 100$ Myr.
Contrast this with the equivalent model in Fig. \ref{fig:tbc} with only viscosity, where it 
takes $\gtrsim 400$ Myr (grey solid curve) 
just to reach the B-C boundary. At first, the mass inflow rate is low, but, as the amount 
of material builds up in the inwardly extended region, the mass inflow rates from ML 
become substantial. In the absence of viscosity, the outer edge of the annulus  
moves inwards. The sharpness of the inner edge appears relatively {{unchanged}} 
because even though
the mass flux is high there due to high surface density, the drift velocities are much lower
compared with the lower surface density extended region. We show an additional
case in Fig. \ref{fig:MLBT} which includes both ML and BT (red curve) plotted at $2\cdot 10^7$ years. For
BT, we have used the more conservative choice of $Y=10^4$. The general effect as expected is even
more rapid induced radial drifts compared to ML alone away from the high-surface density
inner edge of the annulus.


\begin{figure}[t]
 \resizebox{\linewidth}{!}{%
 \includegraphics{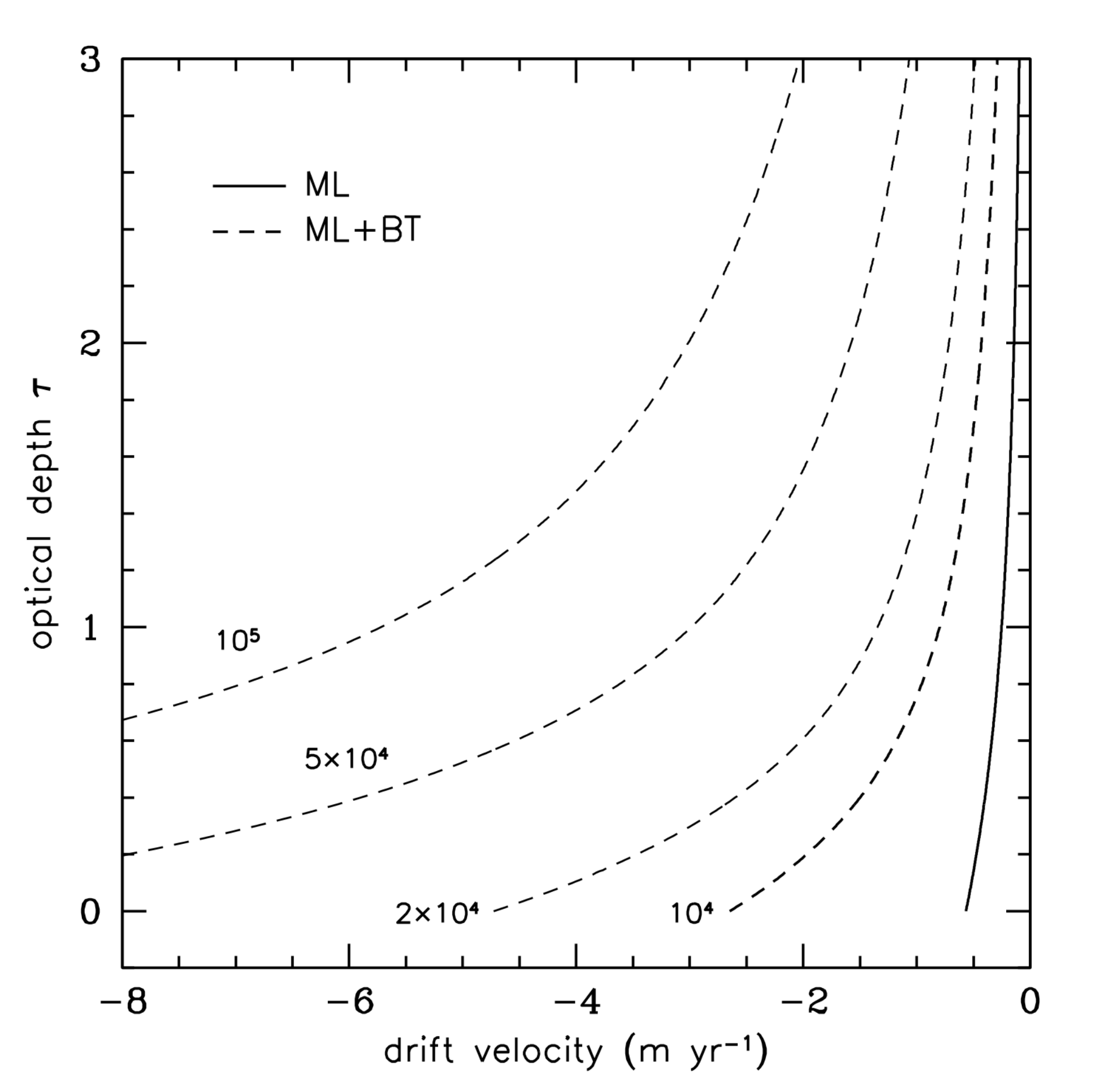}}
 \vspace{-0.3in}
\caption{Magnitude of the inward radial drift velocities induced by mass loading 
(solid curve) and the combination of mass loading and ballistic transport (dashed curves) as a 
function of the optical depth. Curves for ML+BT are labeled with the value of the impact yield $Y$
assumed for ballistic transport. For reference, the radial location at which these velocities
are computed is 120,000 km, the location of the initial inner boundary for the model shown 
in Fig. \ref{fig:MLBT}. The bold dashed curve shows the ML+BT inward drift velocity 
for our standard choice of $Y = 10^4$. 
}
\label{fig:vr}
\vspace{-0.1in}
\end{figure}  

This behavior can be better understood by looking at the magnitude of the induced radial drift
velocities due to ML and BT. In {\bf Figure \ref{fig:vr}}, we show
the magnitudes of $v_{\rm{r}}^{\rm{load}}$ and $v_{\rm{r}}^{\rm{load}}+v_{\rm{r}}^{\rm{ball}}$
as a function of the optical depth for the radial location of the inner edge of the annulus 
of Fig. \ref{fig:MLBT}. At high $\tau$, the inward drift velocities are relatively
low, but as the optical depth decreases, the drift velocities can increase by over an order of
magnitude. For ML, the inward drift of the inner edge occurs because as impacts decrease the specific angular momentum of the material there, it drifts into previously empty regions where impacts at the new radial location decrease its specific angular momentum further, and so on. For BT, owing to the 
overwhelmingly prograde nature of ejecta\footnote{It should be noted that a small fraction 
of the ejecta distribution from cratering impacts is retrograde \citep{CD90}, so that some material
may be thrown inwards, but it does not {{affect}} the global inward mass flow which is produced by the 
overwhelmingly dominant prograde ejecta. Preliminary results find that the EKB distribution is even more
prograde biased \citep{Est18}.}, they carry away from their point of ejection more specific angular 
momentum than is required for a circular orbit there, and arrive at locations further out with less 
specific angular momentum than needed for a circular orbit \citep{DE23}. So inward drifts are induced at 
both locations.

Moreover, the magnitude of the drift velocity for a given $\tau$ 
increases with decreasing $r$ due to the micrometeoroid impactor focusing by the planet, so 
for example the ML curve for $2\cdot 10^7$ years in Fig. \ref{fig:MLBT} is further along than what the ML drift velocity would predict from
Fig. \ref{fig:vr} (solid curve) which is the drift velocity at $r = 120,000$ km. 
These drifts become especially severe for BT if large 
values of the impact yield $Y$ are used in Eq. (\ref{equ:scriptr0}).
So, even though the mass influx $2\pi r\Sigma v_{\rm{r}}$ itself may be relatively low, the
speed at which ring material drifts inwards is fastest for the lowest optical depths and in fact
approaches a limiting value for $\tau \rightarrow 0$ as $2\pi r\Sigma v_{\rm{r}}
\rightarrow 0$. This explains why the region inside the
annulus is quickly populated with material. The clear implication is that any disk
subject to micrometeoroid bombardment will spread inward and lose material faster 
than by viscosity alone,
and that it will continue to do so even when viscosity becomes relatively unimportant. 
Both ML alone and ML+BT ensure that  
a radially extended region of lower surface density and optical depth, like the
C ring, will form inside an initial dense ring annulus\footnote{One can also imagine that a similar
effect may be happening in the inner A ring. Presumably, as Mimas opens a gap, material would seep into the Cassini division in a similar manner. In the context of our models, this requires that it was a relatively recent event \citep[see Sec. \ref{subsubsec:constantflux};][]{Bai19,Noy19}.}.

\begin{figure}[t]
 \resizebox{\linewidth}{!}{%
 \includegraphics{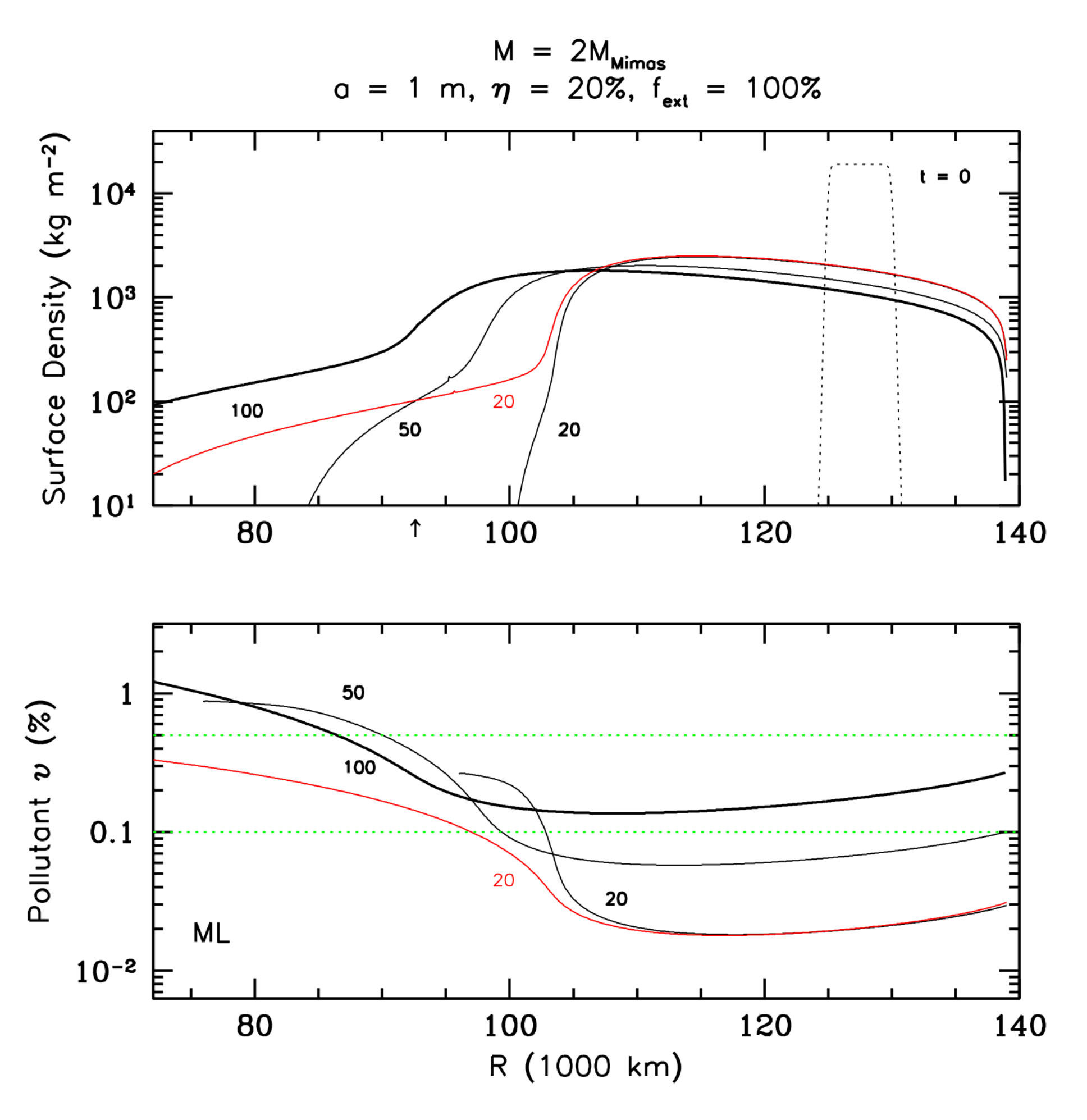}}
 \vspace{-0.3in}
\caption{Simulations that includes ML with viscosity (solid curves) for an initial 
2 M$_{\rm Mimas}$ ring with $a = 1$ m, $f_{\rm{ext}}=100$\%, $\eta = 20$\% and 
the EKB flux. \underline{upper panel}: Surface mass density plotted at 20, 50 and 100 Myr. 
Like Fig. \ref{fig:MLBT}, the formation of a C ring analog occurs relatively rapidly,
reaching the location of the inner C ring boundary (75,000 km) at $\sim 70-80$ Myr. The
red curve shows the evolution for ML+BT with $Y=10^4$ after 20 Myr.
\underline{lower panel}: Evolution of the volume fraction of pollutant. The lower optical C
ring analog darkens very quickly in both cases compared to the higher optical depth regions,
but eventually becomes diluted as icier material spills in from the denser regions. Even so, 
after 100 Myr, the volume fraction of pollutant is consistent with observational values.
Definitions are the same as in Fig. \ref{fig:1M_1m_LH}.}
\label{fig:MLBTmodel}
\vspace{-0.1in}
\end{figure}

In {\bf Figure \ref{fig:MLBTmodel}}, we add viscosity to the effects of ML (black curves) to the
evolution of a 2 M$_{\rm Mimas}$ initial disk (as in Fig. \ref{fig:2M_1m_LH}), plotted at 20, 50 and 
100 Myr. Here 
we choose $\eta = 20$\% rather than our nominal $\eta = 10$\% because of the relatively short time scales 
involved compared to the model with viscosity only (Table \ref{tab:models}). 
Comparing with Fig. \ref{fig:2M_1m_LH}, we see that the differences are quite pronounced compared
to when only viscosity is included. Strong viscosity in the high surface density regions
allows the ring to spread inwards and outwards, but after $\sim 20$ Myr we see a 
low surface density region forming which 
spreads to the current location of the inner edge 
of the C ring ($\sim 75,000$ km) in about $\sim 70-80$ Myr. After 100 Myr, mass is still being lost
beyond the outer edge of what would be the A ring. We also include an ML+BT model (red curve) plotted
at 20 Myr using $Y = 10^4$ which shows inward spreading about a factor of $\sim 5$ faster than ML
alone\footnote{Note that the error in the angular momentum conservation is only $\sim 2\%$ for this model.}. A longer evolution including BT would eventually cause the outer edge of the ring to
drift inwards as the magnitude of the induced inflow overwhelms the viscosity. This
effect is not entirely realistic because of the approximate nature of our inclusion
of BT.  When treated properly, we expect the mass inflow will restructure the rings to
conserve angular momentum. Until this is done, we cannot say definitively what the behavior at the
outer edge will be in the presence of BT \citep[though see Fig. 9.6,][also \citealt{Dur89}]{Est18}.

The evolution of the volume fraction of pollutant (bottom panel) shows that the contrast between 
the high and low optical depth regions is strong in both cases. This is well know 
from earlier pollution evolution studies where the low optical {{depth}} regions darken much more quickly 
than high optical depth regions despite the absorption rates being higher when $\tau$ is large 
\citep[][Sec. \ref{subsubsec:polevol}]{CE98,Est15}. Another well known
effect is plainly seen, namely that the {{radial drift}} 
of less polluted material {{from the higher optical depth regions}} into the low optical 
depth regions dilutes it and the volume fraction actually decreases. Comparing the ML
curves at 50 and 100 Myr shows that the rings are less polluted in the region between 80,000
and 100,000 km than they were at 20 Myr. This effect would be more pronounced 
in the ML+BT case. We see that, after 100 Myr, the evolution has reached a non-icy volume fraction 
in the higher surface density regions that {{is}} consistent with the observed values in the B and A rings.

Adding the effects of ML (and BT) indeed appears to form 
a low surface density (optical depth) region that extends from the inner edge of the high surface 
density (optical depth) edge to the inner boundary. This idea has been suggested previously 
\citep[see, for example,][]{ED11,DE23}, but has not until now been explicitly
demonstrated. The simulations we conduct here imply that the high-density inner
edge, such as the inner B ring edge, would continue to spread 
and increase the surface density of the lower density C ring. By virtue of the B ring's less polluted state,
this overly dilutes the non-icy material fraction in the C ring. We believe that this is not realistic 
and is also due to the limitations of our approximate treatment of BT in these simulations
which we now explain. 

An important effect in full BT simulations, where more than just the BT induced radial drifts are 
included, is that sharp edges like the inner B ring edge are sculpted and maintained by BT 
\citep{Dur92,Est15}. The B ring inner edge is maintained at its present width through a balance 
between viscosity and BT because of the prograde-biased ejecta distribution, even as the edge drifts 
inwards as a unit over longer timescales \citep{Lat12,Lat14b,Est18}. The rate of drift of the edge
is much slower than the rate at which ring material flows across it. A by-product of this
in detailed BT simulations is the formation of a linear ramp of moderate-to-low
optical depth that connects the B ring edge to the low optical depth C ring. The ramp is produced
by a small excess over direct mass effects of surface density changes caused by differential 
radial drift when integrated over a power-law ejecta speed distribution 
\citep[see Section 5a and Appendix B of][]{Dur92}. When pollution evolution is
combined with full BT, we find that the difference in pollution across the edge
is sustained for longer times \citep{Est15} than it is in our current more approximate
treatment. Capturing these effects thus requires full BT simulations in 
which case angular momentum will also be conserved as well.

Clearly, detailed modeling will be required in the future 
to delineate the formation and maintenance of the C ring at low optical depth. 
Our current simulations at least suggest that the C ring formed from the B ring and that it 
can feed mass into the regions close to the planet at a rate regulated 
by the current mass density in the C ring and by the mass flowing from the B ring into the 
C ring region. With more detailed BT, we expect that the inner edge of the high-density
region in Fig. \ref{fig:MLBTmodel} will remain sharp, and we speculate that this may allow 
for a mass inflow that achieves some sort of quasi-balance with the B ring inflow. 
Other effects, not currently included, such as ring rain, may also remove material from near 
the C ring/inner B ring edge.

\begin{figure}
 \resizebox{\linewidth}{!}{
 \includegraphics{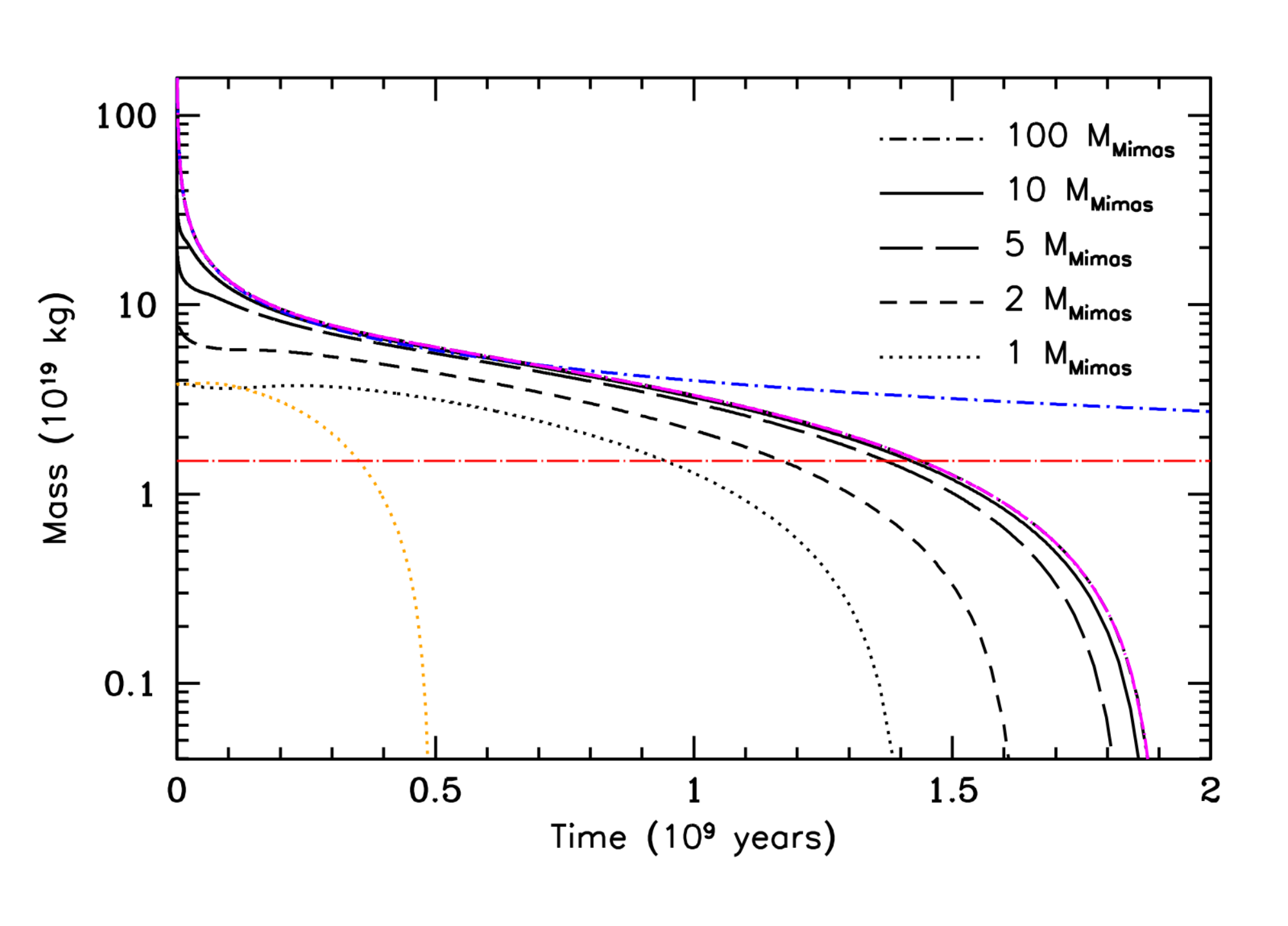}}
 \vspace{-0.3in}
\caption{Evolutions of rings with initial disk masses of $1-1000$ M$_{\rm Mimas}$ under the influence
of viscosity and ML. In these simulations, 
the EKB flux and $\phi = 0$ are used.
Simulations are stopped when the ring mass reaches $10^{-2}$ M$_{\rm Mimas}$. The rings no longer
have an asymptotic mass as time increases. Instead, with a given constant meteoroid influx, they have 
a finite lifetime, which has an asymptotic value as the initial ring mass increases. The
1000 M$_{\rm Mimas}$ curve, plotted in magenta, lies right on top of the 100 M$_{\rm Mimas}$ curve
and cannot be readily distinguished from it. The red dashed curve indicates the current mass of the 
rings, and the blue curve is the 100 M$_{\rm Mimas}$ model when only transport by viscosity is included, 
as in Fig \ref{fig:salmon}. We show one simulation for an initial mass of 1 M$_{\rm Mimas}$ (dotted 
orange curve) where we add BT (with $Y = 10^4$). This simulation suggests that BT decreases the ring 
lifetime by a significant factor.}
\label{fig:supdate}
\vspace{-0.1in}
\end{figure} 

\begin{figure}[t]
 \resizebox{\linewidth}{!}{
 \includegraphics{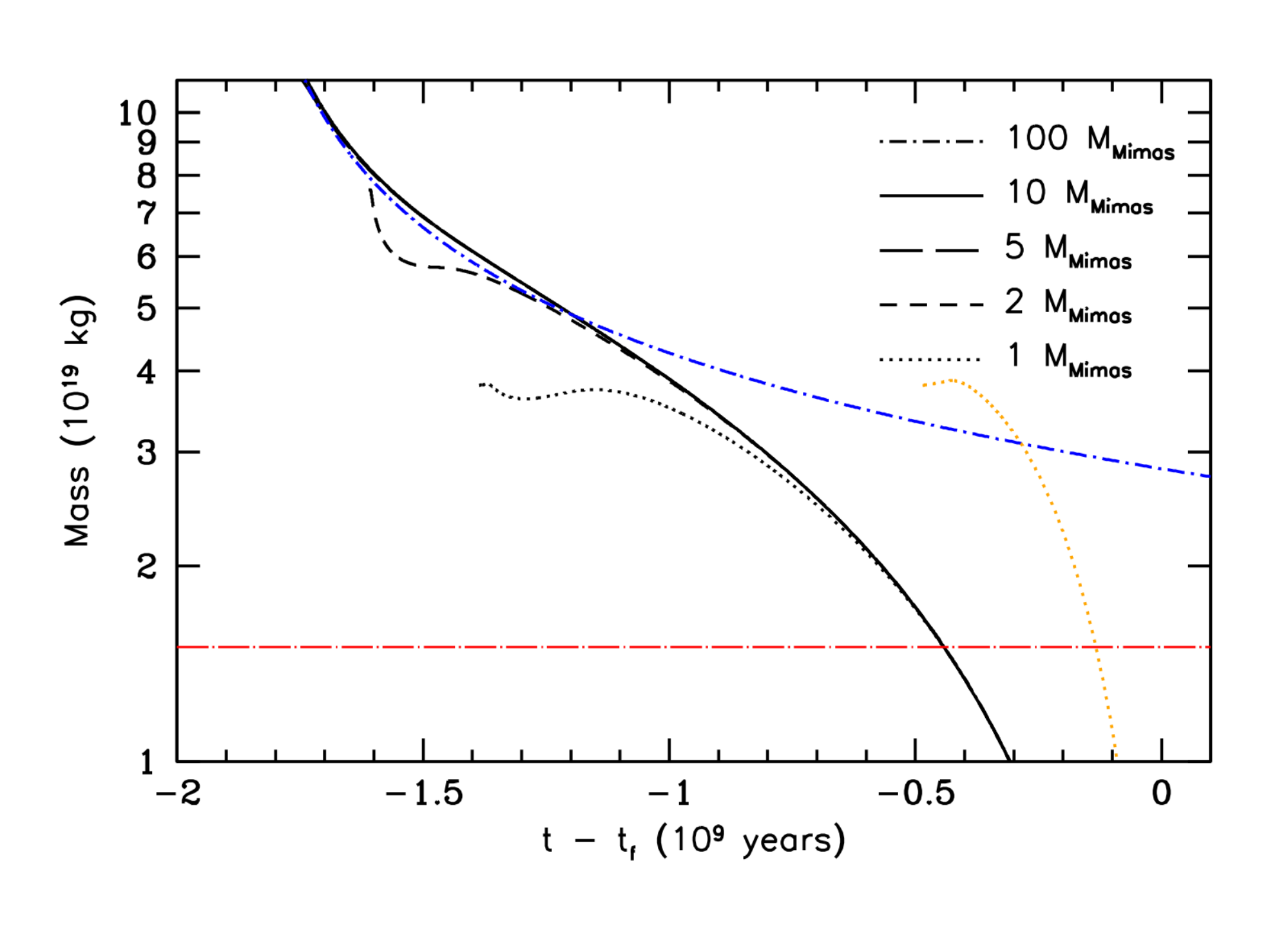}}
 \vspace{-0.3in}
\caption{{{A different visualization of Fig. \ref{fig:supdate} which demonstrates at what disk mass the rings transition from a viscosity dominated regime to one in which ML dominates. Here we plot time as $t - t_{\rm{f}}$, where $t_{\rm{f}}$ is the time at which the disk mass reaches $10^{-2}$ M$_{\rm{Mimas}}$. The blue curve is as before the viscosity-only 100 Mimas mass case. The ML curves begin to veer away from the latter curve when the disk mass is $\sim 6\cdot 10^{19}$ kg, or roughly 1.6 M$_{\rm{Mimas}}$. Using this format, it is more easily seen that all disk masses eventually follow the same trajectory evolving to smaller masses. The ML+BT case (orange curve) is also plotted which shows that these disk models for a given $Y$ will belong to a different family of curves.}} }
\label{fig:supdate_alt}
\vspace{-0.1in}
\end{figure}

\begin{figure}[t]
 \resizebox{\linewidth}{!}{%
 \includegraphics{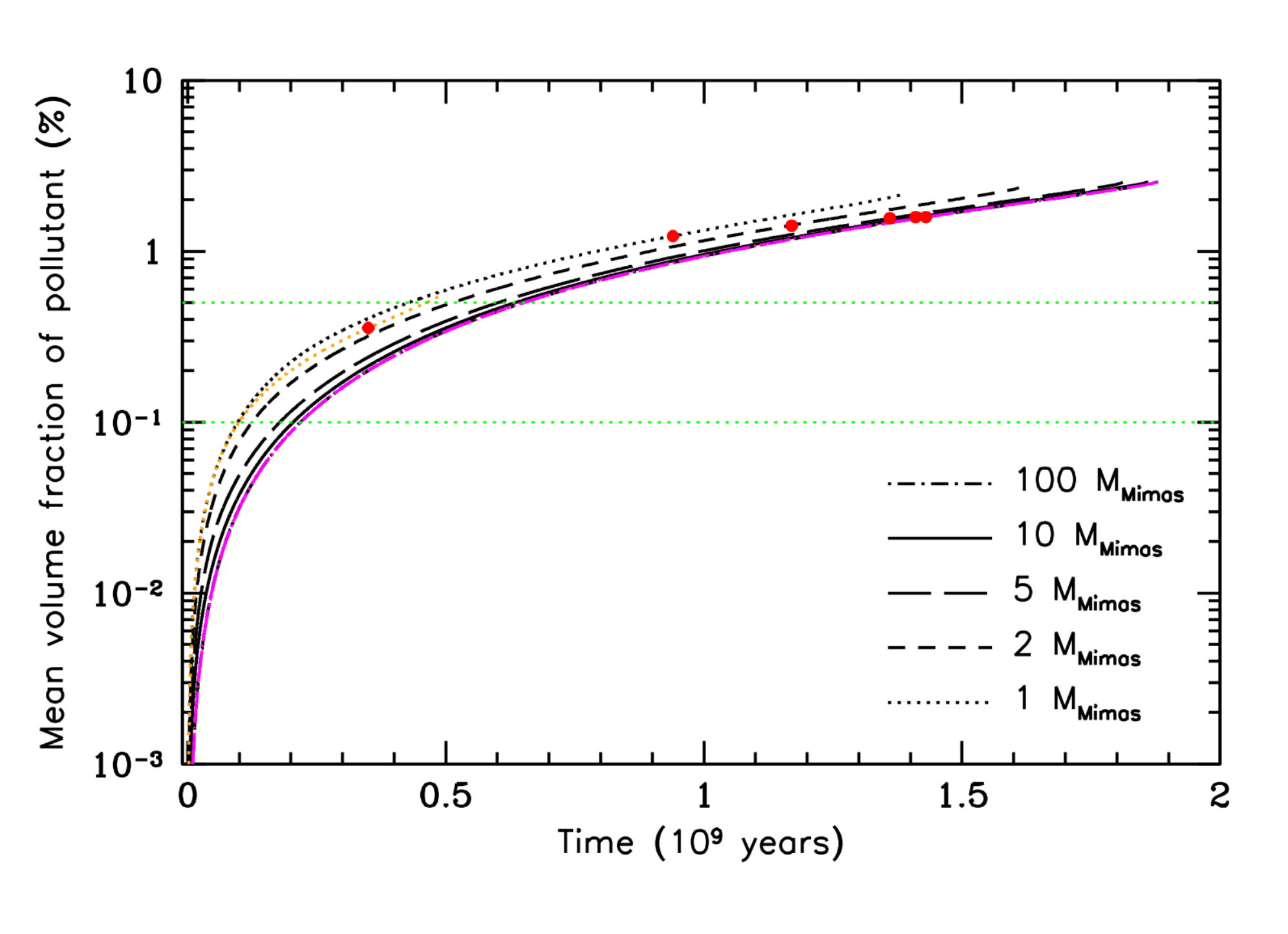}}
 \vspace{-0.3in}
\caption{The mean (mass weighted) volume fraction of pollutant as a function of time for initial disk masses 
from $1-1000$ M$_{\rm Mimas}$ evolving under the influence of viscosity and ML. In these simulations, 
the EKB flux, $\phi = 0$ and $\eta = 10\%$ are used. Simulations are stopped when the ring 
mass reaches $10^{-2}$ M$_{\rm Mimas}$. The green dashed lines mark the observed values of the non-icy
volume fractions in the current A and B rings. {{The red dots indicate where the Saturn's current ring mass is achieved}}. By 100 Myr the lowest masses already satisfy the 
observational constraints, and by 200 Myr all masses do. These simulations indicate that the level of 
accumulated pollutant is limited by the finite lifetimes. As in Fig. \ref{fig:supdate}, the 1000 M$_{\rm Mimas}$ (magenta curve) lies on top of the 100 M$_{\rm Mimas}$ curve. The dotted orange curve corresponds to the 1 M$_{\rm{Mimas}}$ case with BT ($Y=10^4$) included. The darkening is less due to the faster evolution, but {{shows that a model with BT can match both the current ring mass and current pollution level constraint simultaneously.}} 
Recall that our choice of $\eta$ is conservative. Increasing to 20\% would
shift these curves upwards by a factor of two.}
\label{fig:mupdate}
\vspace{-0.1in}
\end{figure}

One result that is strongly suggested by our simulations is that ML and BT will eventually dominate over viscosity in driving the rings' dynamical evolution.
In {\bf Figure \ref{fig:supdate}}, we plot the evolution of ring mass 
under the effect of viscosity and ML only for initial disk masses of 1, 2, 5, 10
and 100 M$_{\rm Mimas}$ (cf. Fig. \ref{fig:salmon}). For these models, we adopt the EKB flux and zero porosity, and have assumed the same initial conditions for the annulus as used by \citet[][see Sec. \ref{subsec:nummethod}]{Sal10}. 
The simulations were halted when the disk mass reached $10^{-2}$ 
M$_{\rm Mimas}$. The black curves show that the rings do not slowly converge asymptotically 
in mass over long times, as in the viscosity-only cases shown in Fig. \ref{fig:salmon}. For
comparison, the blue curve shows the viscosity-only case for 100 M$_{\rm Mimas}$. 
Instead, with ML included, the 100 M$_{\rm Mimas}$ 
disk has the longest but still finite lifetime of $\sim 1.88\cdot 10^9$ years. The lifetime of 
the 10 M$_{\rm Mimas}$ case is only slightly shorter. The lifetime then decreases more 
rapidly with mass, with the lifetime of the 1 M$_{\rm Mimas}$ case being $\sim 1.38\cdot 10^9$ years. In all cases, the current mass of the rings is achieved sooner than these times. For example, the 1 M$_{\rm{Mimas}}$ mass model reaches 0.4 M$_{\rm{Mimas}}$ in $\sim 900$ Myr. 

Instead of having an asymptotic final mass, we find that the rings have 
an asymptotic lifetime as the initial ring mass increases. To demonstrate this, the solid 
magenta curve is for a 1000 M$_{\rm Mimas}$ initial disk mass which is nearly indistinguishable from the 100 M$_{\rm Mimas}$ curve. This is not surprising because rapid viscous evolution leads to identical masses for these two cases after only a few million years\footnote{The independence of the result for increasingly large initial disk masses can be inferred from the work of \citet[][{{see also \citet{CC14}}}]{Sal10}.}. 
Eventually though, the dynamical effects of MB become the leading factor in determining the mass inflow into the planet\footnote{Note that with ML (or BT) {{positioning the initial annulus closer to the planet}} 
for the lower initial masses leads to shorter life times simply because mass loss to the planet occurs sooner.}. {{Once the rings fall below a threshold mass,}} 
their dynamical evolution transitions from a viscosity-dominated regime to one dominated by ML and BT. From Fig. \ref{fig:supdate}, the threshold mass when considering ML alone is {{$\sim 6\cdot 10^{19}$ kg, or $\sim 1.6$ M$_{\rm{Mimas}}$ (which occurs at $\sim 6.7\cdot 10^8$ years}} for {{for an initial mass}} $M \gtrsim 100$ M$_{\rm{Mimas}}$). {{This can be seen more clearly in Figure \ref{fig:supdate_alt} where we plot the same disk models in terms of $t - t_{\rm{f}}$, where $t_{\rm{f}}$ is the time that the disk reaches $10^{-2}$ M$_{\rm{Mimas}}$.}}

This result suggests, based on ML alone,
that a primordial massive ring would likely have become low mass {\postrevisionbf{one}} much earlier on in Solar
System history, especially if the flux was much higher in the past. We have also included
an ML+BT model (orange dotted curve) with $Y=10^4$ for comparison. Even though the decrease 
in lifetime (about a factor of three to $\sim 480$ Myr) is consistent with what we might 
expect, our approximate treatment does not allow us to take this number with confidence\footnote{The error in angular momentum conservation for this simulation is $\sim 22\%$.}. 
However, it is clear from Figs. \ref{fig:MLBT},
\ref{fig:vr}, and \ref{fig:MLBTmodel} that BT greatly accelerates the evolution in the
same sense as ML alone by increasing the radial inward drifts. If the yield is closer to
$10^5$, the timescale would decrease by another factor of $\sim 5$. {{Moreover, because BT is even more dominant than ML, the threshold mass below which the effects of MB will dominate the dynamics over viscosity will increase as well.}}

{\bf Figure \ref{fig:mupdate}} summarizes the mean volume fraction of pollutant accumulated 
as a function of time for the ML plus viscosity models shown in Fig. \ref{fig:supdate}. 
As in previous plots, the green dashed lines mark the range of observed volume fraction 
of non-icy material in the A and B rings \citep{Zha17a,Zha17b}. In these {{simulations}}, 
we have assumed $\eta = 10$\%. The lowest mass cases begin to satisfy the observational 
constraints by $\sim 100$ Myr, while all masses satisfy them by $\sim 200$ Myr. 
By this time, even the largest initial mass models have decreased to $\sim 2.5$ M$_{\rm Mimas}$. 
As in the previous figure, the 1000 M$_{\rm Mimas}$ case shown by the magenta curve 
lies right on top of the 100 M$_{\rm Mimas}$ case.  All these curves end 
as the rings reach $10^{-2}$ M$_{\rm{Mimas}}$, so for each mass there is a 
similar finite level of pollution achieved {\postrevisionbf{($\lesssim 3$\%,}}  
though see below). {{These models with ML alone though are darker than the observed A and B ring non-icy fractions once Saturn's current ring mass is achieved (red dots).}} The corresponding case for 1 Mimas mass with 
BT ($Y=10^4$) is shown by the orange dotted curve. {{The faster evolution}} 
further limits the level of pollution {{compared to ML}}, 
but {{demonstrates that the constraints of Saturn's current ring mass and the current level of population can be satisfied simultaneously when BT is included.}} 
Recall though that we use a very conservative value for $\eta$ which may not be appropriate for impacting micrometeoroids composed primarily of silicates. Increasing $\eta$ can offset the faster evolution; for example increasing to 20\% as we did in Fig. \ref{fig:MLBTmodel} would shift the volume fraction curves upwards by roughly a factor of two. 

Given these results, the similarity between the current ring mass 
\citep[][marked by the red dashed line]{Ies19} and the 
asymptotic mass derived for ancient rings only subject to viscosity \citep{Sal10}
is probably just a coincidence. The asymptotic ring mass simply does not apply
to a ring subject to the radial inward drifts caused by micrometeoroid bombardment.
The remaining ring lifetime, as derived from  
measurements of the current mass inflow into Saturn \citep{Hsu18,Wai18,Mit18,Odo19}
and as estimated from other recent theoretical treatments of mass loading and ballistic transport 
\citep{DE23}, is generally consistent with the trends as seen in Fig. \ref{fig:supdate}.

%

\section{Discussion}
\label{sec:conclusion}
 
 
\subsection{Simulations with Viscosity and Direct Deposition}
\label{sec:viscanddd}
\subsubsection{Ancient Rings}
\label{sec:ancconcl}

In our simulations of primordial/ancient rings, we examine two extremes for the amount of
pollutant that is retained from an impact, $\eta = 10$\% and $\eta=100$\% and find that, regardless of the 
initial ring mass, the rings always end up considerably darker than observed for the Cassini 
measured EKB flux. For the previously assumed cometary flux \citep{Gru85}, the current
pollution levels could be consistent with the observations only for the $\eta = 10$\% case. 
Overall, the volume fractions for the old cometary influx range from $\sim 0.6-6$\%, while they 
are $\sim 5-50$\% for the EKB flux for surface densities that would be
consistent with the current A and B rings. Since the EKB flux is a measured value, it would appear to rule out a primordial/ancient origin for the rings based on meteoroid deposition of pollutants alone.
The amount of pollutant acquired over the age of the Solar System for a given $\eta$ is roughly independent 
of the initial ring mass, except in the lowest initial mass cases, which acquire more. This is because 
the rings spend the majority of their lifetime at low masses where they are most susceptible to pollution 
by MB.

The rings also appear to evolve to an asymptotic mass \citep[as in][]{Sal10}, but this is not 
surprising since only viscosity is responsible for mass and angular momentum transport 
in these simulations. Massive rings evolve very quickly to a low mass state due to the high viscosity, 
and much of the disk mass is removed in short order by losses inwards to the planet or 
outwards beyond the Roche limit. This lost material
also carries away pollutant, with the amount lost typically $\sim 30 - 50$\% 
of what was delivered over the age of the Solar System.
This asymptotic mass is higher than the limit determined by
Salmon et al. ($\sim 0.4$ M$_{\rm Mimas}$) {{simply}} because the rings {{have accumulated a significant amount of additional mass due to direct deposition over the simulation time (Sec. \ref{subsec:longterm}).}} 
{{Because of the constant mass deposition rate}}, eventually {{over very long time scales the rings will evolve}}
to a state where the mass gained from MB and lost via viscous transport become {{similar}}. 
 
We also explore simulations where the flux at LHB was a factor of 10 larger under the reasonable
assumption that micrometeoroid flux was higher in the past. The flux is assumed to be ten times
higher until LHB and decays exponentially to its current value today. Both for the
cometary and EKB flux cases, the level of pollution accumulated is higher, a little more than 
a factor of two
in the cometary case, but a slightly smaller factor of $\sim 1.8$ for EKB. The difference is that
much more mass is deposited in the rings up to the LHB, giving them higher surface density compared to the nominal cases, and thus higher viscosity which leads to more rapid evolution. Even though the EKB case gains much more mass, the faster evolution means that the rate of material removal (and thus more pollutant) from the grid compared to the rate of mass deposition (the rate at which they are being polluted) is higher in the EKB case relative to the cometary one.

\subsubsection{Young Rings}
\label{sec:youngconcl}

Our next set of simulations start with a lower initial ring mass and only evolve for a few 100 Myr, now
using the observed 
value for the EKB flux, 
as given in \citet{Kem23}, in the conservative limit of 
$\eta = 10$\%. The mass influx rates of material falling into Saturn from the rings based on 
Cassini measurements during the last 22 orbits of the Grand Finale \citep{Wai18,Hsu18,Mit18}
imply that the rings were more massive in the recent past. For this reason, we chose to explore 
models with initial masses of $\sim 1-3$ M$_{\rm Mimas}$. 

All such simulations produce the observed volume fraction of pollutants in the A and B rings on the time scale of
$\sim 100-300$ Myr, with higher masses naturally taking somewhat longer to darken to the same state as lower 
masses. A higher $\eta$ will make darkening that much easier and would suggest even younger ages 
for the rings. For the given initial conditions, larger initial mass disks spread further inwards
maintaining a relatively sharp edge due to {{the viscosity being a strong function of $\Sigma$}}, 
but do not
reach the planet in the simulation times, while models with $\gtrsim 2$ M$_{\rm Mimas}$ can shed a Mimas 
mass or more outside the Roche limit to perhaps contribute to the formation of an icy moon.
We note that different initial conditions (in the location of the initial annulus) only lead to 
differences in how much mass is shed, not to the level of pollution.  

\subsection{Young Rings Including Mass Loading and Ballistic Transport}
\label{sec:mlandbt}

The inclusion of ML and BT produces radially inward drift velocities 
that increase both with decreasing optical depth, and with decreasing
$r$ because of gravitational focusing of the micrometeoroid flux by the planet. So the material 
at the lowest optical depths, which occur 
at the base of the initial ring annulus, drifts inward faster than the material at high optical depth. 
As a result, ML and, to even greater extent, BT quickly lead to {{the}} formation of a low surface 
density, low optical depth ring interior to the main ring annulus. The drifts are always inwards due to 
ML because the meteoroid mass decreases the specific orbital angular momentum of the ring material.
The drifts are also always inward due to BT because the ejecta from meteoroid impacts are
predominantly prograde. Both mechanisms naturally produce a structure resembling the C ring that extends 
down to the planet, where material is lost to the parent body much more rapidly 
than by viscosity alone.

Viscous simulations for an initial mass of 2 M$_{\rm Mimas}$ that include  either ML alone or both
ML and BT do show much more rapid evolution. {{The}} formation of a C ring is a robust feature. We choose 
a slightly higher $\eta=20$\% 
because of the shorter run times, but the darkening time remains roughly the same. The low optical depth region 
darkens very quickly relative to the higher optical region, but, as more and more mass fills the C ring region, 
the ring material becomes diluted by icier material flowing into it from the higher optical depth ring. 
Both of these effects are basic features of pollution under BT \citep[e.g., see][]{CE98}. 
However, the rapid rate at which bright material flows from high to low optical depth regions in these
simulations is exaggerated by the simplified treatment of BT in this paper. 
Important structural aspects of BT evolution (which incidentally are required for angular momentum conservation), 
such as maintenance of the B ring inner edge and the formation of the linear ramp at
its base \citep{Dur92,Lat14a,Lat14b,Est15}, are not properly captured by our approximations. 
Even though material drifts across the edge, the edge itself should maintain a steady
sharpness, even as it migrates inward at a reduced pace \citep{Lat12,Est18}.
We expect that in a full BT simulation, the C ring will still form but the edge between the low
and high optical depth regions will stay relatively sharp and not smear out 
indefinitely in width so that the region interior 
to it will remain at lower optical depths. This will be the subject of future modeling.
 
In the viscous evolution of a massive ring, the ring approaches an asymptotic mass over the age of the
Solar System  independent of its initial mass (see Fig \ref{fig:salmon}; see also \citealt{Sal10}).
As shown in Fig. \ref{fig:supdate}, our simulations demonstrate that a disk subject 
to micrometeoroid bombardment no longer has an asymptotic mass at long 
times, but instead, given a sustained meteoroid influx, has an asymptotic lifetime 
as the initial ring mass increases. 
In fact, when ML is included in a viscous evolution, there is no {{perceivable}} difference in the 
lifetime of rings with initial masses of 100 and 1000 M$_{\rm Mimas}$, whereas an initial 10 M$_{\rm Mimas}$ 
mass disk has a notably shorter lifetime, and the lifetime decreases systematically {{with decreasing mass}}. 
Including BT in addition to ML will undoubtedly make the lifetimes even shorter, 
up to more than an order of magnitude, depending on the assumed ejecta impact yield $Y$. However, we have 
less confidence in the quantitative accuracy of this last result because of the simplifying approximations 
made for BT used in this paper.

\subsection{Some Caveats and Speculations}
\label{sec:close}

\subsubsection{Ring End States}
\label{subsubsec:endstate}

The results of Sec. \ref{subsec:mlandbt} (Fig. \ref{fig:supdate}) indicate that when 
ring {{simulations include}} ML a primordial massive ring may have an asymptotic lifetime much less
than the age of the Solar System. Including BT significantly shortens this lifetime. However,
all our simulations including ML or BT are idealized in the sense that {\it all} material
is eventually ``lost''. More realistic models which account properly for conservation of angular momentum due to BT will likely leave some material behind in the disk \citep[see, e.g.,][]{Dur89,Est18}, albeit it at low optical depth/surface density where the effectiveness of mass and angular momentum transport due to ML and BT are lessened. For instance, a thin, low optical depth ringlet may be subject to MB, but the throw distances of ejecta (defined by their velocities) increasingly lie outside the radially thin ring so that the ejecta would return to their point of origin effectively neutralizing BT. Only ML would play a role.
External torques due to moons \citep[see][for more discussion on this issue]{DE23}, or shepherding of ring annuli by small moonlets might also influence the long term outcome. 
Moreover, such ringlet systems would be free to continue to darken from meteoroid bombardment beyond the values seen in Fig. \ref{fig:mupdate}. 
 
Our results thus lead us to speculate that the eventual ``end state'' of 
a massive ring
evolving under the combined effects of viscosity, mass loading, and ballistic transport may ultimately resemble something like the ring systems of Uranus and Neptune. 
This is in agreement with \citet{CC12} who first posited that these bodies may have initially had massive rings, but then could not explain why 
these planetary ring systems were well below their asymptotic mass. For instance, a similar calculation 
for Uranus as those shown in Fig. \ref{fig:salmon} yields an asymptotic mass of $\sim 0.06$ M$_{\rm{Mimas}}$,
or $\sim 2\cdot 10^{18}$ kg,  which is more than two orders of magnitude larger than the estimated 
mass of the Uranian rings \citep{GP87,ME02,CC03}. This mystery can apparently be solved, though, if one 
considers that MB can drive ring dynamics once viscosity is ineffective. 
 
\subsubsection{Micrometeoroid Flux}
\label{subsubsec:constantflux}

In all of our simulations, we have implicitly assumed that the micrometeoroid flux was constant at its Cassini measured value or was higher in the past. The latter is a reasonable assumption given that there would have been much more source material around up to the LHB \citep{Zah03}. Recall we found only one case, based on viscous evolution only for $\eta = 10$\% and the cometary flux, in which the rings current level of pollution can be compatible with ancient rings (Sec. \ref{subsec:longterm}). Given our findings from Sec. \ref{subsec:mlandbt}, and our discussion in Sec. \ref{subsubsec:endstate}, though, this result is not 
realistic. We cannot say for certain, though, that this has been the case in the more
recent past. One could consider that it may be possible that the flux may have been
weaker, or inactive for periods over that time potentially allowing for rings that could
be considerably older than what our models predict. 

We argue that this is unlikely based on a combination of our simulations here, and on
detailed ballistic transport calculations which offer an argument in favor of a continuous flux 
with only minor variation. Simulations show that structural evolution of the 
rings due to BT requires a roughly constant flux over 100's to 1000's of gross erosion times to form 
and/or maintain 
observed structures such as the inner B ring edge and the roughly linear ramp that 
connects the edge to the C ring \citep{Est15,Est18}. The 
gross erosion time $t_{\rm{G}} = \Sigma/Y\dot{\sigma}_{\rm{im}}$ is
the time it would take a ring region to completely erode away if all locally produced ejecta
were lost, and none were gained \citep{Dur84}. We can compare this to the pollution time to
darken the B ring ($\upsilon \sim 0.3$\%) to its current state 
\citep[see Sec. \ref{subsubsec:polevol},][]{Kem23,DE23}

\begin{equation}
    \Delta t_{\rm{pol}} \simeq \frac{\rho_{\rm{pol}}}{\rho} \upsilon \frac{Y}{\eta} t_{\rm{G}},
\end{equation}

\noindent
which for our range of $Y$ and $\eta$ gives $\sim 10^2 - 10^4$ t$_{G}$, in rough agreement with 
structural times {{mentioned above}}. 

The structural time for the inner B ring and C ring ramp though 
are not necessarily the same as the exposure or formation age of the rings. As has been shown 
previously \citep{Lat12,Lat14b,Est18}, the relative sharpness of the inner B ring edge can be 
maintained, even if it migrates inwards as a unit. This implies it would have formed 
further away from Saturn than its current radial location at $\sim 92,000$ km. It should be
noted that micrometeoroid bombardment and BT cannot {\it create} the inner B ring edge, only sculpt 
and maintain it against viscosity \citep[][Sec. \ref{subsec:mlandbt}]{Dur92,Est15,Est18}. 
Therefore, an initial relatively sharp inner edge is required which is 
facilitated in our models because the initial massive ring annulus has a sharp edge {{due to the viscosity's strong dependence on surface density}} 
(Sec. \ref{subsubsec:gensol} and \ref{subsec:longterm}). We 
saw that in our simulations as the annulus evolves, a C ring structure is formed {{emerging from the low optical depth foot of the annulus}}, but the inner edge
begins to smear out (Fig. \ref{fig:MLBTmodel}). Including the structural effects of BT will work 
to prevent this. If the
bombarding flux were to weaken or become inactive for any reason, however, BT's hold on
the inner B ring edge would be broken and the edge would spread due to viscosity, and BT would 
probably not be able to recreate it once active again. Since the inner B ring edge remains sharp, this would 
suggest that the flux has been relatively constant at least over time scales that are consistent 
with our young ring models.

A similar structure is present at the inner A ring edge. The linear ramp that 
connects the A ring to the Cassini Division has the same radial extent and slope as its B 
ring counterpart \citep[see Fig. 1,][]{Est15}. As implied above, ML and BT cannot create the 
Cassini Division, but it can maintain the A ring inner edge once it forms and develop the ramp feature. 
Its presence in the context of our models requires then that it must have formed relatively 
recently. 
Recent work suggests that it could have formed due to the rings' interaction with the inward migration 
of a Mimas moon as recently as $\sim 4$ Myr ago, and may eventually disappear in the next $\sim 40$ Myr \citep{Bai19,Noy19}. Such a scenario is consistent with our models, with the clear implication that 
the structural age of the inner A ring is younger than that of the inner B ring. We plan to 
incorporate such a scenario into our detailed 
models as part of our future study \citep{DE23}.

\subsubsection{Composition of Mass Inflow}
\label{subsubsec:composition}

During the Cassini Grand Finale, the CDA \citep{Hsu18}, which was sensitive to tens of nanometer-sized grains, 
measured a total flux arriving at Saturn of $\sim 320-1200$ kg s$^{-1}$ of which 70\% fell near the midplane 
and the remaining 30\% at ring rain latitudes. The non-icy mass fraction range from $\sim 8-30$\%, 
or $\upsilon \sim 3-12$\%, {{with the lower bound close to C ring composition}}. 
The depletion in water of these nanograins at the ring rain latitudes appears consistent with
the amount of charged water products needed to account for the latitudinal pattern of H$_3^+$ emission 
in Saturn's atmosphere \citep{Odo19} if
one were to reconstitute the CDA flux to C ring composition \citep{DE23}. Hsu et al. attribute these nanograins
to high-speed ejecta that spiral along magnetic field lines and mainly originate from the C ring ($\sim 50$\%),
but with a significant fraction from the B ring ($\sim 40$\%) where ring
composition can be much icier. So the total amount of depleted water may be higher, but likely still consistent with what is needed to account for the ring rain phenomenon. Other factors, such as the dissociation of water molecules and recombination contributing to the production of the ring atmosphere of O$_2$ may account for some of this loss \citep[see][and references therein]{Cuz09}.

On the other hand, the more than an order of magnitude larger inflow to Saturn claimed to be 
seen by INMS is only $\sim 25$\%
water by mass \citep{Wai18}, or equivalent to a volume fraction of $\upsilon \sim 50$\% of non-icy material, 
far larger than mean ring composition. It is important to note that, unlike CDA 
which is directly sampling the end point trajectories of the relatively small fraction of high-speed ejecta that can originate from all parts of the main rings, \citet{Wai18} 
infer that the INMS measured inflow mostly originates 
near the ring midplane close to the planet in an narrow equatorial band and 
requires a regular transfer of material from the
C ring to the D ring to maintain it. That is to say, in the context of ML and BT,
INMS is sampling the very end of the ``conveyor belt'' of the inflow produced by micrometeoroid impactors and the bulk of their ejecta distributions\footnote{The fraction of high-speed ejecta CDA measures accounts for only about $\sim 10^{-4}$ of the global ejecta production rate across the rings which is dominated by micron-size, lower-velocity particles \citep{Hsu18,DE23}.}.
Given that the mean C ring composition is $\sim 6$\% non-icy material by mass 
\citep[$\upsilon \sim 2$\%,][]{Zha17a}, this observation is puzzling.
It would imply that the total mass flow should be $\sim 6\cdot 10^4 - 5.6\cdot 10^5$ kg s$^{-1}$ if it were reconstituted to ring composition, considerably higher than observed. BT and ML could continue to provide the {\it observed} INMS mass inflow, but it would require perhaps unrealistically high ejecta {{yields of $Y\gtrsim 10^6$}} to account for $\sim 90$\% fraction
of the missing water, {\it and} the bulk of that ejecta cannot contribute to the angular momentum outflow, and thus the mass inflow.

This puzzling observation might lead to speculation that somehow micrometeoroid impacts 
are preferentially removing the non-icy component, but leaving the volatile behind 
suggesting the possibility that micrometeoroid bombardment and BT is cleaning the rings rather than polluting them. Ring ``cleaning'' has also been suggested by \citet{Cri19} as a possible explanation for the anomalous composition of the INMS inflow {{and ring rain, and that the rings may have started out much more non-icy than today \citep[see][]{CB21}}}. 
Assuming there were some global process for redistribution of the vaporized water, one would then have
to answer the question of how is it that the non-icy volume fraction near the planet ($\sim 50$\%) is 
so decidedly different from the main rings ($\sim 0.1-2$\%).
It is worth reiterating that the INMS equatorial inflow originates close to Saturn whereas the CDA measurement, which is consistent with ring composition, samples ejecta material from across the rings. The latter is thus not consistent with a ring cleaning mechanism.
Nevertheless, if ring cleaning were at work, then at the current observed rate, the rings would become pure ice in the next $\sim 10^5-10^6$ years. One
would also have to explain how this process can produce exactly the same distribution of non-icy 
material in the rings that traditional micrometeoroid bombardment and BT theory naturally predicts - low$-\tau$ regions like the C ring and Cassini Division are polluted faster, and are darker than high-$\tau$ regions like the A and B rings. 

Given these complications, we consider this scenario highly unlikely. Nevertheless, we 
can for the moment entertain the notion, and ask what implications it has for the rings. It would imply that the initial ring composition was much more non-icy, with a fraction perhaps 
similar to (and probably constrained by) the bulk composition of the Saturnian satellite system as a whole. This would not affect much the remaining lifetime we have calculated here because the mass inflow rates due to ML and BT would remain essentially unchanged. The age of the rings and their initial 
mass might then be constrained not by the time it takes for pure ice rings to evolve to their current state, but evolving the rings backwards in time to a compositional state similar to the mid-sized moons. Indeed, 
assuming a mass loss rate to the planet of $\sim 10^3-10^4$ kg s$^{-1}$ comparable to
the INMS inflow, $\sim 75$\% of which is 
non-icy material, and assuming that the liberated water fraction is somehow 
recondensed over their radial extent, it 
would suggest that the rings had a composition with mass fraction of non-icy material between $\sim 30-70$\%, 
and with initial masses of $\sim 1-3$ M$_{\rm Mimas}$, no more than a few $100$ Myr ago, not unlike the conclusions from our more standard pollution and evolution models. These times would be even shorter using the observed INMS rates, but the initial masses are also likely larger due to some mass loss 
due to viscous evolution earlier on. 

We believe a much simpler explanation exists for the anomalous composition of the INMS inflow that does not 
require that we invoke an uncomfortably high ejecta yield $Y$ or create a missing water problem.
\citet{Zha17a} found that in order to match both the thermal emission and the anomalous opacity in the C ring 
between $\sim 78000-87000$ km, there would need to be a large amount of dense, non-icy material embedded in 
the cores of the larger icy-mantled ring particles. 
They estimated the additional ring mass in this so-called ``rubble belt'' to be on the order of $10^{17}$ kg, or a mass fraction as high as $\sim 80$\%, 
which is similar to the non-icy fraction of the equatorial inflow observed by INMS \citep{Wai18}. It may be then that INMS is seeing evidence for this reservoir of non-icy material, which 
we posit has continued to make its way inwards as part of the inflow and is now manifest in the innermost 
ring regions\footnote{\citet{Hed13} found that the concentration of contaminant in the C ring seems to steadily increase as one approaches the planet in their analyses using Cassini VIMS data.}. How this embedded material is being liberated remains to be studied. For instance, material closer to the planet may be more easily ground down due to collisions. Moreover, micrometeoroid impact velocities are highest close to the planet which gives higher probability for smaller ring particles to be disrupted. Such material would naturally explain why the current makeup of the inflow is not representative
of the rings as a whole. 
We will consider this, as well
as the cleaning scenario above as the subject of future work that involve detailed BT simulations of
ring evolution.

\subsubsection{Implications for Ring Origin Scenarios}
\label{subsubsec:originscenarios}
 
All giant planets have ring systems, but only Jupiter lacks any substantial rings, ringlets or arcs. The reason for this is not clear, but one 
notable architectural difference between Jupiter's and the rest of the giant planets' satellite systems are the family of mid-sized moons deep within the planet's potential well, where stochastic processes may be the rule rather than deterministic ones \citep[e.g., see Fig. 1,][]{ME03a}\footnote{Neptune likely had a mid-sized system earlier in its history. It was shown by \citet[][see also \citealt{Mck84}]{Gol89} that Triton's capture would have broken up or scattered any pre-existing moons between $\sim 5-100$ planetary radii with a collision probability near unity.}. The latter may be a prerequisite for ring systems. If these planets had primordial or ancient rings, we have speculated above that they would look highly evolved (Sec. \ref{subsubsec:endstate}). 


However the simulations in this paper, coupled with Cassini observations, appear to argue
against Saturn's {\it current} rings being primordial. This does not
rule out that Saturn had rings previously, perhaps initially massive, that may have led to the
formation of the mid-sized moons \citep{CC12}. It would simply say that {\it those} rings would long have dissipated. \citet[][{{see also \citealt{CB21}}}]{Cri19} have recently argued that the rings
of Saturn could be old, but appear to look young, so that their formation and exposure
age are not the same thing. But, these authors do not take into account that ML and BT can drive the rings'
dynamical evolution over viscous spreading as we have done here. 
The effects of ML and BT do not allow the 
formation and exposure ages to be decoupled, even if the flux has varied in the past. {\it Any} contribution 
from an extrinsic micrometeoroid flux over time will influence the dynamics of the rings. Based on 
the observations, Saturn's rings appear to be quite young, no more than a few $100$ Myr, and we have argued in this paper that their formation age and exposure age are most likely similar. If they are not, then we may  
need to revise our current understanding of the regular structures in the rings that detailed models 
tell us are due to ML and BT (Sec. \ref{subsubsec:constantflux}).

This leaves open the question of how its rings could have formed so recently and what this implies about the origin and recent evolution of Saturn's inner icy moons.
Given the very low probability of a heliocentric interloper of sufficient size being available to break up a body, or by itself to form a ring of several Mimas masses, a suggestion might be that some event \textit{within} the Saturn {{system}} itself may have served as the catalyst for recent ring formation. 
As mentioned above, families of mid-sized moons 
deep within the potential well of their parent giant planet are more prone to stochastic 
processes, perhaps always teetering precariously on some saddle point just 
requiring a little ``push'' to set things awry.

Recently, \citet{Cuk16} advocated a recent dynamical instability within the inner Saturnian system which was initiated when an outward evolving moon entered an
evection resonance in which the moon's orbit precession rate about
Saturn becomes commensurate with Saturn's orbital motion. The resulting instability may have led to the destabilization, disruption and reaccretion of the system $\sim 100$ Myr ago, similar to the estimated age of the rings. In this picture, collisions between, and fragmentation of {{differentiated}} icy moons progressively closer to the planet means that the rings may even be composed of fragments from multiple satellites. 
Such a scenario is only beginning to be studied in proper detail \citep[e.g.,][]{Teo23}.

 
\section{Main Conclusions}
\label{sec:mainconclusions}

%

We have conducted numerical simulations of an evolving ring in order to determine the initial
conditions consistent with
the set of key Cassini observations (Table \ref{tab:observe}), which, 
taken together, constrain the age of Saturn's rings 
to be $\lesssim$ a few 100 Myr. These key observations include the
ring mass \citep{Ies19}, the volume fraction of non-icy pollutants in the rings \citep{Zha17a,Zha17b},
and the extrinsic micrometeoroid flux at Saturn \citep{Kem23}. The observations also
constrain the remaining lifetime of the rings to be of the same order as their age (or less) due to 
mass loss to the planet \citep{Hsu18,Wai18,Odo19,DE23}. For these simulations, we added the effects of
micrometeoroid bombardment to a 1-D viscous evolution model similar to that of  \citet{Sal10},
with the same realistic viscosity treatment that accounts for the enhancement of 
the viscosity by gravitational instabilities \citep{Dai01}, and we followed the evolution of
the volume fraction of accumulated pollutant due to micrometeoroid deposition for primordial/ancient ring 
origin scenarios and for cases where the rings have lower mass and originate more recently.
We then introduced into the simulations the mass inflow in the rings due to mass loading 
by the meteoroids and, in a simplified way \citep{DE23}, the inflow caused by the ballistic 
transport of meteoroid impact ejecta. 

The implications of our results for {{the}} ring evolution and ring lifetime
are profound:
 
1) Primordial or ancient ring models, where the rings persist over time spans
similar to the age of the Solar System, are excluded on the basis of
micrometeoroid deposition and pollution alone, except for the case 
of a very low $\eta$ and a constant value of micrometeoroid influx for meteoroids 
of cometary origin. Observations show that cometary micrometeoroids are not 
currently the dominant impactor population. Evolutions using the observed EKB
micrometeoroid population produce unacceptably high levels of ring pollution.

2) When one uses the currently measured micrometeoroid influx rate and the 
observationally determined EKB nature of the impactor population, the current 
state of ring pollution suggests a relatively recent origin, up to a few 100 Myr
from an original annulus of debris with a mass $\sim$ 1 to 3 M$_{\rm Mimas}$. 

3) Inclusion of the radial mass inflow due to micrometeoroid bombardment
\citep{DE23} induced by mass loading and ballistic transport  has two important 
results: (a) A lower optical 
depth region inevitably forms interior to an initial high density annulus of debris. 
This occurs because the inflow speeds induced by mass loading and by
ballistic transport are larger for lower optical depths, so the base of the
inner edge of an initial annulus spreads inward much faster than the
top of the edge. A structure resembling the C ring forms in only a matter of 
tens of millions of years. (b) The inward drifts induced by mass loading and 
ballistic transport cause a dense ring to have a relatively short lifetime 
once these processes {{become}} dominant over viscous spreading and drive the rings' dynamical evolution. Instead of 
an asymptotic final mass at a Solar System lifetime for all initial ring masses,
as in the case for pure viscous evolution, mass loading alone enforces a finite 
asymptotic upper limit on ring lifetime as {{the}} initial ring mass increases. 
This asymptotic lifetime is $\lesssim 2$ Gyr 
for very large initial masses. For smaller initial masses, the lifetimes can be 
much smaller, only about $\sim$ 1.4 Gyr for 1 M$_{\rm Mimas}$, {{while}} the same ring reaches the currently observed Saturn's rings mass in $\sim 900$ Myr. When we include
our approximate treatment of ballistic transport, the lifetimes become a factor
of several up to over an order of magnitude 
shorter, depending on the assumed impact yield $Y$.

4) Arguments \citep[see, for example,][]{DE23} now converge from several directions to 
indicate that Saturn's rings are quite young ($\sim$ few 100 Myr) and have a similarly 
short future lifetime as a dense ring system. This leaves us in search of a viable formation 
mechanism.

5) Ballistic transport is clearly a dominant mechanism for the global evolutions of dense 
rings once viscosity weakens. This paper used an approximate treatment based on the 
steady-state model of \citet{DE23}. Global {{simulations with an accurate}} 
treatment of 
ballistic transport are needed to understand the full implications of micrometeoroid 
bombardment for {{the global ring evolution}} 
and fate.

We speculate, given the implications of the results of this work, 
that even an initially massive ring subject to meteoroid bombardment over very long time scales may eventually become highly polluted and reach a low density, tenuous state similar to the ring systems of Uranus and Neptune where mass loading 
and ballistic transport are no longer efficient evolution mechanisms. This is in agreement with
\citet{CC12} who first suggested that these ice giants may have initially had massive rings {{from which their regular moons formed}}.


\vspace{0.0in}
\begin{center}
{\bf Acknowledgements}
\end{center}
\vspace{0.0in}

We thank Jeff Cuzzi for many useful insights regarding Cassini observations, and their
implications for the work presented here. We thank Kevin Zahnle, Dale Cruikshank, and Matt Hedman for useful discussions, and especially those with Sebastien Charnoz which helped to inspire this work. We also thank
Pierre-Yves Longaretti, and an anonymous reviewer for their detailed and thoughtful reviews that have improved the exposition of this paper greatly. This work was supported by a grant from NASA's Cassini Data Analysis Program (PRE).
 



\vspace{0.2in}
\bibliographystyle{model2-names.bst}\biboptions{authoryear}
\bibliography{my.bib}







\end{document}